\documentclass{aa}
\usepackage{txfonts}
\usepackage{array}
\usepackage[dvips]{graphicx}
\usepackage{natbib}
\usepackage{longtable}
\bibpunct{(}{)}{;}{a}{}{,}
\def\vec#1{\ensuremath{\mathchoice{\mbox{\boldmath$\displaystyle#1$}}
{\mbox{\boldmath$\textstyle#1$}}
{\mbox{\boldmath$\scriptstyle#1$}}
{\mbox{\boldmath$\scriptscriptstyle#1$}}}}
\def\tens#1{\ensuremath{\mathsf{#1}}}

\begin{document}

\title{Parallaxes and proper motions for 20 open clusters as based on the new Hipparcos catalogue}

\author{Floor van Leeuwen}
\institute{Institute of Astronomy, Madingley Road, Cambridge, UK}

\date{Received 19 Nov. 2008  / Accepted 2 Feb. 2009 }

\abstract{A new reduction of the astrometric data as produced by the Hipparcos
mission has been published, claiming that the accuracies for nearly all stars brighter than magnitude $\mathrm{Hp}=8$ are improved, by up to a factor 4,
compared to the original catalogue. As correlations between the underlying abscissa residuals have also been reduced by more than an order of 
magnitude to an insignificant level, our ability to determine reliable 
parallaxes and proper motions for open clusters should be improved.}
{The new Hipparcos astrometric catalogue is used to derive mean parallax and 
proper motion estimates for 20 open clusters. The HR-diagrams of the nearest 
clusters are compared and combined to provide future input to sets of observational isochrones.} 
{Three different methods are applied, according to the proximity 
of the cluster, to compensate, where needed, for projection effects, spread in position along the line of sight, and the internal velocity dispersion of the cluster stars.}
{The new parallaxes have accuracies between 2 and 2.5 times higher than 
what had been derived from the original Hipparcos catalogue. At least two to three groups of clusters, mostly of similar ages, are observed to each occupy their own specific space in the HR diagram. A significant discrepancy in distance moduli from those obtained with isochrone-based main-sequence fitting remains, in particular for one of these groups, containing the Pleiades, NGC~2516, and Blanco~1. The difference in absolute magnitudes between this group and one containing the Hyades and Praesepe clusters appears to be correlated with systematic differences in the Str{\"o}mgren $\Delta c_0$ index between those groups. The same dependency has been known to exist for a long time, and is again confirmed by the Hipparcos data, in variations in absolute magnitudes for field stars of the same effective temperature.}
{The positions of the cluster HR diagrams are consistent within different groups of clusters shown for example by the near-perfect alignment of the sequences for the Hyades and Praesepe, for Coma~Ber and UMa, and for the Pleiades, NGC~2516, and Blanco~1. The groups are mutually consistent when systematic differences in $\Delta c_0$ are taken into account, where the effect of these differences on the absolute magnitudes has been calibrated using field-star observations.}


\titlerunning{Parallaxes and proper motions for 20 open clusters}
\maketitle

\section{Introduction}

The data accumulated by the Hipparcos mission allowed for the determination of large numbers of accurate absolute-parallax measurements for nearby stars \citep{esa97,perry97L,fvl07Val}. The new reduction \citep{fvl07} (\textsc{VL7B} from hereon), for example, provides some 27~000 stars with parallaxes known to better than 10~per~cent, and over 10~000 known to better than 5~per~cent. One of the many things learnt from this mission is the relatively large intrinsic spread in luminosities, even among stars which appear very similar in spectral type and photometric colour indices. An examination of G8V and K0V stars, for example, showed an intrinsic dispersion of 0.4 in absolute magnitudes for stars of the same colour index $\mathrm{B-V}$ (\textsc{VL7B}). A smaller, but still very significant, level of spread in absolute magnitudes is also observed for stars which were selected by \citet{nicol81} to be very similar in colour indices within the Geneva photometric system, the so-called photometric boxes (\textsc{VL7B}). Such a spread contrasts markedly with the very narrow main sequences observed for the best-studied and nearest open clusters, the Hyades \citep{perry98,mdl02}, Pleiades \citep{mitchel57,fvl86} and Praesepe \citep{crawford69}. The ability to superimpose the HR diagrams of such clusters on the HR diagram of the nearby stars, using parallax-based distance moduli rather than isochrone fitting, could provide quantitative information on some of the dependencies that cause the observed spread among the absolute magnitudes of apparently very similar field stars. Predictions on those dependencies are available from theoretical isochrones as based on a combination of models of stellar structure, stellar evolution and stellar atmospheres. However, observational evidence is required to test and eventually validate such models. The possibility of doing so has been hampered by the difficulty to determine with sufficient accuracy the parallaxes of star clusters. Even with the release of the Hipparcos data in 1997, some questions were left, most noticeably because of correlations between the errors on the underlying data from which the astrometric parameters for the clusters had been obtained. Doubts have in particular been raised on the effect that small areas of the sky with concentrations of bright stars may have locally on the measured parallaxes.

The new Hipparcos reduction has improved this situation in two ways: the 
error-correlation level on the underlying data has been reduced to an 
insignificant level, and the formal accuracies for most stars, and in 
particular the brighter stars, have been significantly improved. Both
of these improvements have been made possible through a detailed study of the
satellite dynamics and incorporating the findings of that study in the 
reconstruction of the along-scan attitude, which provides the reference frame 
for all the Hipparcos astrometric measurements. A potential weakening of the
overall rigidity of the reconstructed catalogue. which may be caused by 
small areas with many bright stars, has also been dealt with. This so-called 
``connectivity'' issue has been explained further in \textsc{VL7B} and 
\citet{fvl07Val} (\textsc{VL7V} from hereupon). Given these improvements in both formal errors and overall reliability of the data, a renewed analysis of cluster 
parallaxes seemed very needed, not least as an apparent discrepancy for the 
Pleiades parallax as determined from the Hipparcos data has been used
on several occasions to allegedly demonstrate errors in that catalogue
\citep{pinso98,soder98,pan04,zwahl04,southw05,soder05}. The reasons used in
those papers to explain the Hipparcos results should be unreliable for the 
Pleiades cluster, such as error correlations and poor connectivity, have been eliminated by the new reduction (\textsc{VL7B}, \textsc{VL7V}). 

For the open clusters studied here we make a few basic assumptions. The
first is that all cluster-member stars share the same space motion, have
the same age and originate from the same chemical composition ``cloud''. 
The stars are bound to the cluster, which means they will have a small 
internal dispersion in their velocities, decreasing towards the halo 
of the cluster. However, in particular the older clusters also tend to 
be surrounded by a wider halo of escaped former cluster members, which
can complicate membership determinations. Furthermore, some of the
youngest clusters can not easily be distinguished from fairly accidental
concentrations in OB associations, and may not always be bound. However,
in all cases the assumption of a shared overall velocity and distance 
has been applied.

The determination of astrometric parameters for clusters poses different 
problems as well as possibilities depending on the distance of the cluster.
For the nearest cluster, the Hyades, the group is effectively 
three-dimensionally resolved, while its proximity causes it to cover
a relatively large area (about 35 degrees diameter) on the sky. This 
creates the possibility of using the proper motions of individual cluster 
members to resolve the cluster along the line of sight. Intermediate-distance
clusters can not be resolved this way, but astrometric studies still need to take into account projection effects on the proper motions, and a dispersion along the line of sight in the parallaxes of the individual cluster members. In
defining the membership and determining the mean proper motion, the internal
velocity dispersion also has to be taken into account, though this will 
often not be available as an independent measurement. Beyond about 250~pc, the
dispersion of the parallaxes resulting from the finite size of the cluster
becomes small with respect to formal errors on the measured parallaxes of the
individual stars, and the same applies to the proper motion accuracies and
the contribution from the internal velocity dispersion. The stars in these
clusters can be treated as all having effectively the same parallax and proper motion, which can then be derived directly from a combined solution of the underlying data. This is the combined-abscissae method that was first presented by \citet{vLDWE}, and applied for all clusters by \citet{fvl99a} and 
\citet{robic99}. In those studies, however, the emphasis was on compensating
for the error correlations in the underlying data, as well as in the 
proper combination of the two data streams that were used to create the old
catalogue. These complications no longer apply when using of the new 
catalogue to derive astrometric parameters for open clusters.

This paper has been organized as follows. Section~\ref{sec:data} presents a
summary of the data, with particular emphasis on its use in deriving 
star-cluster astrometric parameters. Section~\ref{sec:makarov} reviews briefly attempts made by \citet{makar02,makar03} of applying corrections to the original Hipparcos astrometric data in the context of cluster-parallax determinations. Section~\ref{sec:techn} presents the different techniques applied to clusters at different distances. Section~\ref{sec:groundbased} presents an overview of other, isochrone independent distance modulus determinations for open clusters, and in particular introduces Hipparcos data to determine accuracies of these methods. This is followed by Section~\ref{sec:clust250}, where the results for individual clusters within 250~pc are presented, and Section~\ref{sec:clust500} for clusters between 250 and 500~pc. In Section~\ref{sec:disc} the new set of distance moduli as derived from the parallaxes is compared with a range of earlier determinations, and the HR-diagram positions of cluster main sequences as obtained through ground-based photometry are compared. Section~\ref{sec:spacevel} briefly summarizes the space velocities of the clusters as derived from the new parallax and proper motion data. Finally, Section~\ref{sec:concl} presents conclusions from the current study. 
 
\section{The Hipparcos data} 
\label{sec:data}

This section provides a summary of those aspects of the Hipparcos data that
are of direct relevance to determining astrometric parameters for open 
clusters. For a more detailed description of the reductions that led to the new catalogue the reader is referred to \textsc{VL7B}, while \textsc{VL7V} provides a report on the validation tests applied to the data in the new catalogue.

The Hipparcos data originate from two fields of view, separated by a basic angle of 58 degrees, approximately scanning the sky along great circles. The transiting images seen through the two fields of view were projected on the same focal plane. The Hipparcos astrometric data have been derived from transit times over a modulating grid in this focal plane. Through the reconstruction of the along-scan attitude, these transit times are translated to scan phases. In the original reductions these scan phases are projected and measured on a reference great circle, with one such circle defined for each orbit (10.6~h period) of the satellite. The abscissae provided thus refer to a mean over all observations of a star over that period of time. These observations will therefore generally contain the combined information from between one and 8 transits of the field of view. In the new reduction the abscissae provided refer to individual field of view transits and are measured relative to the instantaneous scan direction. This allows for an improved recognition of faulty measurements, which usually are specific to just one field-of-view transit, and the ability to correct for small-scale grid distortions as a function of the across-scan coordinate.

With respect to the data presented in the first catalogue \citep{esa97}, a number of very significant changes and overall improvements have been made. All improvements stem form the way the along-scan attitude is modelled, and how dynamic peculiarities of the satellite motion have been taken into account. There were two peculiarities in the satellite motion that were known but the
extent and effect of which had been underestimated in the original reductions:
(1) hits by very small particles, causing rotation-rate discontinuities, and 
(2) discrete satellite-non-rigidity events due to temperature changes affecting 
at least one of the solar panel mountings, causing so-called ``scan-phase 
discontinuities''. In the original reductions by the two consortia, FAST 
\citep{koval92} and NDAC \citep{lindeg92}, these events were left mostly unrecognized and effectively smoothed over. This caused locally strongly correlated errors in the reference frame as created by the along-scan attitude reconstruction, as well as a non-Gaussian noise added to the abscissa measurements. The consequences of this ``smoothing over'' were made worse still by the use of the great-circle reduction method \citep{vdmar88,vdmar92}, which referred all measurements over a 10.6~hour period to projections on a single reference great circle. This was initially needed in order to deal with the relatively large errors (compared to the Hipparcos target accuracies) on the \textit{a priori} astrometry provided by the Hipparcos Input Catalogue \citep{turon92,esa92}. The great-circle reduction process caused multiple replication of local attitude errors. The effects of a local scan-phase discontinuity could therefore be felt at many different positions on the reference great circle, all separated by an integer number times the basic angle between the two fields of view. A further and more detailed analysis of these effects can be found in \citet{vLDWE}, \citet{fvl05a} and \textsc{VL7B}.

\begin{figure}
\includegraphics[width=8.9cm]{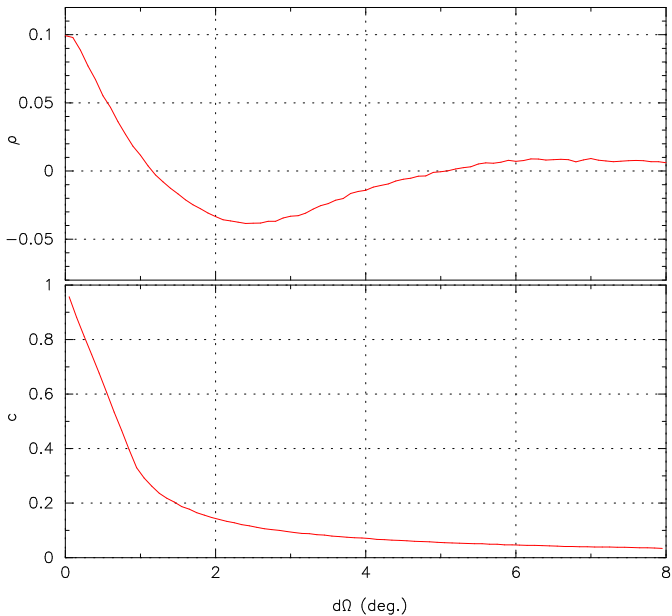}
\caption[]{Top: Abscissa-error correlation level ($\rho$) for all field-transit 
abscissa residuals (FTARs) with formal errors below 6 mas, as a function of 
separation along the scan direction. Bottom: Coincidence fraction
$c$ of field-transit measurements as a function of separation on the sky.
The coincidence data are based on a random selection covering 10~per~cent of 
all single stars in the new catalogue. (from \textsc{VL7V})}
\label{fig:absc_frm_corr}
\end{figure}
\begin{figure}
\includegraphics[width=8.9cm]{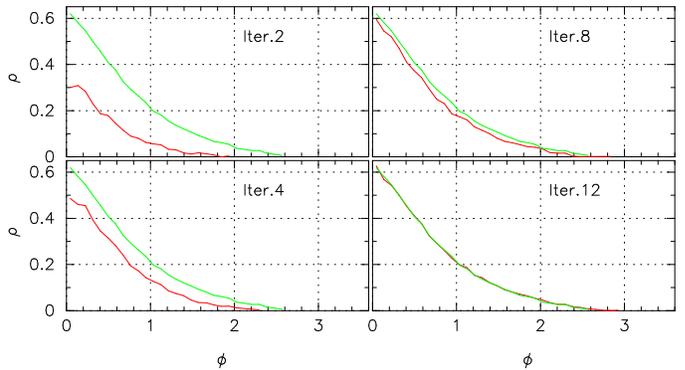}
\caption[Parallax update correlations]{Correlations ($\rho$) in the 
differences between the parallax determinations in the old and new solutions
as a function of stellar separation on the sky ($\phi$, in degrees). The figures show how initially the catalogue still tries to fit to the original 1997 catalogue, and correlations are small. After several iterations the memory of the original catalogue is lost and the correlations in the original catalogue become fully visible. By iteration 12, this process has stabilized, and memory of the "errors" in the original catalogue has been effectively lost. The upper curve in each figure is the situation for the final iteration. A comparison with the abscissa correlations (Fig.~\ref{fig:absc_frm_corr}) shows clearly that the source of these correlations is entirely in the original reduction \citep{esa97}.}
\label{fig:parcorr}
\end{figure}

In the new reductions the peculiarities in the dynamics of the satellite have 
been incorporated fully in the attitude model \citep{paper4}. The great-circle reduction process has been replaced by a global iterative solution, similar to what is planned for the Gaia mission \citep{omullane07}. The iterations are between the reconstruction of the along-scan attitude and the construction of the star catalogue, both using the abscissa residuals as input data. A calibrated set of instrument parameters provides the link between the two systems. The result is a reduction in the overall noise level for the reconstructed along-scan attitude by about a factor five, and a reduction of the abscissa-error correlations by at least an order of magnitude. In the new reduction, the error-correlation level for field-transit abscissae has been examined by \textsc{VL7V} (Fig.~\ref{fig:absc_frm_corr}). Even for relatively bright stars, correlation levels stay in all but the most extreme cases well below 0.1. To see how these error correlations may accumulate to correlated errors in the astrometric parameters of neighbouring stars, the coincidence factor has been introduced \citep{fvl99b}. Considering the scan of a single field of view as contained within a single full rotation of the satellite, the coincidence factor determines how often the measurements of two stars are contained within the same scan, as a fraction of the total number of scans for each star. This fraction drops rapidly as a function of the separation of the stars on the sky, as can be seen in the lower graph of Fig.~\ref{fig:absc_frm_corr}. Any correlations between astrometric parameters for neighbouring stars would have to be the result of the combined effects of the abscissa-error correlations and the coincidence fraction. Given the low level of the correlations and the small range of the coincidence factors, no significant level of correlation is expected for the astrometric parameters of neighbouring stars in the new reduction. Similarly, a comparison between the new and the old reduction exposes the correlations in the astrometric parameters for the old reduction. Following these correlations as iterations for the new solution proceeded showed how gradually the memory of the old solution was lost (Fig.~\ref{fig:parcorr}).

The reduction of about a factor five in the noise level on the reconstructed
along-scan attitude leads to a significant improvement in the formal errors
on the parallax determinations for the brighter stars. In the published 
catalogue these accuracies were set by the reconstructed-attitude noise, while 
in the new reduction they are limited by primarily the photon noise on the 
abscissa data. For some of the brightest stars, this has resulted in 
accuracy improvements by up to a factor five. Also the reliability of the
formal errors is much improved, as the errors on the field-of-view transit data can be directly related to the integrated photon count of the underlying 
measurements. Small corrections to the formal errors were derived and applied according to the across-scan position of a transit in the field of view, and
the colour index of the star, in particular affecting the very red stars. Altogether, the new reduction provides a significant improvement in accuracy as well as overall reliability of the Hipparcos astrometric data, all of which should benefit the determination of astrometric parameters for open clusters.  

\section{Corrected Hipparcos data?}
\label{sec:makarov}
At this point it is worth noting the work done by Valeri Makarov on trying to remove the correlated errors caused by problems in the attitude modelling from the abscissa residuals as presented in the 1997 reduction \citep{makar02,makar03,makarov06}. This correction was based on an idea first presented by \citet{fvl99b}, which mainly focussed on the issue of connectivity between the two fields of view, and how this can be affected by the presence of locally dense groups of bright stars. The idea is that if such a group, like the core of the Pleiades cluster, is dominating the information for the attitude reconstruction, then if the astrometric data for this group of stars contains systematic errors, these errors would also reflect in the abscissa measurements for stars measured simultaneously with the Pleiades stars, but in the other field of view. By then examining the abscissa residuals of such measurements, one may be able to detect these systematic errors in, say, the Pleiades field. A simple test on the abscissa residuals of stars that shared some measurements this way with the Pleiades showed, however, that the unit weight standard deviation of the residuals for coinciding measurements was in fact slightly smaller than for the non-coinciding measurements \citep{fvl99b}. In other words, there is no signal contained in those residuals. The reason that Makarov was still able to produce a different parallax for the Pleiades cluster based on such data is that the coinciding stars tend to be much fainter than the Pleiades stars, and add substantial amounts of photon noise to the "corrected" data. The actual correction factor and other criteria as used by Makarov were quite arbitrary \citep[see also][]{fvl05c}, and, combined with the photon noise contributions, could be adjusted to provide an expected result. The new reduction has taken care of the attitude modelling effects Makarov was trying to adjust, and it has done so in the only correct way, i.e.\ by re-analysing in detail the satellite dynamics and implementing this in an iterative solution. The results from the new reduction therefore effectively supersede the earlier results obtained by Makarov, even when these appear to give a "more correct" (that is, as expected by some) result. 

\section{Techniques for determining the astrometric parameters}
 \label{sec:techn}
In this section we look at techniques used to extract astrometric parameters for open clusters under different conditions of observing. First a set of exact relations is derived, followed by a first-order approximation, and finally the combined-abscissae method, as used for distant clusters, is recalled.

\subsection{Exact solution}
We define a fiducial reference point which we associate with the centre of 
the cluster. Its position vector with respect to the Sun is defined as $\vec{R}$, which can be expressed in, for example, equatorial coordinates as: 
\begin{equation}
\vec{R} = R\cdot \left[\begin{array}{r}\cos\alpha\cos\delta\\\sin\alpha\cos\delta\\\sin\delta\\ \end{array}\right].
\end{equation}
Similarly, the velocity vector $\dot{\vec{R}}$ can be expressed as:
\begin{equation}
\dot{\vec{R}} = \left[\begin{array}{rrr}\cos\alpha\cos\delta & -\sin\alpha &
-\cos\alpha\sin\delta \\ \sin\alpha\cos\delta & \cos\alpha & -\sin\alpha\sin\delta \\
\sin\delta & 0 & \cos\delta \\ \end{array}\right]\,\cdot\,\left[\begin{array}{l}
\dot{R} \\ R\,\dot{\alpha}\cos\delta \\ R\,\dot{\delta}\\ \end{array}\right],
\label{equ:spacVel1}
\end{equation}
where the coefficients on the right relate to the observed radial velocity 
$V_\mathrm{rad}$ and proper motion $(\mu_{\alpha*},\mu_\delta)$:
\begin{eqnarray}
\dot{R} & = & V_\mathrm{rad} \nonumber \\
R\,\dot{\alpha}\cos\delta & = & \kappa\,\frac{\mu_{\alpha*}}{\varpi} \nonumber \\
R\,\dot{\delta} & = & \kappa\,\frac{\mu_\delta}{\varpi}.
\label{equ:spacVel2}
\end{eqnarray}
Here $\kappa=4.74047$, and is the transformation factor from 1~mas~yr$^{-1}$
at 1~kpc to 1~km~s$^{-1}$. As was introduced with the release of the
Hipparcos catalogue, we use $\mu_{\alpha*} \equiv \mu_\alpha \cos\delta$.

What we are interested in is how the proper motions for cluster members 
will vary as a result of position in the cluster. For this purpose we
include Eq.~\ref{equ:spacVel2} in Eq.~\ref{equ:spacVel1} and invert it to 
give  
\begin{equation}
\left[\begin{array}{l}V_{\mathrm{rad},i}\\
\kappa\,\mu_{\alpha*,i}/\varpi_i \\ 
\kappa\,\mu_{\delta,i}/\varpi_i \\\end{array}\right] = 
\left[\begin{array}{rrr} 
\cos\alpha_i\cos\delta_i & \sin\alpha_i\cos\delta_i & \sin\delta_i \\
-\sin\alpha_i & \cos\alpha_i & 0 \\ 
-\cos\alpha_i\sin\delta_i & -\sin\alpha_i\sin\delta_i & \cos\delta_i \\
\end{array}\right]
\,\cdot\,\dot{\vec{R}},
\label{equ:spacVel3}
\end{equation}
where the index ``$i$'' has been added to indicate reference to different
positions in the cluster, with $i=0$ being the values for the assumed cluster 
centre. We now write Eq.~\ref{equ:spacVel1} and Eq.~\ref{equ:spacVel2} as
\begin{equation}
\dot{\vec{R}} = \tens{A}^{-1}_0\,\vec{v}_0,
\end{equation}
where $v = (V_\mathrm{rad},\kappa\mu_{\alpha*}/\varpi,\mu_\delta/\varpi)$,
and similarly Eq~\ref{equ:spacVel3} as
\begin{equation}
\vec{v}_i = \tens{A}_i \dot{\vec{R}} = \tens{A}_i\tens{A}^{-1}_0\,\vec{v}_0,
\label{equ:local_vel}
\end{equation} 
which fully describes the relation between the observed components of the
space velocity as a function of position on the sky and distance of the 
star. 

\subsection{Convergent point}
\label{subsec:conver}
The velocity vector of the cluster points to a position referred to as the 
convergent point:
\begin{equation}
\dot{\vec{R}} = \dot{R}\left[\begin{array}{l}\cos\alpha_c\cos\delta_c \\
\sin\alpha_c\cos\delta_c \\ \sin\delta_c\\\end{array}\right].
\end{equation}
The common cluster-velocity components of all cluster member proper motions
point towards this convergent point. The position of the convergent point 
depends on the actual velocity vector, of which one component, the radial 
velocity, is measured directly, and the other component, the proper motion,
is scaled with the parallax of the cluster. Therefore, its determination on
the sky allows for a measurement of the cluster distance. This is the 
traditional method used for determining the distance of the Hyades cluster.
With the accuracies available in the Hipparcos data, this is no longer
the best method available for the Hyades, but still the same effects can be
put to use in different ways, to learn about the spatial distribution in
the cluster.

Equation~\ref{equ:spacVel3} can be rotated such that the proper motion is
aligned with the direction of the convergent point, by multiplying it 
left and right by the matrix
\begin{equation}
\left[\begin{array}{rrr} 1 & 0 & 0 \\ 0 & \cos\psi & \sin\psi \\ 0 & -\sin\psi 
& \cos\psi \\\end{array}\right], 
\end{equation}
where $\psi=\arctan(\mu_{\delta,p}/\mu_{\alpha *,p})$. The additional
index ``p'' shows that the proper motion used here is predicted,
based on the assumed space velocity of the cluster and the position of the star on the sky. This rotation then creates the predicted proper motion direction towards the convergent point. Applying the same rotation to the observed proper motion gives the observed proper motions in the direction of the convergent point and perpendicular to it. The rotation also needs to be applied to the errors and error-correlation level for the proper motions. The length of the observed proper motion vector in the direction of the convergent point is a measure of the distance of the star, given the assumed distance of the cluster centre. The predicted value for the perpendicular component is zero. The observed perpendicular proper motion components can therefore be further investigated for the internal velocity dispersion of the cluster. A rotation of the observed proper motions to the direction of the convergent point and a re-scaling as described above, provides so-called reduced proper motions \citep{altena66}, which are useful for membership recognition in the Hyades cluster.

\subsection{First-order approximation}
\label{sec:nearby}
For clusters more distant than the Hyades, the differences between the stellar
coordinates and those of the cluster centre are sufficiently small to be 
approximated in the relevant trigonometric functions. Thus, we develop
the matrix $\tens{A}_i$ in Eq.~\ref{equ:local_vel} as
\begin{equation}
\Delta\tens{A}_i \approx \frac{\partial\tens{A}_0}{\partial\alpha}\,\Delta\alpha_i
+\frac{\partial\tens{A}_0}{\partial\delta}\,\Delta\delta_i,
\label{equ:matr_der}
\end{equation}
where $\Delta\alpha_i=\alpha_i-\alpha_0$ and 
$\Delta\delta_i=\delta_i-\delta_0$. Substituting Eq.~\ref{equ:matr_der} in
Eq.~\ref{equ:local_vel} gives
\begin{equation}
\Delta\vec{v}_i = \vec{v}_i - \vec{v}_0 \approx \biggl(\frac{\partial\tens{A}_0}{\partial\alpha}\,\Delta\alpha_i
+\frac{\partial\tens{A}_0}{\partial\delta}\,\Delta\delta_i\biggr)\,\tens{A}_0^{-1}\,\vec{v}_0.
\end{equation}
The two matrix products give
\begin{equation}
\frac{\partial\tens{A}_0}{\partial\alpha}\,\tens{A}_0^{-1}=\left[\begin{array}{rrr}
0 & \cos\delta_0 & 0 \\ 
-\cos\delta_0 & 0 & \phantom{-}\sin\delta_0 \\
0 & -\sin\delta_0 & 0 \\
\end{array}\right],
\end{equation}
and
\begin{equation}
\frac{\partial\tens{A}_0}{\partial\delta}\,\tens{A}_0^{-1} = \left[\begin{array}{rrr}
0 & 0 & \phantom{-}1 \\ 
0 & \phantom{-}0 & 0 \\
-1 & 0 & 0 \\
\end{array}\right].  
\end{equation}
The first-order corrections for the proper motions and radial velocities
of cluster members as a function of position on the sky are thus given by
\begin{eqnarray}
\Delta V_{\mathrm{rad},i} &\approx& \phantom{-}\Delta\alpha_i\,\frac{\kappa\,\mu_{\alpha *,0}}{\varpi_0}\,\cos\delta_0  + 
\Delta\delta_i\,\frac{\kappa\,\mu_{\delta,0}}{\varpi_0}, \nonumber \\
\Delta\mu_{\alpha *,i} &\approx& \phantom{-}\Delta\alpha_i\,\bigl(\mu_{\delta,0}\sin\delta_0 
- \frac{V_{\mathrm{rad},0}\,\varpi_0}{\kappa}\,\cos\delta_0\bigr), \nonumber \\
\Delta\mu_{\delta,i} &\approx& -\Delta\alpha_i\,\mu_{\alpha *,0}\,\sin\delta_0 - 
\Delta\delta_i\,\frac{V_{\mathrm{rad},0}\,\varpi_0}{\kappa}, 
\label{equ:firstappr}
\end{eqnarray}
where
\begin{eqnarray}
\Delta\mu_{\alpha *,i}&\equiv&\mu_{\alpha*,i}\,\frac{\varpi_0}{\varpi_i} - \mu_{\alpha,0} \nonumber\\  
\Delta\mu_{\delta,i}&\equiv&\mu_{\delta,i}\,\frac{\varpi_0}{\varpi_i} - \mu_{\delta,0}.
\label{equ:delta_pm}
\end{eqnarray}
Using $\Delta\varpi=\varpi_i-\varpi_0$, this can be written as
\begin{eqnarray}
\Delta\mu_{\alpha*,i} &\approx& -\frac{\Delta\varpi_i}{\varpi_0}\,\mu_{\alpha*,0} + (\mu_{\alpha *,i} - \mu_{\alpha *,0}),
\nonumber \\
\Delta\mu_{\delta,i} &\approx& -\frac{\Delta\varpi_i}{\varpi_0}\,\mu_{\delta,0} + (\mu_{\delta,i}- \mu_{\delta,0}).
\label{equ:relsecpar}
\end{eqnarray}
When the proper motion is large and the cluster is relatively nearby, the first term on the right will create an additional proper motion dispersion along the direction of the cluster proper motion. This effect is referred to as the relative secular parallax \citep{fvl83}. In the extreme case of the Hyades cluster, the proper motions can provide precise measurements of $\Delta\varpi_i$. The precision of these determinations is ultimately
limited by intrinsic noise on the proper motions that results from the
internal motions in the cluster. The range within which the above approximations can be used is set by the amplitudes of the proper motion and radial velocity, relative to the formal errors on the measured proper motions.

When combining the astrometric data of different cluster members, the
formal errors and parameter correlations need to be taken properly into
account. There are additional noise contributions to the proper motions
from the internal velocity dispersion and to the parallaxes from the
finite size of the cluster. The main catalogue for the new reduction 
provides for each solution the upper triangular matrix $\tens{U}$
with which the astrometric parameters are to be multiplied when combining
data from different stars. This multiplication normalizes the errors on
the parameters. In fact, the inverse of the matrix $\tens{U}$ is a
square root of the noise variance matrix $\tens{N}$:
\begin{equation}
\tens{N}_i = \tens{U}_i^{-1}\cdot(\tens{U}_i^{-1})^T.
\end{equation} 
To this we have to add the ``cosmic'' noise contributions from the 
parallax and proper motion dispersions. The parallax dispersion is derived
from the typical positional spread of stars in an open cluster, for which a 
value of $\sigma_r=3$~pc is assumed. The transformation to a dispersion on the 
parallax is given by
\begin{equation}
\frac{\sigma_{\varpi,p}}{\varpi_0} = \frac{\sigma_r}{r_0}.
\label{equ:disp_dist_par}
\end{equation}
Thus, as a cluster becomes more distant, this effect rapidly looses 
significance. Even for the Pleiades the effect is small, at about
$\sigma_\varpi\approx0.2$~mas.

The additional noise on the proper motions originates from the internal
velocity dispersion, which has a typical value of 
$\sigma_V\approx0.8$km~s$^{-1}$ in the cluster centre, and the relative 
secular parallax, for which the contributions are described by 
Eq.~\ref{equ:relsecpar}. In detail, both contributions depend on the 
projected distance from the cluster centre, reflecting the space density 
distribution of the cluster. In a cluster like the Pleiades both effects 
are of order 1~mas~s$^{-1}$, but for most other clusters the effects are 
smaller to much smaller. The contribution to the noise matrix for the 
parallax and proper motion measurements as due to the actual distance of
a cluster star is given by
\begin{equation}
\tens{N}_p = \left[\begin{array}{rrr}\sigma_{\varpi,p}^2 & 
-\sigma_{\varpi,p}^2\frac{\mu_{\alpha*,0}}{\varpi_0} & 
-\sigma_{\varpi,p}^2\frac{\mu_{\delta,0}}{\varpi_0} \\ 
-\sigma_{\varpi,p}^2\frac{\mu_{\alpha*,0}}{\varpi_0} &
\sigma_{\varpi,p}^2\bigl(\frac{\mu_{\alpha*,0}}{\varpi_0}\bigr)^2 &
\sigma_{\varpi,p}^2\bigl(\frac{\mu_{\alpha*,0}\mu_{\delta,0}}{\varpi_0^2}\bigr) \\
-\sigma_{\varpi,p}^2\frac{\mu_{\delta,0}}{\varpi_0} &
\sigma_{\varpi,p}^2\bigl(\frac{\mu_{\alpha*,0}\mu_{\delta,0}}{\varpi_0^2}\bigr)  &
\sigma_{\varpi,p}^2\bigl(\frac{\mu_{\delta,0}}{\varpi_0}\bigr)^2
\end{array}\right],
\end{equation} 
where $\sigma_{\varpi,p}$ represents the dispersion in parallaxes as 
caused by the spread in distances along the line of sight for cluster members
(Eq.~\ref{equ:disp_dist_par}). The noise contribution from the internal motions
is simply given by
\begin{equation}
\tens{N}_q = \left[\begin{array}{rrr} 0 & 0 & 0 \\
0 & \sigma_{\mu,q}^2 & 0 \\
0 & 0 & \sigma_{\mu,q}^2 \\
\end{array}\right],
\end{equation}
where $\sigma_{\mu,q}$ is the internal proper motion dispersion resulting
from the internal velocity dispersion. The total noise as applicable to
the parallax and proper motion of a cluster member when considering its 
contribution to the cluster parallax and proper motion is given by the sum
of the three matrices:
\begin{equation}
\tens{N}_t = \tens{N}_i + \tens{N}_p + \tens{N}_q = \tens{W}_t^{-1}\cdot
(\tens{W}_t^{-1})^T.
\end{equation} 
Multiplication by the upper-triangular matrix $\tens{W}_t$ of the equations and observations for each cluster member, describing the corrections to a common cluster parallax and proper motion, provides a set of fully normalized observation equations that can be solved for in a least squares solutions.

\subsection{Distant clusters}
\label{sec:distant}

For distant clusters, where the formal errors on the proper motions and 
parallaxes of the individual members are significantly larger than the 
corrections described above, the cluster parameters may be derived from
a combined analysis of the abscissa residuals for the member stars. Thus,
we repeat the astrometric parameter solutions for the cluster stars, but now 
solve for a single proper motion and parallax for all cluster members. 

In order to make this common-parameter solution possible, the abscissa
residuals for stars involved need to be re-calculated, to provide residuals
relative to a chosen reference parallax and proper motion, which can be
seen as the first estimate of the cluster parallax and proper motion.
The abscissa residual for observation ``$j$'' of star ``$i$'' as provided
in the epoch astrometry file in \textsc{VL7B} is represented by 
$\mathrm{d}a_{ij}$. A vector $\vec{z}_i$ represents the cluster parallax and 
proper motion, and similarly $\vec{z}_c$ the reference values for the cluster. 
The corrected abscissa residuals are then given by
\begin{equation}
\mathrm{d}a'_{ij} = \mathrm{d}a_{ij} + (\vec{z}_i-\vec{z}_c)\,\cdot\,
\frac{\partial a_{ij}}{\partial\vec{z}}.
\end{equation}
The initial choice for $\vec{z}_c$ does not affect the linear solution for the
astrometric parameters, which includes positional corrections for each star
and a single parallax and proper motion for all stars together. The advantage
of solving for the cluster parameters in this way is the decrease in
degrees of freedom for the solution. This allows for a better recognition of
outliers among the measurements and thus a potentially more robust
solution. To compensate for abscissa-error correlations, as was described by \citet{vLDWE}, is no longer required using the new reduction data. Neither is there a need to combine the data from two partly correlated reductions. The process of deriving cluster parameters this way has become significantly simpler and because of that also more reliable, as any of the correlation corrections that had to be applied for processing the data from the 1997 release was also a source of uncertainty in the estimates of the astrometric parameters and their formal errors as derived for the clusters.

\section{Some ground-based distance modulus or parallax determination methods, independent of isochrone fitting}
\label{sec:groundbased}
In this section three methods that have been used for isochrone-independent cluster-distance determinations are reviewed by incorporating the new Hipparcos data. Some of the methods described below derived or used relations between observed parameters (generally photometric colour indices) and physical parameters of the observed stars, using stars with ground-based parallaxes, or open clusters and their assumed distances, for calibration. The uncertainty has been in distinguishing the error contributions from the calibration objects (such as the measured parallax) from the cosmic noise on such calibrations. The independent absolute parallaxes obtained from the Hipparcos data allow for an assessment of these noise levels. 
\subsection{Crawford's calibrations of the Str{\"o}mgren photometric system}
\label{sec:crawford}
Here an overview is presented on a set of calibrations that have been (and are still) used for luminosity calibrations based on Str{\"o}mgen photometry. A more recent calibration (but still pre-Hipparcos) is also available \citep{jordi97}, but concerns more the calibration of relations between specific indices, and recognizes that there are significant uncertainties in empirical photometric luminosity calibrations. First calibrations incorporating the Hipparcos data have been presented by \citet{domingo99} for late A-type stars and \citet{jordi02} on preliminary results for F, late A and B type stars.

The empirical calibration of absolute magnitudes through observations in the Str{\"o}mgen photometric system by Crawford was done in three stages, on the F type stars using the Pleiades, $\alpha$~Per and Coma~Ber for establishing calibration slopes, and nearby stars with ground-based  parallaxes for zero-point determination \citep{crawford75}, on the B-type stars \citep{crawford78} and A type stars \citep{crawford79} using distances of the Pleiades and $\alpha$~Per as determined through the F-star calibration. Thus, ultimately, the zero point of the calibration of the system rests on the ground-based parallaxes of 17 nearby F-type stars, while the slopes has been determined by means of the combined data from open cluster stars. 

\begin{figure}[t]
\centering
\includegraphics[width=8cm]{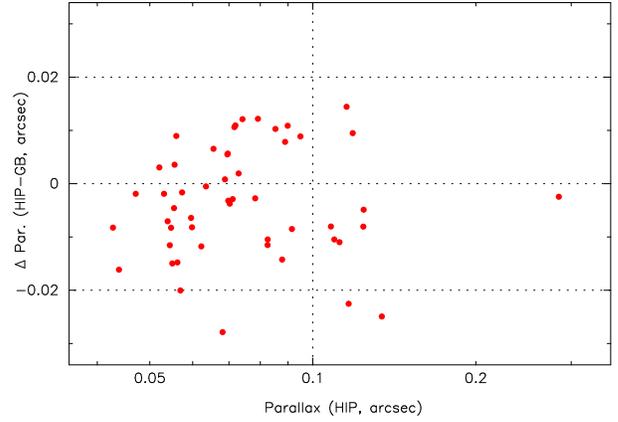}
\caption{A comparison between the ground-based parallax values as used by \citet{crawford75} and parallaxes for the same stars as obtained from the new Hipparcos reduction. The comparison shows a systematic difference of 3.4~mas and a noise on the ground-based parallaxes of 10~mas.}
\label{fig:crawf_par}
\end{figure}
As the ground-based parallaxes used by Crawford had relatively large errors, he applied a Lutz-Kelker correction \citep{lutzk73} to compensate for the bias in the derived distance moduli and thus the absolute magnitudes. Such corrections are provided for 48 of the 51 stars. Using the new Hipparcos parallaxes, the Lutz-Kelker correction is not needed, as parallax errors are for nearly all stars below 1~per~cent. The direct comparison between the parallaxes (Fig.~\ref{fig:crawf_par}) shows a systematic difference of $\varpi_\mathrm{HIP}-\varpi_\mathrm{GB}=-3.4\pm1.4$~mas, and a noise level on this selection of ground-based parallaxes of 10.3~mas. Very similar results were obtained by \citet{jordi02} based on the 1997 reduction of the Hipparcos data. There it was also shown that the mean offset for the nearest stars, as used by Crawford, the correction was still larger, at $-5$~mas. A comparison between the absolute magnitudes as given by Crawford before ($\mathrm{M_V(\varpi)}$) and after the Lutz-Kelker correction ($\mathrm{M_V(cor)}$) and those derived using the new Hipparcos parallaxes ($\mathrm{M_V(HIP)}$) gives the following results:
\begin{eqnarray}
\mathrm{M_V(HIP) - M_V(\varpi)\phantom{c} } &=& \phantom{-}0.090\pm0.041,\nonumber \\
\mathrm{M_V(HIP) - M_V(cor)} &=& -0.093\pm0.041,
\end{eqnarray} 
where in the first solution the noise level was determined as 0.29, and in the second solution as 0.31 magnitudes. The offset in the first solution is caused entirely by the systematic differences of 3.4~mas between the ground-based and Hipparcos parallaxes. After applying the correction for the systematic offset in the ground-based parallaxes, the observed difference is a (rounding-off) residual of $-0.013\pm0.041$ magnitudes before application of the Lutz-Kelker correction. The latter "correction" seems to do no good to the data, possibly because corrections of this kind are derived for, and are strictly only applicable \textit{statistically} to \textit{groups} of stars with the same relative errors on the measured parallaxes. The relation between the absolute magnitude, $\mathrm{H}\beta$ and $\Delta c_1$ is now given by
\begin{eqnarray}
\mathrm{M_V(HIP)} &=& (3.965\pm0.046) - (1.545\pm0.101)\,\mathrm{d}\beta \nonumber\\
&+&(1.43\pm0.26)\,{\mathrm{d}\beta}^2 - (11.3\pm0.7)\,\Delta c_0,
\label{equ:mv_calibr}
\end{eqnarray}
where $\mathrm{d}\beta \equiv 10\cdot(\mathrm{H}\beta-2.64)$, and for these nearby stars the reddening has been assumed zero ($\Delta c_0 = \Delta c_1$). The slope for $\Delta c_1$ is effectively the same as determined by Crawford, both from using field stars and from an application to NGC~752, where the turn-off for the main sequence is found in the F-star region. The latter confirmed the relation between $\Delta c_0$ and the surface gravity, $\log g$. An alternative solution was made in which the data for the Pleiades \citep{crawford76} and $\alpha$~Per \citep{crawford74} was incorporated, with two additional terms to account for the distance moduli of these clusters, but with $\Delta c_0=0$, as was also done by Crawford in his determination of the distance moduli of those clusters. This gave the following solution:
\begin{eqnarray}
\mathrm{M_V(HIP)} &=& (3.958\pm0.041) - (1.668\pm0.076)\,\mathrm{d}\beta \nonumber\\
&+&s(0.92\pm0.18)\,{\mathrm{d}\beta}^2 - (10.5\pm0.6)\,\Delta c_0,
\label{equ:mv_calibr_clust}
\end{eqnarray}
and a nominal distance modulus for the Pleiades of $5.621\pm0.058$ (against 5.54 found by Crawford) and for $\alpha$~Per of $6.045\pm0.061$ (6.1 found by Crawford). However, what is important here is that even after application of Eq.~\ref{equ:mv_calibr_clust}, the remaining uncertainty on $\mathrm{M_V}$ for the field stars is still 0.20 magnitudes (compared to 0.29 magm.\ in Crawford's original calibration, and 0.25~mag.\ in Jordi's calibration), while for the cluster stars alone the remaining noise is significantly smaller, at around 0.14 magnitudes. The distance moduli found here for those two clusters make the assumption that the stars in those clusters are most like the average field star, but as long as the source behind the variation of the field-star magnitudes has not been quantified, this assumption has little meaning. The actual uncertainty in the distance moduli for the two clusters is therefore likely to be significantly higher, and should provisionally be put at 0.20 magnitudes.
 
In the solution to Eq.~\ref{equ:mv_calibr} also dependencies on $\Delta m_0$ and cross terms of $\Delta c_0$ and $\hat{\beta}$ were tested, but the result was not significant. The parameter $\Delta m_0$ is correlated with abundance variations \citep{nissen70a, nissen70b, gustafsson72, crawford75}, which suggests that such variations are not a significant contribution in the remaining spread in absolute magnitudes.
\begin{table}[t]
\setlength{\extrarowheight}{2pt}
\caption{Comparison between absolute magnitude calibrations for different mean values of $\overline{\beta}$. }
\centering
\begin{tabular}{|r|rrr|}
\hline
$\overline{\beta}$ & $\mathrm{\overline{M_c}}(\varpi)$ &  $\mathrm{\overline{M_c}(cor)}$ & $\mathrm{\overline{M_c}(HIP)}$ \\
\hline
2.604 & $4.77\pm0.23$ & $4.64\pm0.22$ & $4.677\pm0.044$ \\
2.631 & $4.21\pm0.25$ & $4.03\pm0.26$ & $4.115\pm0.038$ \\
2.648 & $3.92\pm0.30$ & $3.69\pm0.30$ & $3.830\pm0.043$ \\
2.671 & $3.69\pm0.52$ & $3.43\pm0.58$ & $3.529\pm0.050$ \\
2.711 & $3.53\pm0.12$ & $3.33\pm0.17$ & $3.237\pm0.087$ \\
\hline
\end{tabular}
\begin{list}{}{}
\item The first three columns are from \citet{crawford75}
\item The final column from the current paper, using the least squares solution that provided the coefficient in Eq.~\ref{equ:mv_calibr_clust}.
\end{list}
\label{tab:crawffstars}
\end{table}
Table~\ref{tab:crawffstars} shows the results of applying Eq.~\ref{equ:mv_calibr_clust} to the mean $\beta$ values derived by Crawford for the five intervals he used for his analysis. The $\mathrm{M_V}$ values derived from the new analysis differ by up to 0.14 magnitudes from the values derived by Crawford for $\mathrm{M_c(cor)}$. 

Conclusions on the Crawford calibrations are the following. The systematic errors on the ground-based parallaxes for the calibration stars were partly offset by the Lutz-Kelker correction, which by itself does not seem to do any good to the data. The Crawford calibration can, as a result be offset by up to 0.14 magnitudes. However, the remaining noise level for the calibration standards is much larger than what would be caused by measurement inaccuracies. It may be partly caused by duplicity, in which case the zero point of the calibration is also affected. When deriving distances of individual objects or star clusters using this calibration, this noise needs to be accounted for in the final accuracy figure until it can be quantified and related to independent observed parameters. For the calibrations of the A and B stars there are now available large numbers of calibration stars with well determined parallaxes, which should replace the current extrapolation based on the F stars and cluster  distances derived therefrom. This concept is further evaluated in Section~\ref{sec:provcalib}. For the current paper the important aspect is the difference between the determination precision, usually quoted as accuracy, for the cluster-distance determinations, and the actual accuracy, which also takes into account the observed cosmic spread intrinsic to the calibration relation.

\subsection{Photometric boxes in the Geneva photometric system} 
\label{sec:nicolet}
The method of the photometric boxes is based on the hypothesis that if two stars  have nearly the same colours or derived parameters, then they will also have  nearly the same physical and atmospheric properties \citep{nicol81,creze80}. This was applied by \citet{nicol81} to photometric data in the Geneva system for determining the distance moduli of 43 open clusters. The comparisons in the boxes were done directly between stars with ground-based parallax determinations and cluster stars, always sharing the same boxes. The accuracies for these parallaxes were still a major factor in these calculations, which can now be eliminated by substituting those data with the Hipparcos results. What can be measured is the difference in apparent magnitude between cluster members and parallax stars sharing the same photometric boxes. In the original solution two compatible noise sources had to be accounted for, the uncertainty of the measured parallaxes and the uncertainty in the hypothesis stated above. Using the Hipparcos parallaxes, the uncertainty in the hypothesis dominates fully. Nicolet estimated this to be at a level of 0.3 magnitudes.  
 
\begin{table}[t]
\centering
\caption{Data from \citet{nicol81} for the cluster NGC~752.}
\begin{tabular}{|r|rr|rr|}
\hline
Id & HD & HIP&$\varpi$ (mas)  & $\Delta m$ \\
\hline
 88 & 160915 &  86736 &  $56.65\pm0.24$ & 6.834 \\
123 & 109085 &  61174 &  $54.70\pm0.17$ & 6.815 \\
139 & 139664 &  76829 &  $57.35\pm0.16$ & 6.957 \\
235 &   7439 &   5799 &  $42.76\pm0.30$ & 6.165 \\
259 &  22001 &  16245 &  $46.12\pm0.13$ & 6.565 \\
266 &  22001 &  16245 &  $46.12\pm0.13$ & 6.430 \\
273 &  43318 &  29716 &  $26.89\pm0.31$ & 5.727 \\
273 & 201772 & 104738 &  $30.34\pm0.34$ & 5.987 \\
304 &  70958 &  41211 &  $37.57\pm0.33$ & 6.188 \\
401 &   4813 &   3909 &  $63.48\pm0.35$ & 6.876 \\
402 & 189245 &  98470 &  $47.06\pm0.41$ & 6.295 \\
401 & 191862 &  99572 &  $36.10\pm0.41$ & 6.079 \\
401 & 196378 & 101983 &  $40.55\pm0.27$ & 7.032 \\
402 &  40136 &  28103 &  $67.21\pm0.25$ & 7.144 \\
403 & 193307 & 100412 &  $32.24\pm0.47$ & 6.144 \\
403 & 221356 & 116106 &  $38.29\pm0.54$ & 5.820 \\
414 & 198188 & 102762 &  $17.11\pm0.88$ & 4.759 \\
416 &  88742 &  50075 &  $43.77\pm0.41$ & 6.459 \\
416 & 210918 & 109821 &  $37.57\pm0.33$ & 6.186 \\
428 & 104731 &  58803 &  $39.49\pm0.28$ & 6.420 \\
430 & 221356 & 116106 &  $38.29\pm0.54$ & 5.905 \\
432 &  23754 &  17651 &  $56.73\pm0.19$ & 7.115 \\
\hline
\end{tabular}
\begin{list}{}{}
\item The identifier in column~1 comes from a proper motion study of this cluster by \citet{ebbig39}. 
\item The identifiers in columns~2 and 3 apply to the comparison star.
\item Column 4 gives the parallaxes from the new Hipparcos reduction.
\end{list}
\label{tab:nicolet}
\end{table}
The detailed data for one cluster, NGC~752, are provided by \citet{nicol81}. They are partly reproduced here in Table~\ref{tab:nicolet}, together with the new parallax data. Based on these data, an estimate of the distance modulus for the cluster can be obtained:
\begin{equation}
\mathrm{dm} = 10 - 5\log\varpi +\Delta m + \varepsilon,
\end{equation}   
where $\varepsilon$ is the error on $\Delta m$, which was estimated by Nicolet to have a $\sigma$ value of 0.3 magnitudes, and has now been determined to be around 0.27 to 0.28 magnitudes. The (photometric) distance modulus is determined as $8.25\pm0.06$. However, this is assuming that NGC~752 in its photometric properties behaves as the average of the comparison stars. In this respect, a star cluster, with its own very narrow distribution of luminosities, would be the equivalent of a single field star, and could itself in principle be anywhere within the $\Delta m$ distribution. 

The photometric boxes method could be explored further though. Using the now much more extensive volume of photometric data and the Hipparcos parallax information for many of these stars should allow for a much closer examination of the differential effects within these boxes, and the way these may be related to luminosity variations. This in turn could lead the way to an improved photometric calibration of distance moduli for distant star clusters.
 
\subsection{Differential parallax determinations with the MAP}
\label{sec:gatew}
\begin{figure}[t]
\centering
\includegraphics[width=9cm]{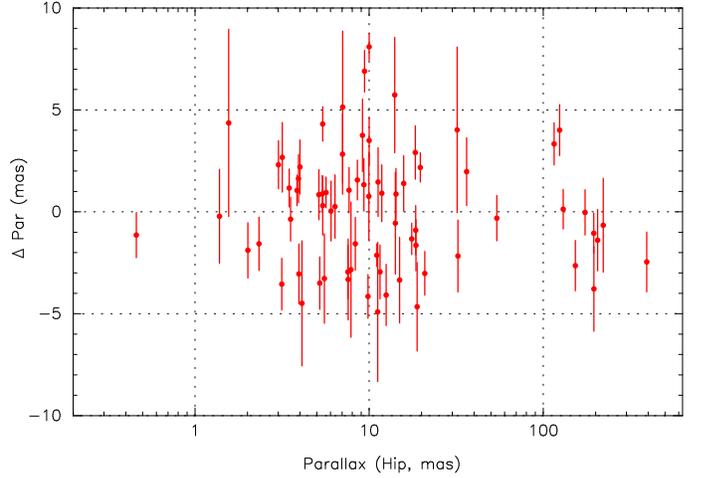}
\caption{Differences between MAP-based and Hipparcos parallaxes for 71 stars. The error bars contain the contributions of the formal errors on the Hipparcos parallaxes and the precisions as given for the MAP data. }
\label{fig:gatewlog}
\end{figure}
Before starting the analysis of the Hipparcos data for the open clusters, it is worthwhile to investigate the results obtained with differential (small-field) parallax measurements, in particular those obtained with the Multichannel Astrometric Photometer (MAP) \citep[see for example][]{gatew90} and Hubble Space Telescope, both of which use the same techniques to calibrate from measured relative to estimated absolute parallaxes. Of particular interest are those results obtained before the release of the Hipparcos catalogue, as these provide the most likely measurements not biassed by an assumed outcome. A comparison of this kind was done by \citet{gatew98} using the first release of the Hipparcos catalogue, where it was noticed that an additional noise component was present, but the origin and amplitude was not specified. Ground-based parallax measurements obtained mostly with the MAP were found for a total of 71 stars that have also data entries in the new Hipparcos catalogue. The errors as given for those measurements are explained by Gatewood as being precision rather than accuracy, and this becomes quite clear when examining the statistics of the differences with the Hipparcos measurements, shown in Fig.~\ref{fig:gatewlog}. Without adding absolute-calibration noise, the weighted mean of the differences and associated standard deviation are given by 
\begin{eqnarray}
\langle\Delta M\rangle &=&-0.501 \\
\sigma\Delta M &=& \phantom{-}2.425.
\end{eqnarray}
With correct weights, the expected value for the standard deviation equals 1.0. Adding a noise contribution to account for calibration uncertainties to the precision reduces the unit-weight standard deviation. A value close to 1.0 is reached when adding a calibration noise of 2.37~mas:
\begin{eqnarray}
\langle\Delta M\rangle &=&-0.114 \\
\sigma\Delta M &=& \phantom{-}1.002 \\
\sigma(\langle\Delta M\rangle) &=& \phantom{-}0.333.
\end{eqnarray}
The added noise should be regarded as an average over the studies included here, and may well vary significantly depending on sources available for calibration. However, it seems unlikely that these parallaxes will have absolute accuracies any better than at best 1.5 to 2.0~mas. The source of this calibration uncertainty is likely the generally small number of calibration stars available and the large intrinsic spread in absolute magnitudes within either an interval in colour index or a spectral class. For the present paper, these results are important when comparing the Hipparcos results with MAP-based parallax determinations for the same clusters.

\section{Nine clusters and a moving group within 250~pc}
\label{sec:clust250}
In this section the results for clusters within 250~pc are presented. In order
of derived distance, these are the UMa moving group, the Hyades, Coma~Ber, Pleiades, IC~2602, IC~2391, $\alpha$~Per, Praesepe, NGC~2451 and Blanco~1. NGC~2451, however, does not correspond to the group of stars that was originally given this designation \citep{roeser94}. The cluster identified as Cr~359, assigned by \citet{loktin94} a distance within the same range, has been rejected, as it does not show from the Hipparcos data in either proper motion or distribution on the sky \citep[see also][]{robic99}. The indications within the Hipparcos data for the existence of a cluster at 170~pc in Ophiuchus, as suggested by \citet{mamajek06}, are too weak.

In all cases where HR diagrams are presented, mean reddening corrections for the clusters have been applied according to what appears from the consulted literature to be a reliable estimate. In all cases these corrections are small. No individual corrections for the differential reddening in the Pleiades and $\alpha$~Per, or to data of individual field stars, have been applied. Corrections for metallicity variations have not been applied. For the small variations shown by the clusters examined here, there is no observational confirmation of significant luminosity variations with varying metallicity (see further Section~\ref{sec:provcalib}). 
 
\subsection{The Hyades}
\label{sec:hyad}

A major study of the Hyades cluster as based on the first Hipparcos
catalogue was presented by \citet{perry98}, and further studies based on the same data by \citet{madsen00}, \citet{madsen01} and \citet{bruijne01}. Differences between the results of these studies show one of the problems associated with most open-cluster studies, the statistical variations in results due to the relatively small number of member stars available, further complicated by uncertainties of membership. Combined with different selection criteria, results can look sometimes even significantly different. A major complication for the Hyades cluster, as well as other similar clusters, is the apparent presence of a halo of stars ``evaporated'' from the cluster. These can easily disturb estimates of the distance, space velocity and internal velocity dispersion of a cluster.

\begin{figure}
\includegraphics[width=9cm]{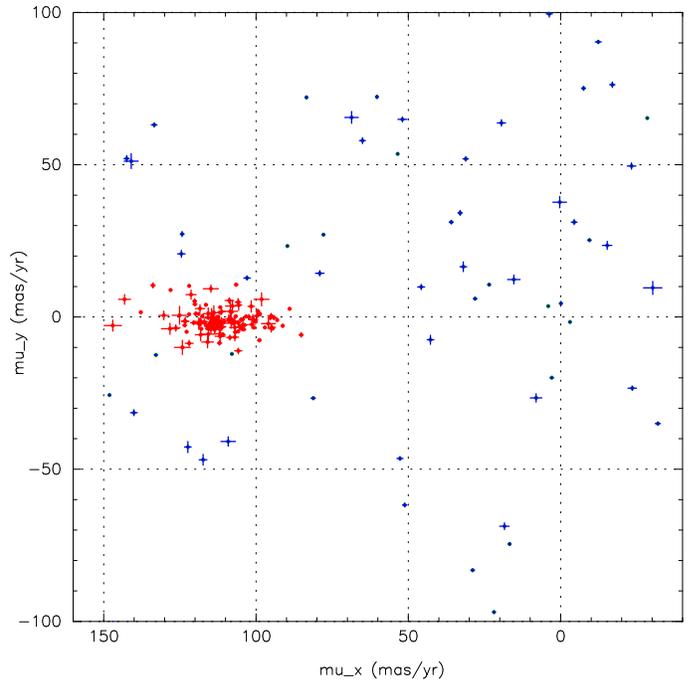}
\caption[]{Reduced proper motions in the selection field for the Hyades
cluster. In this diagram, $\mu_x$ and $\mu_y$ represent the proper motion in 
the direction of the convergent point and the perpendicular direction
respectively. The concentration of stars around (-115,0) (in red) are the cluster members. The distribution is stretched out in $\mu_x$ due to the 
significant effect of the cluster depth. This information is used
to determine differential positions in the cluster along the line-of-sight}
\label{fig:hrpm}
\end{figure}
\begin{figure}
\includegraphics[width=9cm]{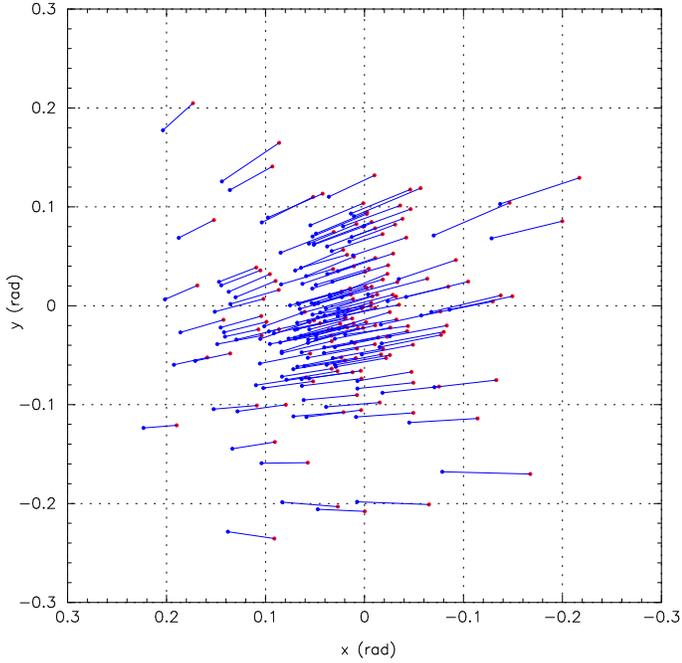}
\caption[]{Map of the selected 150 Hyades stars, showing the current 
positions (points to the right of each line) and those extrapolated by
100~000 years (left-side points)}
\label{fig:hmap}
\end{figure}
Based on the new reduction the Hyades cluster membership was reassessed, and 
similarly the internal dispersion in the proper motions as a function of the 
projected distance to the ``centre'' of the cluster, where it should be 
understood that in a loosely bound, and not very rich cluster like the Hyades, 
the centre is not very well defined.
The position of the projected centre was assumed to be (in the ICRS)
\begin{eqnarray}
\alpha_0 &=& 4^h~26^m,\\
\delta_0 &=& 16\degr~54'.
\end{eqnarray}
The catalogue has been searched for potential Hyades members within a radius
of $18.54\degr$ from this position. Furthermore, the distance from the Sun for
the cluster centre was assumed to be 46.7~pc. Within this area on the sky, stars with parallax values were 
selected such that
\begin{equation}
16.2~\mathrm{mas} < (\varpi\pm 3\sigma_\varpi) <31.5~\mathrm{mas},
\end{equation} 
or between 31.6 and 61.7 pc distance from the Sun. The chosen limits imply that 
a complete sphere with radius 15 pc around the assumed cluster centre is
covered. A total of 270 stars were thus selected. The reduced proper motions of the selected stars were then examined for further membership selection. The reduced proper motions (Fig.~\ref{fig:hrpm}) describe the observed proper motions as the components towards, and perpendicular to, the convergent point of the cluster, and corrected for the angular separation from the convergent point 
(Section~\ref{subsec:conver}) \citep[see also][]{altena66}. Based on the 
reduced proper motions 150 stars were selected as probable members
(one additional star, HIP~20895, was rejected based on a comparison between
the measured parallax and the kinematic distance). 
Figure~\ref{fig:hmap} shows these stars and their extrapolated displacements
over the next 100~000 years that would result from the observed proper 
motions. The mean space velocity relative to the Sun for the selected stars has been derived from the observed parallaxes and proper motions only, applying Eq.~\ref{equ:spacVel3} in a least-squares solution, and is found to be 
\begin{equation}
\dot{\vec{R}} = \left[\begin{array}{r}-5.99 \\ 44.73 \\ 4.93 \\\end{array}\right] \pm \left[\begin{array}{r}0.25 \\ 0.61 \\ 0.14 \\\end{array}\right]~\mathrm{km~s}^{-1},
\end{equation}
with a unit-weight standard deviation of 1.86. The latter is a clear 
indication that the observed proper motions are significantly affected 
by internal motions. As a result, the formal errors on the proper motions
are insufficient for representing their actual noise level. The formal errors on the input data to the solution do include the errors on the measured parallaxes.

\begin{figure}
\includegraphics[width=9cm]{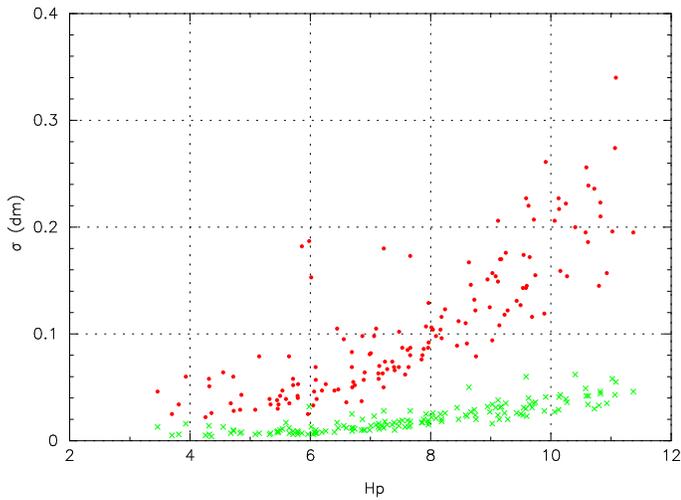}
\caption[]{Comparison between the formal errors on the distance moduli for the
Hyades stars as measured directly through parallaxes (dots) and as derived from the space velocity and the observed proper motions (crosses).}
\label{fig:hsdmc}
\end{figure}
The space velocity, together with the observed proper motions and their
covariances can be used to derive, by means of Eq.~\ref{equ:local_vel}, the
kinematic (relative) parallaxes for the cluster members. The advantage of 
this for the Hyades cluster stems from the relatively large proper motion,
combined with small formal errors, and can clearly be seen when comparing 
the error estimates on distance moduli derived from the measured and kinematic
parallaxes (Fig.~\ref{fig:hsdmc}).
The mean parallax of the selected stars was calculated from the kinematic
parallax estimates, and after projecting their distances onto the 
line-of-sight for the cluster centre, to avoid any (though quite small) 
projection-induced bias. The value thus obtained is
\begin{equation}
\varpi_\mathrm{Hyad}=21.53\pm2.76~~~(\pm0.23)~~~\mathrm{mas}, 
\label{equ:Hyades}
\end{equation}
where the first error shows the standard deviation, representing the spread
of stars in the cluster. The formal errors on the measurements hardly 
contribute to this value, being overall much smaller. The second error in 
Eq.~\ref{equ:Hyades} gives the error on the mean. Thus, based on the error on 
the mean, the weighted mean distance of the 150 plausible Hyades members
selected here is $46.45\pm0.50$~pc, which is equivalent to a distance modulus
of $3.334\pm0.024$. The determination by \citet{perry98} gave a distance of
$46.34\pm0.27$~pc, based on a somewhat different selection of stars and 
reduction methods.

\begin{figure}
\includegraphics[width=8cm]{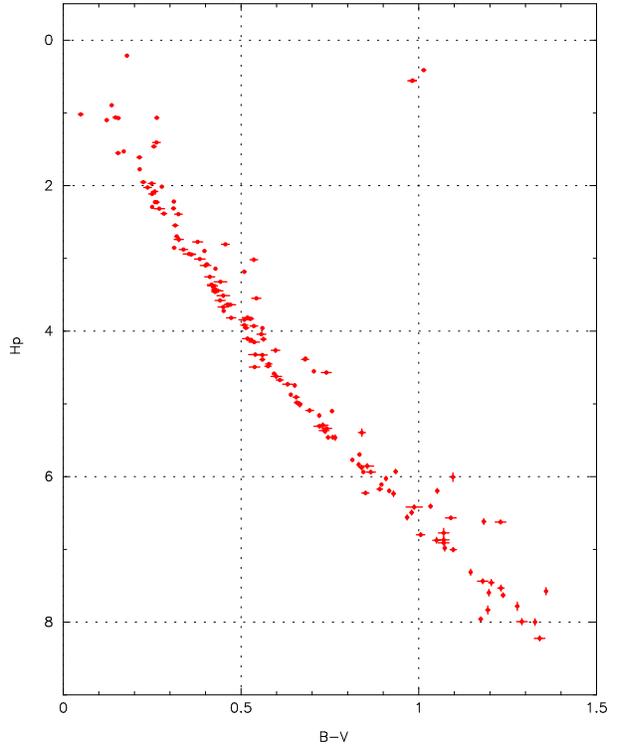}
\caption[]{The Hyades HR diagram for 150 probable members, using distance 
moduli for individual stars as based on kinematically derived parallaxes.}
\label{fig:hhr}
\end{figure}
For the current study, the most important result is the position and
shape of the Hyades HR diagram, representing an observed isochrone for stars
of a specific age and chemical composition. The comparison between the 
diagrams as derived using the observed or the kinematic parallaxes has been
shown on various previous occasions \citep{madsen99, madsen00,bruijne01} and \textsc{VL7B}, so here is shown only the diagram using the distance moduli for individual stars as 
based on their kinematic parallaxes (Fig.~\ref{fig:hhr}). Clear from this 
figure is the narrow and well-defined main sequence, as well as the 
scattering of double stars up to 0.75~mag above it.  

\subsection{Coma~Ber and UMa}
\label{sec:coma}

\begin{figure*}
\centering
\includegraphics[width=13cm]{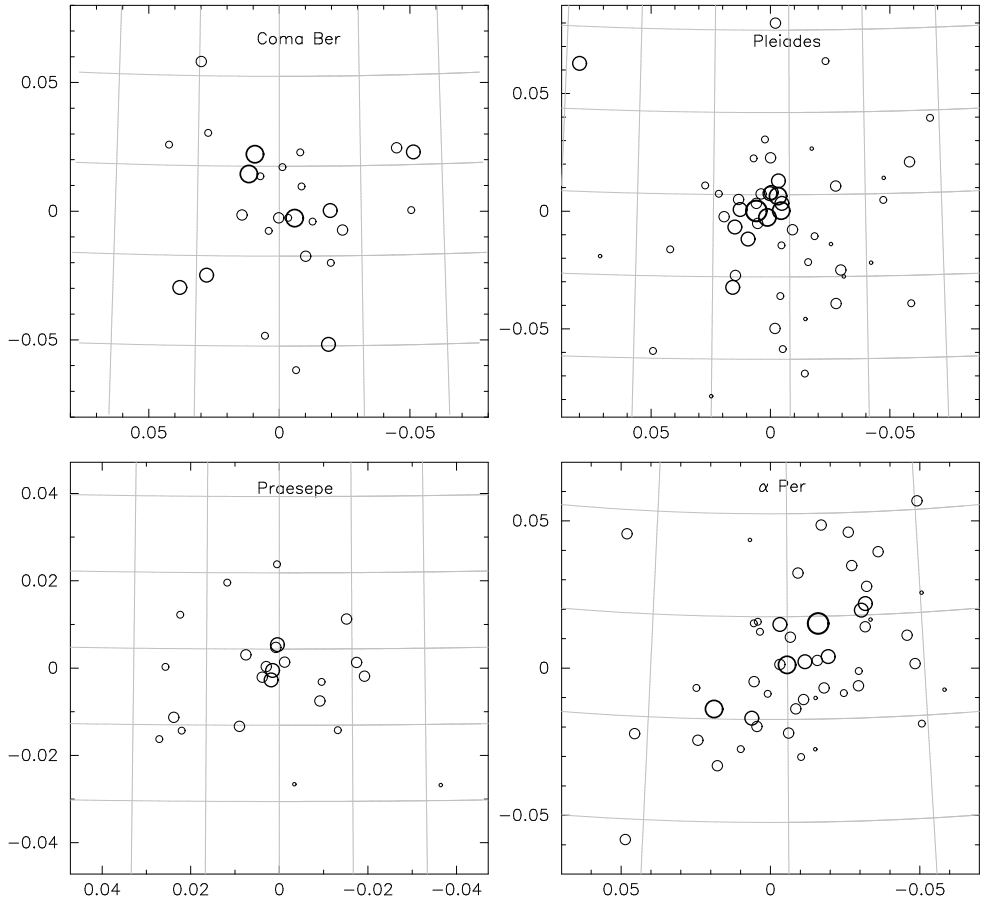}
\caption[]{A tangential projection of the distribution on the sky for
members of the clusters Coma~Ber (top left), Pleiades (top right),
Praesepe (bottom left) and $\alpha$~Per (bottom right). The coordinates are in radians, relative to the assumed cluster centre.}
\label{fig:mapcb}
\end{figure*}
 
\begin{figure*}
\centering
\includegraphics[width=14cm]{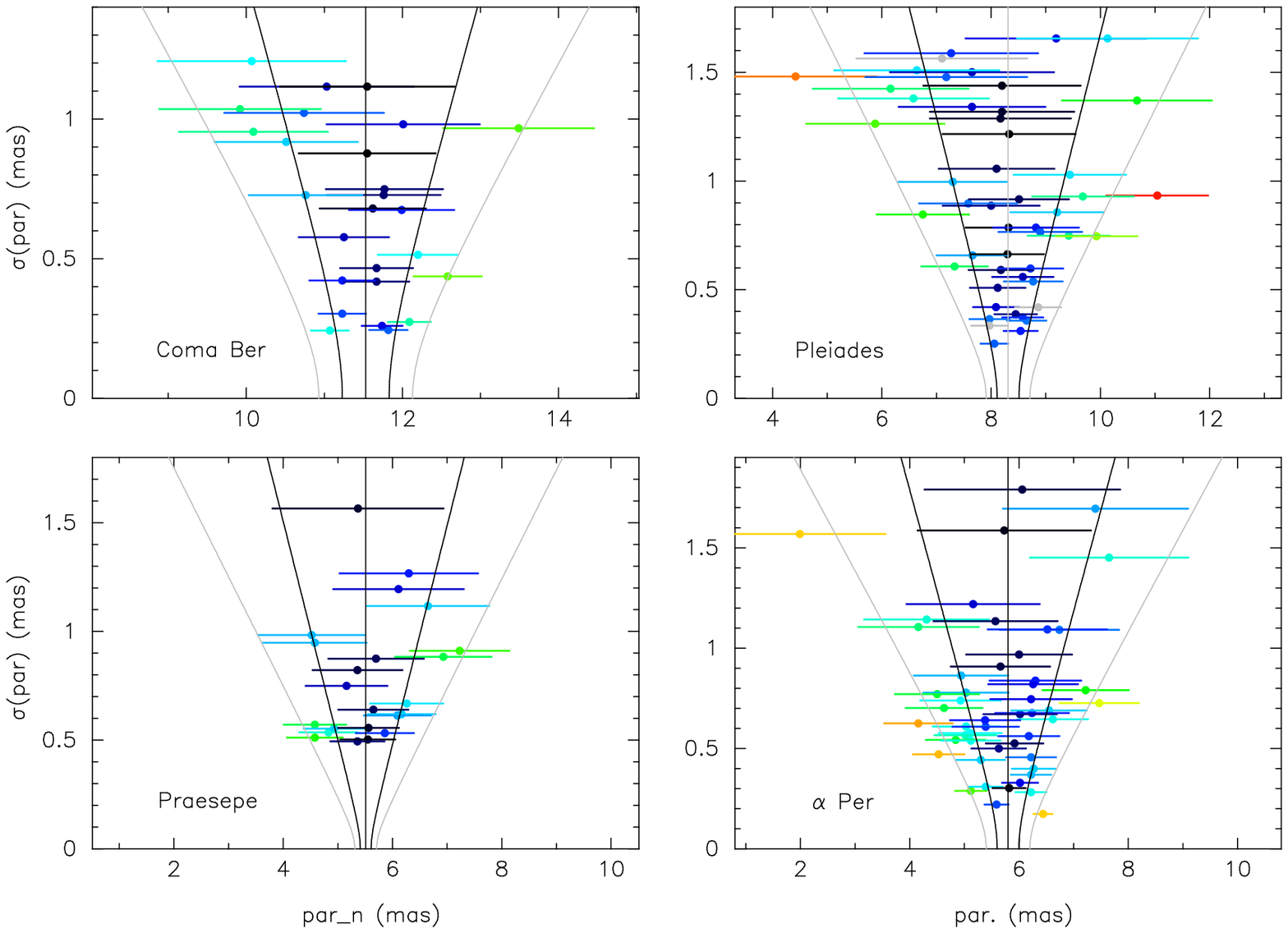}
\caption[]{The distribution of the parallaxes against their formal errors for
the clusters Coma~Ber (top left), Pleiades (top right), Praesepe (bottom left)
and $\alpha$~Per (bottom right). The central vertical line
represents the mean parallax as determined in each cluster based on members 
identified in the Hipparcos catalogue. The black lines on either side show the $1~\sigma$ error level, including the internal distance dispersion
for the cluster. The outer grey lines similarly represent the $2~\sigma$ level.
This display of the data should best allow the reader to evaluate the 
consistency of the measured parallaxes for the member stars in each cluster.}
\label{fig:parcb}
\end{figure*}

Coma~Ber is a sparse cluster at high galactic latitude, considered older than the Pleiades and younger than the Hyades and Praesepe. \citet{crawfordb69} derived a distance modulus of 4.5 for this cluster, and showed in the same paper the results for a small number of members of the core of the UMa moving group, which have characteristics very similar to the Coma~Ber cluster. For this group a distance modulus of 1.8 is stated. Both distance determinations used the calibrations of the parameters $c_1$ and $\beta$ as presented by \citet{stroemg66}. That calibration is in fact based on the one presented by \citet{fernie65}, who used open cluster data for which distances had been derived from on UBV photometry by \citet{johnson61}. Among those clusters was Coma~Ber, for which a distance modulus of 4.5 was given. All distances given in that paper were tied-in to a distance of 40~pc for the Hyades, which we now know should be 16~per~cent larger, at 46.45~pc. It is therefore reasonable to assume that the distance and distance-moduli estimates based on this material may have been underestimated. It would in fact be more surprising if they were correct. 

There are few independent distance determinations for Coma~Ber. Based on the so-called photometric boxes in the Geneva photometric system, and thereby effectively calibrating the cluster distance on ground-based parallaxes of nearby stars, \citet{nicol81} derived a distance modulus of $4.76\pm0.15$. A differential parallax determination by \citet{gatew95} gave a value of $4.34\pm0.09$ for the distance modulus. This is an internal error, and considering the discussion in Section~\ref{sec:gatew}, an error of the order of 0.3 to 0.4 magnitudes for the absolute value may be more realistic. Based on UBV photometry, \citet{cameron85a} derived a distance modulus of 4.65, while a more recent photometric distance determination by \citet{pinso98} gave a value of $4.54\pm0.04$. The parallax determinations using the first release of the Hipparcos data gave values of $4.70\pm0.04$ \citep{robic99} and $4.77\pm0.05$ \citep{fvl99a}.  
 
Eight members of the UMa group were analyzed for the present paper, with individual parallax accuracies in the new Hipparcos reduction between 0.14 and 0.41~mas, or relative errors between 0.35 and 1.0~per~cent. The selected stars are from the list of \citet{crawfordb69} (only those with single five-parameter astrometric solutions), including HD~113139 (HIP~63504). The average distance modulus for these stars as derived from the new Hipparcos catalogue is $2.016\pm0.051$, placing this group about 10~per~cent further away than was derived by \citet{crawfordb69}. The main reason for including these stars in the present paper and in connection with the Coma~Ber cluster is the similarity in age between the two groups. The UMa stars are not examined with the cluster-analysis tools shown in Section~\ref{sec:techn}, but rather as individual stars, taking full advantage of the high relative accuracies of the individual parallaxes. 

\begin{figure*}[t]
\centering
\includegraphics[width=17.8cm]{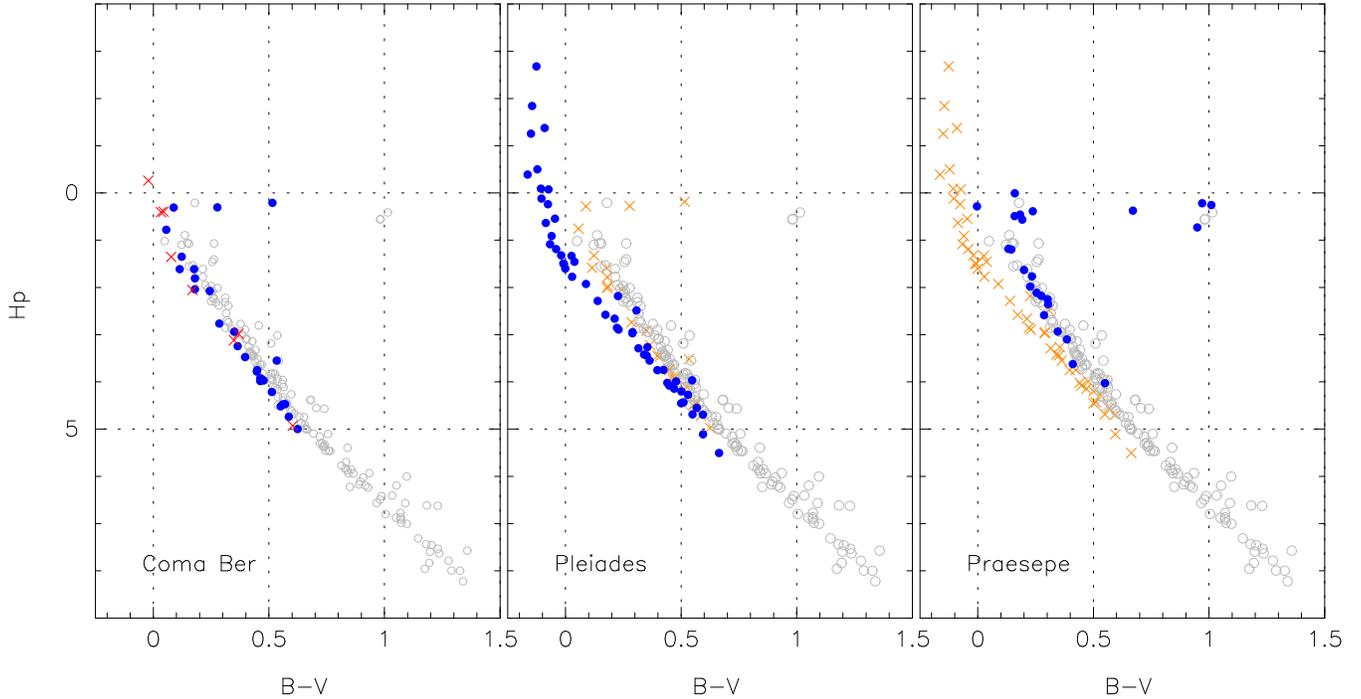}
\caption{The HR diagrams for Coma~Ber (left), Pleiades (centre) and Praesepe (right), shown as full dots in each diagram. The Hyades stars are shown in the background as open circles. In addition. The Coma~Ber diagram also shows the UMa stars (crosses), the Pleiades diagram shows the Coma~Ber stars as crosses, and similarly for the Pleiades stars are shown in the Praesepe diagram.}
\label{fig:hrdprcbpl}
\end{figure*}

A selection of 27 cluster members has been used for determining the parallax and proper motion of the Coma~Ber cluster. There may be two or three more possible members of Coma~Ber in the Hipparcos catalogue, but these deviate either in proper motion or in parallax by more than what could be expected from the formal errors. Excluding these stars had only an insignificant effect on the final results. Deriving the astrometric parameters for this, as well as the remaining clusters in this Section, has been done through an examination of the astrometric parameters as provided for the individual member stars in the new catalogue. The proper motions have been corrected for the differential projection effects described by Eq.~\ref{equ:firstappr}, using a radial velocity as specified in Table~\ref{tab:distant}, -1.2~km~s$^{-1}$. The cluster members are spread sparsely over an area with a diameter of about 7.5 degrees on the sky (Fig.~\ref{fig:mapcb}). Data correlations therefore did not play a major role when the parallax was derived using the 1997 catalogue, and it is thus not surprising that the results from the new reduction confirm and significantly strengthen the original findings by \citet{fvl99a} and \citet{robic99}. The distribution of the parallaxes and their formal errors is shown in Fig.~\ref{fig:parcb}. The summary data for Coma~Ber are presented in Table~\ref{tab:summary}. The parallax for Coma~Ber is 
determined as $11.53\pm0.12$~mas, which implies a distance modulus of
$4.691\pm0.023$. Thus, the same increase of 10~per~cent in distance as observed for the UMa group with respect to the determinations by \citet{crawfordb69}. Considering the calibration noise for ground-based differential parallaxes as determined in Section~\ref{sec:gatew}, the Hipparcos value is in good agreement with the determination by \citet{gatew95}. There is a good agreement, too, with the values derived by \citet{nicol81} and \citet{cameron85a}. The only marginally significant disagreement is with the determination by \citet{pinso98}, which differs by $3.4\sigma$. The HR-diagram for Coma~Ber is shown in Fig.~\ref{fig:hrdprcbpl} together with the results for the UMa stars and those obtained earlier for the Hyades cluster. At the parallaxes determined here, the HR diagrams of Coma~Ber and the UMa group accurately overlap, but are slightly sub-luminous ($~0.15$~mag.) with respect to the Hyades.

\subsection{The Pleiades} 
\label{sec:plei}

More than any other result from the Hipparcos mission, the parallax estimate for
the Pleiades cluster has been causing controversy \citep{pinso98,naray99,fvl99a,fvl99b,robic99,stello01,perciv03,pan04,perciv05,southw05,fvl05c,an07,vallsg07}. 
In many ways this has been highly beneficial, as it has stimulated research projects on the Pleiades. In total there have been at least 15 attempts to determine the parallax or distance modulus of the Pleiades since the publication of the Hipparcos data in 1997. It has also been one of many reasons to continue with a further detailed analysis of the Hipparcos data. With the results of a wide range of studies now available, including the new Hipparcos determination, the solution may not as yet have been identified, but the discrepancy between different approaches has become clearer, and will undoubtedly be resolved in the near future. As has been announced on a few earlier occasions \citep{fvl08}, the new reduction of the Hipparcos data did not change significantly earlier estimates for the cluster parallax.  

The Pleiades cluster covers an area on the sky of about 9 to 10 degrees in 
diameter \citep{fvl80,fvl83}, which is, contrary to statements made by \citet{an07}, much larger than the error-correlation scale length for the 1997 Hipparcos catalogue \citep{fvl99b}. In that area, the Hipparcos catalogue contains about 57 members of the cluster, but only 53 of these can be used for determining the cluster parallax. The parallax determinations for the remaining 4 stars may be disturbed by duplicity or orbital motion. Figure~\ref{fig:mapcb} shows the distribution on the sky of the selected stars. The distribution of parallax measurements and their formal errors is shown in Fig.~\ref{fig:parcb}, with respect to the measured parallax of $8.32\pm0.13$~mas (distance modulus $5.40\pm0.03$), as an internally consistent distribution of data points. In the new reduction, these data points are essentially uncorrelated (VL7B, VL7V). From the examination of the parallax zero point in the new catalogue (VL7V), there is no reason to suspect that all parallaxes in the catalogue are offset by anything more than the offset-detection limit of about 0.1~mas. 
\begin{table}[t]
\caption{Summary of distance modulus determinations for the Pleiades over the past 30 years.}
\centering
\begin{tabular}{|l|rrrr|}
\hline
Reference & DM & $\sigma$ & Type & $\Delta$ \\
\hline
\citet{an07} & 5.66 & 0.05 & 1 &  0.03 \\
\citet{soder05}  & 5.63 & 0.02 & 3 & 0.00 \\
\citet{perciv05} & 5.63 & 0.05 & 1 & 0.00 \\
\citet{southw05} & 5.71 & 0.07 & 2 & 0.08 \\
\citet{zwahl04} & 5.61 & 0.05 & 2 & -0.02 \\
\citet{munar04} & 5.61 & 0.03 & 2 & -0.02 \\
\citet{pan04} & 5.65 & 0.03 & 2 & 0.02 \\
\citet{stello01} & 5.61 & 0.03 & 1 & -0.02 \\
\citet{gatew00} & 5.59 & 0.12 & 3 & -0.04 \\
\citet{naray99}  & 5.58 & 0.18 & 4 & -0.05 \\
\citet{pinso98} & 5.63 & 0.03 & 1 & 0.00 \\
\citet{giann95} & 5.61 & 0.26 & 1 & -0.02 \\
\citet{fvl83} & 5.57 & 0.08 & 1 & -0.06 \\
\citet{nicol81} &5.73 & 0.06 & 1 & 0.10 \\
\hline
\end{tabular}
\begin{list}{}{}
\item Column 4 gives a code for the type of determination: (1) photometric; (2) binary star; (3) differential parallax; (4) convergent point.
\item Column 5 gives the difference from the weighted mean value of 5.63.
\end{list}
\label{tab:pleiades}
\end{table}
\begin{figure}[t]
\centering
\includegraphics[width=8cm]{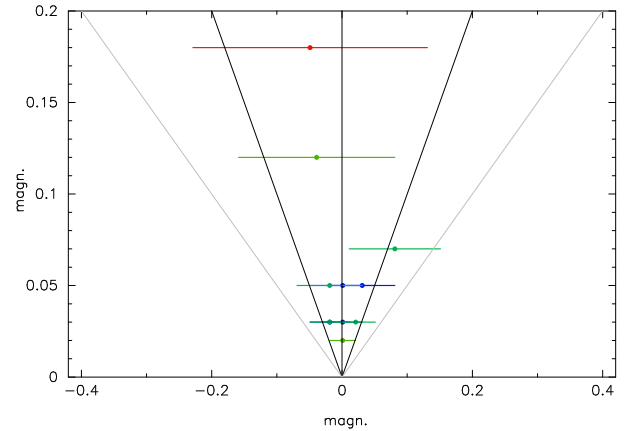}
\caption{Distribution of determinations (since 1997) for the Pleiades distance modulus and its standard error (other than those directly derived from the Hipparcos data). The distribution of the data points strongly suggests the
presence of correlations between the different measurements. The central line represents the mean value of 5.63, the black diagonal lines on either side the $1\sigma$, and the grey diagonal lines the $2\sigma$ error level.}
\label{fig:pleiadesdm}
\end{figure}
A comparison between the Pleiades parallax determination as presented above and some of the many determinations presented in the literature over the past 30 years looks, at first sight at least, alarming. Table~\ref{tab:pleiades} gives a summary for those not-Hipparcos-based observations. A couple of determinations have been left out on purpose: by \citet{gatew90} \citep[superseded by][]{gatew00}; by \citet{makar02} \citep[superseded by][and by the present paper]{fvl07}; by \citet{fox06} and by \citet{cameron85a}, where no formal error is given; and  by \citet{chen99}, which contained serious flaws in the methods used \citep[see also][]{robic99}. The remaining papers do not provide independent observations either, which becomes clear when examining their spread with respect to the quoted errors. Taking only those determinations published after the publication of the Hipparcos data in 1997 (as shown in Fig.~\ref{fig:pleiadesdm}), the weighted mean and standard deviation for 11 observations are determined as 5.63 and 0.58 respectively. The latter value should be close to 1.0 instead. When applying a more realistic formal error to the differential parallax measurements, as described in Section~\ref{sec:gatew}, the uncertainties on the distance-modulus determinations by \citet{gatew00} and \citet{soder05} are likely to be seriously underestimated (finding exactly the "expected value" does not mean that the standard deviation associated with the measuremnent is small), which would further lower the unit-weight standard deviation for the these observations. 

Of the external measurements, those based on photometry only do not constitute independent measurements of the distance modulus, but rather a fitting to a set of models. Completely independent determinations are needed to verify these models. There are only four sets of such observations available: 
\begin{enumerate}
\item The convergent point determination by \citet{naray99}, which has a high error associated with it \citep[see also][]{lmd00, fvl05c};
\item The differential parallax measurements, for which the uncertainties in the calibration to absolute values have been seriously underestimated;
\item One binary system with radial velocity measurements and an astrometric orbit by \citet{zwahl04};
\item The Hipparcos-data based parallax, derived from the measurements of 53 cluster members spread over some 50 square degrees on the sky.
\end{enumerate}
The only meaningful discrepancy with the Hipparcos result (at $3.6\sigma$) is between the results for this one binary star, when its parallax is interpreted as the parallax of the cluster. It drops to below $3\sigma$ when also the uncertainty of the position of the star within the cluster is accounted for. It should be noted that for the data on the individual star in question, there is no significant discrepancy with the Hipparcos data.

\begin{figure}
\includegraphics[width=8cm]{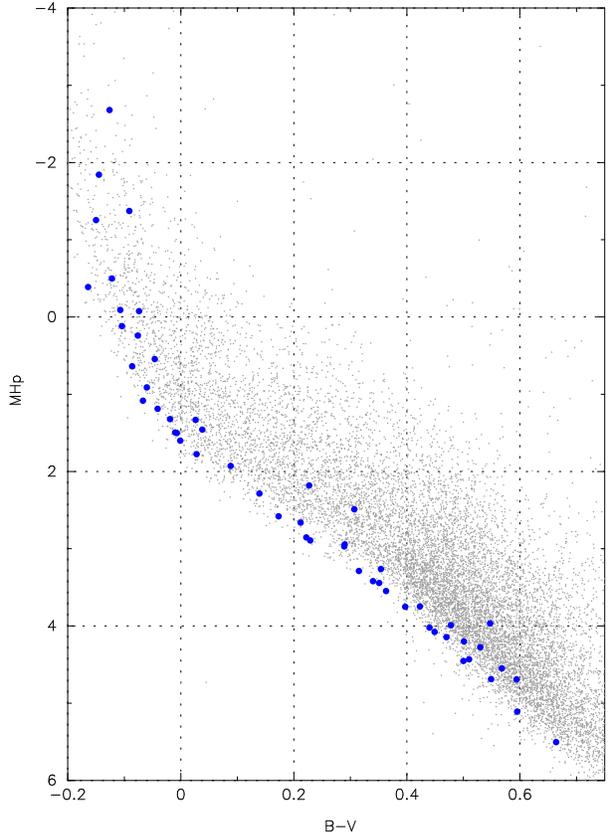}
\caption[]{The HR diagram for the Pleiades (full dots) superimposed on the HR 
diagram for field stars with parallax accuracies better than 7~per~cent. 
The distance modulus applied is the value of 5.40 magnitudes derived in the present study, and the applied reddening is $\mathrm{E(B-V)}=0.04$.}
\label{fig:hrdplei_field}
\end{figure}

It has been stated by \citet{soder98}, and re-iterated by \citet{stello01} and \citet{an07}, that at the parallax found from the Hipparcos data for the Pleiades cluster, the cluster members will occupy an area of the HR diagram in which no field stars, or more specifically no field stars that are assumed to be young, are found. This is not the case, as field stars contained  in the Hipparcos catalogue are observed in the area occupied by the Pleiades. This is shown in Fig.~\ref{fig:hrdplei_field}, where the Pleiades are compared with the data for field stars with parallax accuracies better than 7~per~cent. For those stars reddening corrections tend to be very small. At all areas covered by the Pleiades HR diagram field stars are also found. A detailed study of those stars occupying the same region would be of interest.

There are additional problems in the comparison with the active solar-type stars in \citet{soder98}. The first is that this type of activity as observed in the Pleiades is restricted to stars of spectral type later than about K2 to K3 \citep{fvl82,fvl87}. Observing similar activity among the more massive solar-type stars indicates an age for those stars considerably lower than the Pleiades age. They would therefore be expected to be found in or near star-forming regions, most likely as members of OB associations. However, most of these stars are not linked to clusters or associations. As a consequence, they do not have an independently assessed age. The "young" ages assigned to these stars is almost entirely based on interpretation of their activity \citep{soder98}.

The difference between the distance moduli of the Hyades and the Pleiades does not require an explanation in terms of metallicity. As is shown in Appendix~\ref{sec:provcalib}, the observed differences in luminosity, when accepting the Hipparcos-based parallaxes, are reflected in systematic differences in the photometric indices $\mathrm{B-U}$ and Str{\"o}mgren $c_1$ for F-type stars in the same way as what is observed for a large sample of field stars.

The summary data for the Pleiades are shown in Table~\ref{tab:summary}, and 
a comparison between the HR diagrams of the Pleiades, Coma~Ber and the Hyades
is presented in Fig.~\ref{fig:hrdprcbpl}. Clearly visible in this figure
is the difference in slope of the HR diagram for the Hyades and the Pleiades 
over the colour range 0.0 to 0.5 in $\mathrm{B-V}$. 

\subsection{Praesepe}
\label{sec:praesepe}
The Hipparcos data for Praesepe is relatively poor due to the closeness of the
cluster to the ecliptic plane. Membership is not a major problem for this cluster.  Figure~\ref{fig:mapcb} shows a the map of the 24 Praesepe members included in the Hipparcos catalogue. The area covered by the cluster is about 4\fdg5 in diameter. The parallax measurements for the selected stars are shown in Fig.~\ref{fig:parcb}. The accumulated information on the cluster can be found in Table~\ref{tab:summary}. The distance modulus as derived from the new Hipparcos reduction is, at $6.30\pm0.07$ effectively identical to one determined by \citet{an07} ($6.33\pm0.04$) and fully consistent with the value derived by \citet{perciv03} ($6.24\pm0.04$, using $(V-I)_C$ data). It is larger, but not significantly, than the value of $6.15$ derived by \citet{cameron85a} from UBV data, and the value of $6.10\pm0.18$ derived by \citet{nicol81} using photometric boxes. Also within the same range is the value derived from differential parallax measurements by \citep{gatew94}, $6.42\pm0.33$. These results, in a way, mostly reflect the very strong similarity between the Hyades and Praesepe clusters. As shown in Fig.~\ref{fig:hrdprcbpl}, the HR diagrams of the two clusters closely overlap when distance moduli are determined using the Hipparcos astrometric data. This is as expected, given the difference of 3.03 magnitudes in distance modulus, as derived from the parallax measurements, between the Hyades and Praesepe. This is very close to the value of 3.0 which has been used in the past to determine photometrically the distance of the Praesepe cluster with respect to the Hyades \citep[see for example][]{upgren79}. 

\subsection{$\alpha$~Per}
\label{sec:aper}
\begin{figure}
\includegraphics[width=6cm]{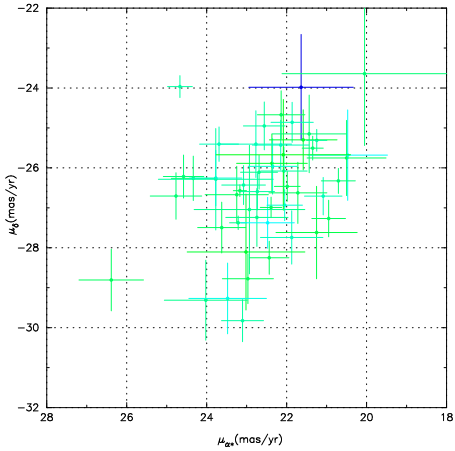}
\caption[]{Distribution of the proper motions for the selected member stars
in the $\alpha$~Per cluster.}
\label{fig:pmap}
\end{figure}

In its distribution on the sky (Fig.~\ref{fig:mapcb}), $\alpha$~Per resembles more a small remnant of an OB association than a bound open cluster. Still there is a very clear clustering in proper motion space (Fig.~\ref{fig:pmap}) and a well-defined HR diagram for the stars selected on proper motions. The Hipparcos catalogue contains 50 stars for which the membership is probable to very likely, providing a well-defined proper motion and parallax for the group (see
Table~\ref{tab:summary}). The radial velocity for this cluster was taken from \citet{petrie69} as -1.6~km~s$^{-1}$, which is based on the measurements of 77 stars in the field of the cluster. One star should be noted here in particular, HD~22192 (HIP~16826, $\psi$~Per), which was considered as a non member by \citet{petrie69}, and also as a possible non-member by \citet{abt78}, but which star is found to have a well determined proper motion and parallax that are very close to the mean cluster values, and which therefore has been considered as a high-probability member in this study. The observed spread in the radial velocities of this star as reported by \citet{petrie69} is larger than would be expected on the basis of the errors assigned to the 10 individual observations. In addition, the mean value quoted for the radial velocity of this star indicates that a straight (not weighted) mean was used, in which case the radial velocity found is $-5.0\pm2.5$~km~s$^{-1}$. For the weighted mean a value of
$-3.1\pm2.6$~km~s$^{-1}$ is found, which is fully within the range of the mean
cluster radial velocity. It is unclear how \citet{petrie69} found an
error on the mean of 1.7~km~s$^{-1}$ for what appears to be the un-weighted 
mean. But why pay all this attention to one star? As a member of the 
$\alpha$~Per cluster it is its brightest B star by more than 0.5 magnitudes, and as such it can play an important role in the determination of the age of this system. The distribution of radial velocities indicates in addition that it 
could be an orbiting binary star.

Compared to the distance-modulus values derived from the 1997 Hipparcos data by \citet{robic99} ($6.40\pm0.08$) or \citet{fvl99a} ($6.31\pm0.08$), the new determination ($6.18\pm0.03$) brings it closer to other, all photometric, determinations. As an example, the value given by \citet{dambis99} is $5.94\pm0.05$, and by \citet{pinso98} it is given as $6.23\pm0.06$. The differences with the two earlier Hipparcos-based determination stem at least partly from the selections of member stars. In the new reduction, the higher accuracies of the proper motions allows for an improvement in selection, and  shows that the selection of members by \citet{robic99} contained at least two non-members (HIP~15160 and HIP~16880) and three stars for which the membership is rather uncertain (HIP~14697, HIP~16340, and HIP~16625). Of these stars, HIP~16340 is by far the brightest, and its much smaller parallax will have affected the earlier results. 

\begin{figure*}
\includegraphics[width=16cm]{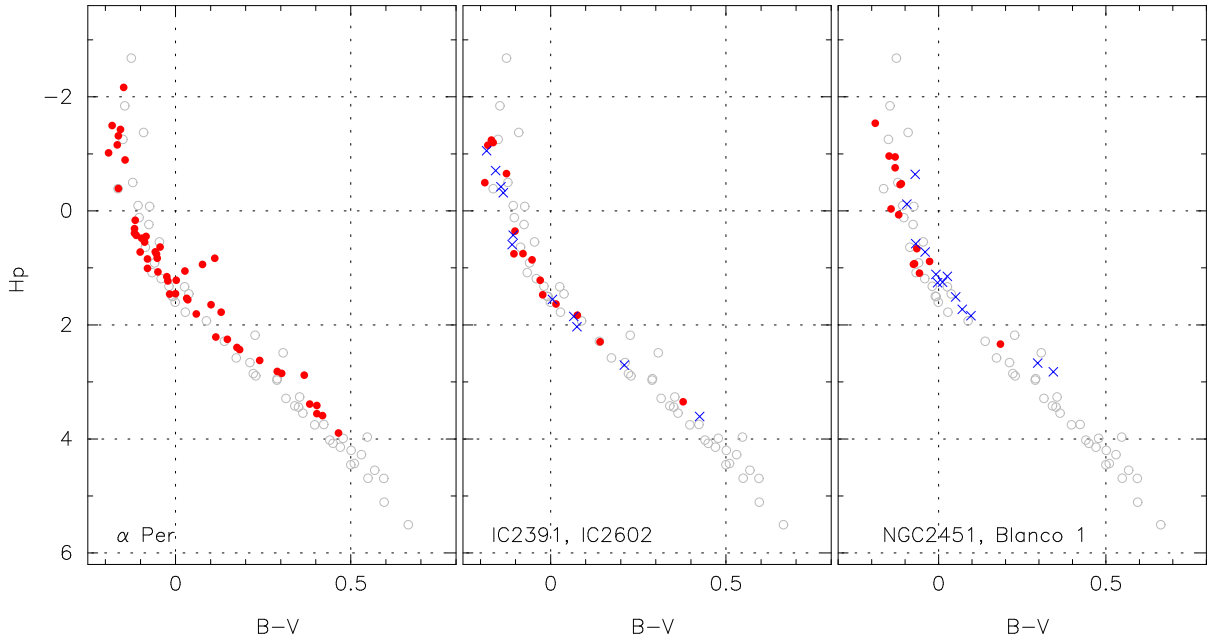}
\caption[]{The HR diagram for: (left) the $\alpha$~Per cluster (dots); (middle) IC~2062 (dots) and IC~2391 (crosses); (right) NGC~2451 (dots) and Blanco~1 (crosses). In each diagram, the Pleiades data are shown as open circles in the background as comparison.}
\label{fig:hdap}
\end{figure*}
The HR-diagram for $\alpha$~Per is shown in Fig.~\ref{fig:hdap} in comparison
with the diagram of the Pleiades cluster. There is an indication that the Pleiades stars on the red-side of $\mathrm{(B-V)}=0.2$ are generally more blue than those of $\alpha$~Per, while for stars on the blue-side the different sequences overlap very well. This raises once again the question as to what part of the main sequence can be used to derive a reliable photometric distance estimate \citep[see also][]{perciv05}, as significantly different results will be obtained for different fitting criteria.

\subsection{IC~2391 and IC~2602}
\label{sec:ic2602}

\begin{figure*}
\includegraphics[width=14cm]{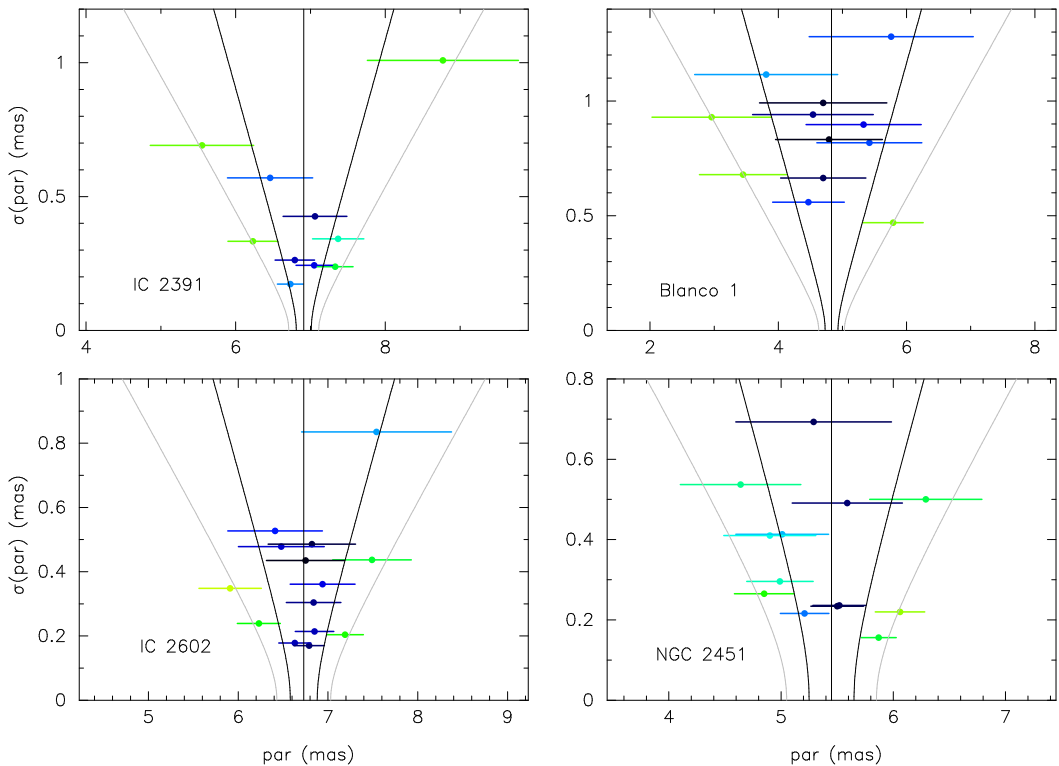}
\caption[]{The distribution of the parallaxes against their formal errors for
the IC~2391 (top left), IC~2602 (bottom left), Blanco~1 (top right) and NGC~2451 (bottom right) members in the Hipparcos catalogue. The central vertical line represents the mean parallax of each cluster, as presented in Table~\ref{tab:summary}. The black lines on either side show the $1~\sigma$ error level, including the internal distance dispersion for the cluster. The outer grey lines similarly represent the $2~\sigma$ level}
\label{fig:paric2391}
\end{figure*}

\begin{figure*}[t]
\centering
\includegraphics[width=13cm]{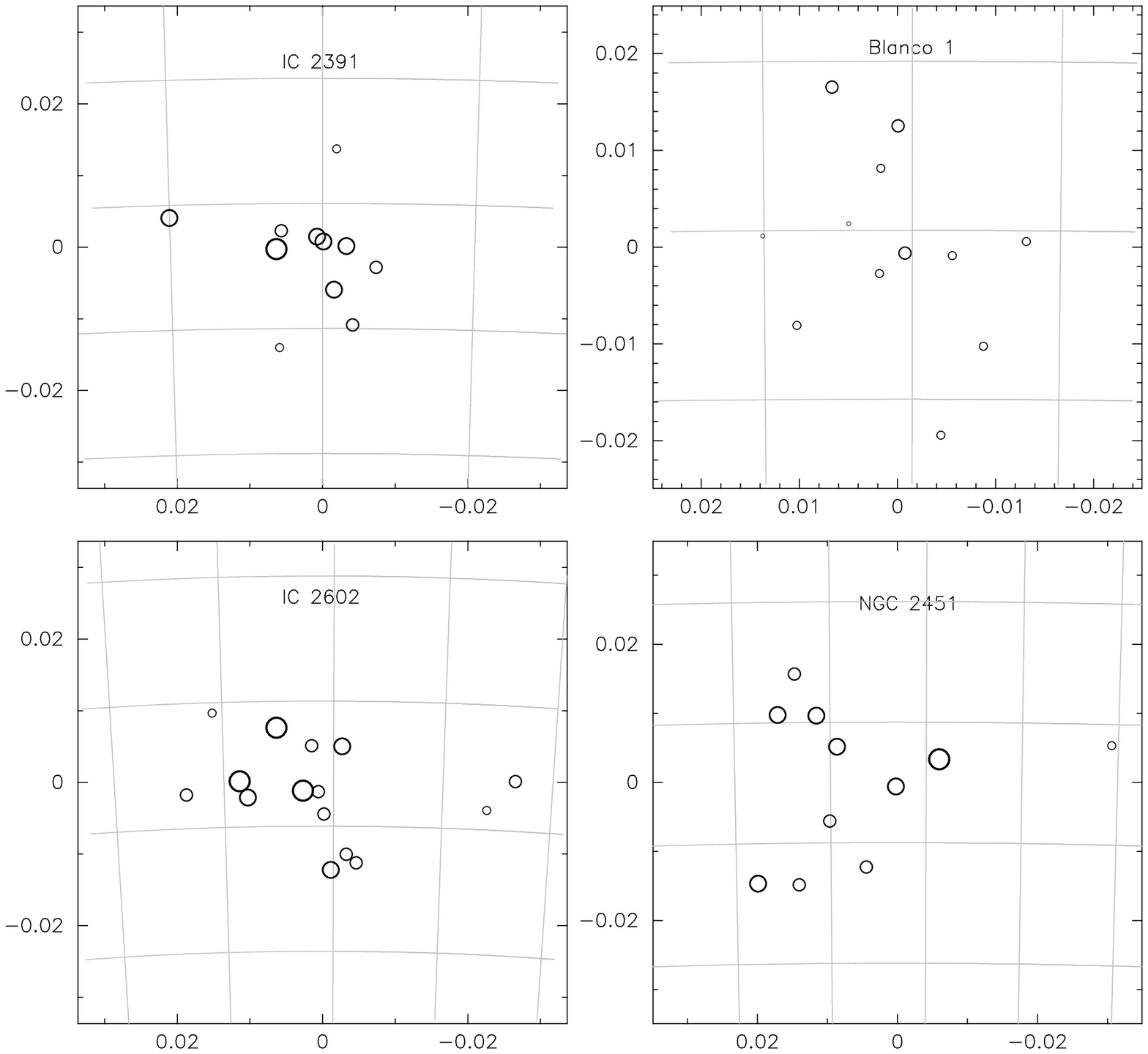}
\caption{A tangential projection of the distribution on the sky for
members of the clusters IC~2391 (top left), IC~2602 (bottom left), Blanco~1 (top right) and NGC~2451 (bottom right). The coordinates are in radians, relative to the assumed cluster centre.}
\label{fig:map_three_clust}
\end{figure*}

The Southern Hemisphere clusters IC~2391 and IC~2602 are sometimes described as being like the Pleiades, though a little younger, and less populous. IC~2602 is embedded in the Scorpio-Centaurus OB association \citep{bertiau58, zeeuw99}. An extensive photometric and astrometric study by \citet{braes61,braes62} defined probable membership and derived a distance estimate of 150~pc (distance modulus 5.88). A combined UBV photometric and spectroscopic study at the same time by \citet{whiteoak61} gave a distance of 155~pc (distance modulus 5.95), using measurements for 40 probable and 29 possible cluster members. Based on $ubvy$ photometry for 33 stars, \citet{hill69} determined a distance modulus of $5.85\pm0.06$, and \citet{nicol81}, using photometric boxes to connect the distance determination directly to stars with measured parallaxes, derived a distance modulus of $5.75\pm0.47$. Finally, \citet{pinso98} reported a distance modulus of 6.02 based on main-sequence fitting.

The proper motion and photometric study of IC~2391 by \citet{hogg60} identified 21 stars are likely members and derived a distance of $153\pm3$~pc, from a distance modulus of $5.92\pm0.05$. Measurements in $ubvy$ and $\mathrm{H}\beta$ for 25 cluster members were obtained by \citet{perry69}, from which was derived a distance modulus of $6.00\pm0.06$. They also derived a mean distance modulus of 5.90 for earlier determinations. \citet{nicol81} determined a distance modulus pf $6.13\pm0.32$ based on the photometric boxes. \citet{maitzen86} conducted a study of CP2 stars in IC~2391. It is interesting to note that four stars identified as cluster members in that study were found not to be so, based on their proper motions, though the parallaxes of these stars are close to the parallax for IC~2391. These stars are HD~74195 (HIP~42536), HD~74762 (HIP~42829), HD~74665 (HIP~42767) and HD~75202 (HIP~43071). Finally, \citet{pinso98} derived a distance modulus for IC~2391 of 5.99, based on main-sequence fitting to theoretical isochrones.

With 11 (IC~2391) and 15 (IC~2602) members in the Hipparcos catalogue respectively, they provide well determined parallax and proper motion estimates in the new reduction (Fig.~\ref{fig:paric2391}). The distance moduli as derived for these two clusters from their parallax measurements are $5.80\pm0.04$ (IC~2391) and $5.86\pm0.03$ (IC~2602), the summary data are presented in Table~\ref{tab:summary}. The distributions of the member stars are shown in Fig.~\ref{fig:map_three_clust}. Figure~\ref{fig:hdap} shows the HR diagrams of IC~2391 and IC~2602 in comparison with the diagram for the Pleiades. The positions of the sequences for the three young clusters coincide well, in the same way as the sequences for the older clusters UMa and Coma~Ber (Section~\ref{sec:coma}) and those of the much older Hyades and Praesepe clusters. With the improvement by a factor two in formal errors on the 
parallaxes for the Pleiades, IC~2391 and IC~2602 (with respect to the 1997
catalogue), this agreement is more substantial now than it was ever before. 
Just as for the Pleiades cluster, the parallax-based distance moduli for IC~2391 and IC~2602 are found to be systematically smaller than has been estimated from main sequence fitting, where values of 5.99 and 6.02 were reported by 
\citet{pinso98} for IC~2391 and IC~2602 respectively. A distance modulus
of 6.02 would imply a parallax of 6.25~mas, which, in view of 
Fig.~\ref{fig:paric2391} appears rather unlikely. 
\begin{table*}[t]
\caption{Summary data for 8 clusters within 250~pc.}
\centering
\setlength{\extrarowheight}{2pt}
\begin{tabular}{|l|rrrr|rrr|rr|r|}
\hline
Cluster & $\alpha,\delta$ & Diam. & $\log(\mathrm{Age})$ &  $V_R$ & $\mu_{\alpha,*}$ & $\mu_\delta$ & $\varpi$ & DM & Dist. & Std\\
 & degr. & Stars & $\mathrm{E(B-V)}$ & (km~s$^{-1}$) & \multicolumn{2}{c}{(mas~yr$^{-1}$)} & (mas) & & (pc) & Rej. \\ 
\hline
Coma~Ber & $186.0$ & 7\fdg5 & 8.78 & $-1.2$ & $-11.75$ & $-8.69$ & 11.53 & 4.69 & 86.7 & 0.87 \\
C1222+263 & $26.0$ & 27 &  0.00 &  & $\pm0.27$ & $\pm0.25$ & $\pm0.12$ & $\pm0.02$ & $\pm0.9$ & 0 \\
& & & & & & & & & &\\
Pleiades & $56.5$ & 8\fdg6 &  8.08 & 5.7 & $20.10$ & $-45.39$ & 8.32 & 5.40 & 120.2 & 1.45 \\
C0344+239 & $24.1$ & 53 & 0.04 &  & $\pm0.28$ & $\pm0.27$ & $\pm0.13$ & $\pm0.03$ & $\pm1.9$ & 0 \\
& & & & & & & & & &\\
Praesepe & $130.0$ & 4\fdg5 & 8.90 &  $33.6$ & $-35.81$ & $-12.85$ & 5.49 & 6.30 & 181.5 & 1.29 \\
C0937+201 & $19.7$ & 24 & 0.01 & & $\pm0.29$ & $\pm0.24$ & $\pm0.18$ & $\pm0.07$ & $\pm6.0$ & 0 \\
& & & & & & & & & &\\
$\alpha$~Per & $52.5$ & 8\fdg0 & 7.55 & $-1.6$ & 22.73 & $-26.51$ & 5.80 & 6.18 & 172.4 & 1.28 \\
C0318+484 & $49.0$ & 50 & 0.09 &  & $\pm0.17$ & $\pm0.17$ & $\pm0.09$ & $\pm0.03$ & $\pm2.7$ & 0 \\
& & & & & & & & & &\\
IC~2391 & $130.0$ & 2\fdg0 & 7.88 & 19.0 & $-24.69$ & 22.96 & 6.90 & 5.80 & 144.9 & 1.12 \\
C0838-528 & $-53.1$ & 11 & 0.01 &  & $\pm0.32$ & $\pm0.31$ & $\pm0.12$ & $\pm0.04$ & $\pm2.5$ & 0 \\
& & & & & & & & & &\\
IC~2602 & $160.2$ & 2\fdg9 & 7.83 & 16.0 & $-17.02$ & 11.15 & 6.73 & 5.86 & 148.6 & 0.99 \\ 
C1041-641 & $-64.4$ & 15 & 0.03 &  & $\pm0.24$ & $\pm0.23$ & $\pm0.09$ & $\pm0.03$ & $\pm2.0$ & 0 \\
& & & & & & & & & &\\ 
Blanco~1 & $1.1$ & 2\fdg5 & 8.32 &  3.5 & 20.11 & 2.43 & 4.83 & 6.58 & 207.0 & 1.24 \\
C0001-302 & $-30.1$ & 13 &  0.01 & & $\pm0.35$ & $\pm0.25$ & $\pm0.27$ & $\pm0.12$ & $\pm12.0$ & 0 \\
& & & & & & & & & &\\
NGC~2451 & $115.3$ & 2\fdg0 & 7.76 & 21.7 & $-21.41$ & 15.61 & 5.45 & 6.32 & 183.5 & 1.42 \\ 
C0743-378 & $-38.5$ & 14 & 0.00 &  & $\pm0.28$ & $\pm0.29$ & $\pm0.11$ & $\pm0.04$ & $\pm3.7$ & 0 \\
\hline
\end{tabular}
\begin{list}{}{}
\item The first column identifies the cluster.
\item Columns 2 to 4 provide external input data, such as reddening, radial velocity, position on the sky.
\item Columns 5 to 7 provide the data obtained in the current study for the astrometric parameters of each cluster.
\item Columns 8 and 9 provide the derived parameters of the distance and distance modulus. 
\item Column 10 gives the solution statistics, the unit-weight standard deviation of the residuals and the percentage of rejected observations. There are always three observations per star.
\end{list}
\label{tab:summary}
\end{table*}

\subsection{Blanco~1 ($\zeta$~Scl)}
\label{sec:blanco1}
The cluster Blanco~1 has in the past been referred to as the $\zeta$~Scl cluster, after the bright star considered to be a cluster member \citep{eggen70,perry78,epstein85,westerl88}. The proper motion and parallax (at $6.49\pm0.25$~mas) measured by Hipparcos, however, clearly indicate that this star is not a cluster member. A suspicion that this star might not be a cluster member was already earlier expressed by \citet{westerl88}. Also confirmed as a non-member is HD~225119 (HIP~291), based on both proper motion and parallax, while the remaining three B star members identified by \citet{westerl88} are all confirmed. As this cluster can now no longer be associated with $\zeta$~Scl, it will be referred to as Blanco~1, after its discoverer \citep{blanco49}. In previous studies of this system, the similarity between its member stars and those of the Pleiades cluster has been noticed repeatedly.

Based on the proper motions in the cluster area, 13 stars were ultimately selected as probable cluster members. A map of those stars can be seen in Fig.~\ref{fig:map_three_clust}. The astrometric parameters derived from the Hipparcos data are presented in Table~\ref{tab:summary}. The distribution of the parallaxes and their formal errors for the member stars is shown in Fig.~\ref{fig:paric2391}.

\subsection{NGC~2451}
\label{sec:N2451} 

The cluster referred to here as NGC~2451 identifies with what has been called
the Puppis Moving Group (PMG) by \citet{roeser94}, and by others as 
NGC~2451a. The reason for this is, that this cluster does not identify with
the group of stars to which originally the name NGC~2451 was assigned. That
group of stars are mainly in the background for the PMG, at about twice the 
distance, and only very few of those stars belong to the PMG.
The PMG is of  approximately the same age as IC~2391 and IC~2602. Membership
determination for this cluster has in the past been somewhat problematic 
\citep[see for example][]{levato75, maitzen86}, and it has been 
suggested that there are in fact two clusters sharing nearly the same line-of-sight \citep{roeser94}, although the existence of the second, more distant cluster, is still uncertain \citep{baumg98}. An example of the problems surrounding this cluster is the list of spectral types for 24 stars in NGC~2451 by \citet{levato75}. When incorporating the Hipparcos data, only 4 of these stars are confirmed as cluster members, 8 stars are definitely not members, while 12 stars are not included in the Hipparcos catalogue. Among the non-members is the K5~II\textit{a} star HD~63032 (HIP~37819), which was still considered a certain member by \citet{levato75}, but which has also been recognized as non-member by \citet{roeser94}. This star had been used in the past as an indicator of the age of NGC~2451. Its proper motion, at $(-10.2\pm0.18, 6.34\pm0.21)$~mas~yr$^{-1}$ and parallax at $2.88\pm0.19$~mas well separates this star from the cluster. The non members in the field of NGC~2451 as contained in the Hipparcos catalogue do not show common parallax and proper motion values that would indicate membership of a second cluster behind NGC~2451. But with the second cluster at possibly twice the distance there may be very few if any actual members of it included in the Hipparcos catalogue. It should be noted that the PPM proper motions as used by \citet{roeser94} are significantly offset in declination, by about 6 to 8 mas~yr$^{-1}$ from the Hipparcos data. The formal internal errors of about 3~mas~yr$^{-1}$ assigned to the PPM proper motions appear to be correct. 

The Hipparcos catalogue contains data for 14 stars that can be identified as 
probable members of NGC~2451 as based on their proper motions and parallaxes. 
The combined parallax data for these stars is shown in Fig.~\ref{fig:paric2391}.
The mean parallax for the 14 selected stars is $5.45\pm0.11$~mas, equivalent to a distance modulus of $6.32\pm0.04$. This is very close to the result obtained by fitting to theoretical isochrones by \citet{platais01}, who give a distance modulus of $6.35$ or $6.38$, depending on the models used. However, it was noted that the model isochrones showed shapes different from that of the observation, which means that slightly different answers can be obtained when changing the weights of the stars contributing to the isochrone fit. The summary data for NGC~2451 are given in Table~\ref{tab:summary}. The HR diagram for the cluster
is shown in Fig~\ref{fig:hdap} together with the data for Blanco~1 and the Pleiades. The HR diagrams of all three clusters appear to coincide 
well when applying the astrometrically derived distance moduli.

\section{Clusters between 250 and 500~pc}
\label{sec:clust500}

The astrometric parameters derived for clusters between 250 and 500~pc are generally less secure, as membership of individual stars can be doubtful, and generally few stars are contained in the Hipparcos catalogue. The clusters selected here are all well established, but do suffer, within the Hipparcos catalogue, from low numbers of (sometimes doubtful) members. A reasonably reliable set of astrometric parameters has been obtained for the following clusters: NGC~6475, NGC~7092, NGC~2516, NGC~2232, IC~4665, NGC~6633, Collinder~140, Trumpler~10, NGC~2422, NGC~3532 and NGC~2547. NGC~2264 is also described, but uncertainties in member identification do not allow for conclusions on its existence and astrometric parameters. The results for these clusters are presented in the sections below, including reference to  other distance determinations for the same systems and comparisons with the results for nearby clusters of approximately the same age. All cluster solutions have followed the "combined abscissae" model as described in Section~\ref{sec:distant}. A comparison was also made for NGC~6475 with the method used for the for the more nearby clusters as described in Section~\ref{sec:nearby}. The observed differences were well below the significance level.

\begin{figure*}[t]
\centering
\includegraphics[width=16cm]{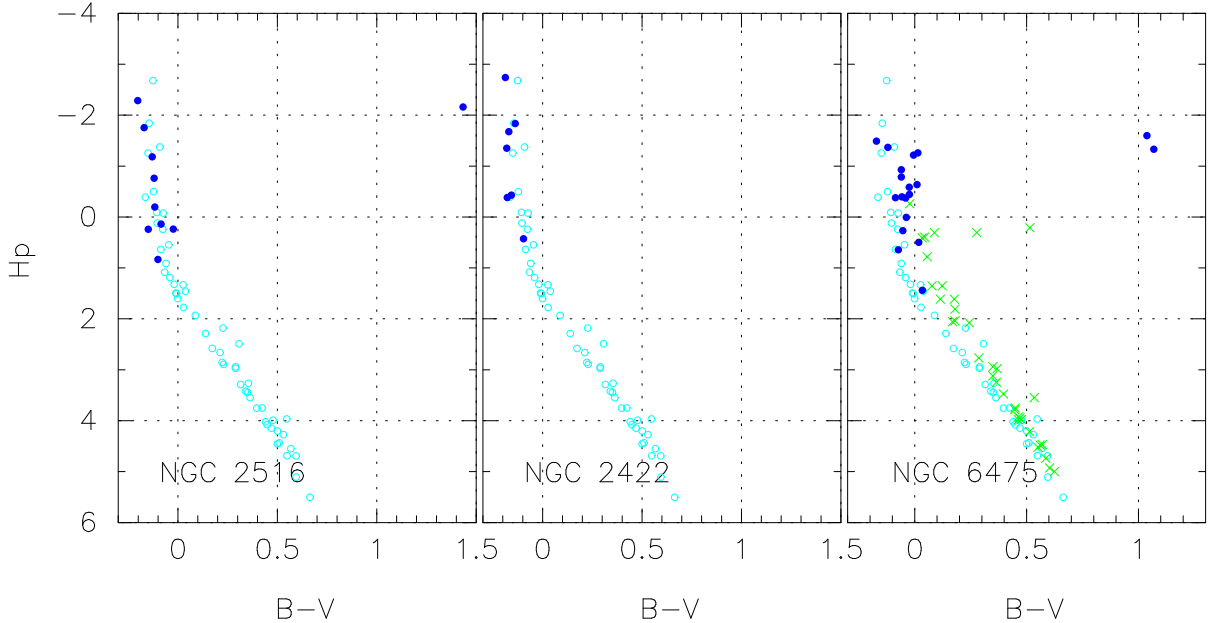}
\caption{The HR-diagram positions of the stars used in the astrometric solutions to NGC~2516 and NGC~2422 (each compared with the Pleiades, open symbols), and NGC~6475 (compared with the Pleiades, open symbols, and Coma~Ber, crosses).}
\label{fig:HRD_N2516}
\end{figure*}
\subsection{NGC~6475 (M7)}
\label{sec:n6475}
NGC~6475 is a well populated cluster, which, with at an estimated age of 200 to 260~Myr, is older than the Pleiades but younger than UMa and Coma~Ber \citep{patenaude78}. Membership has been determined in the past through both proper motions \citep{constantine69} and radial velocities by \citet{buscombe68} and \citet{zentelis83} for the brighter, and \citet{giesek85} for the fainter potential members. A spectral study was done by \citet{abt75} for the brightest members, and combined with UBV photometry by \citet{loden84}. Photometric studies in $ubvy\beta$ were done by \citet{snowden76}, \citet{loden80} and \citet{nissen88}. Recent distance modulus estimates have varied from $7.17\pm0.19$ \citep{nicol81} to 7.01 \citep{nissen88} and 7.08 \citep{pinso98}, where the latter is an apparent distance modulus. The reddening of the cluster was determined by \citet{nissen88} as $\mathrm{E(b-y)}=0.039\pm0.006$, equivalent to $0.053\pm0.008$ in $\mathrm{E(B-V)}$. \citet{pinso98} use a value of $\mathrm{E(B-V)}=0.06$. 

Within the Hipparcos catalogue 17 to 22 stars can be identified as probable members based on their proper motions. They cover an area of about 3.5 degrees in diameter, which is similar, considering the difference in distance, to the area covered by, for example, the Pleiades cluster. For the solution made here 20 stars were finally selected. Solutions with slightly different selections did not produce significantly different results. In particular restricting the extent of the cluster to a 2 degrees diameter only changed the parallax from $3.71\pm0.14$ to $3.70\pm0.14$. In all aspects the solution to the astrometric parameters of the cluster appears healthy, with only one rejected abscissa measurement out of 1328 in total used. Details of the solution are shown in Table~\ref{tab:distant}. The HR-diagram positions of the stars used in the astrometric solution are shown in Fig.~\ref{fig:HRD_N2516}, in comparison with similar data for the Pleiades and Coma~Ber. 

\subsection{NGC~7092 (M39)}
\label{sec:n7092}
A proper motion study by \citet{ebbig40} and radial velocities by \citet{trumpler28} established this group as a cluster with a radial velocity of $-14$~km~s$^{-1}$, and served as the main membership criterion for the following decades. The individual data for the latter study do not appear to have been published, but some derived information was presented by \citet{eggen51}, who also determined a photometric parallax of 4.5~mas. Another early photometric study of M39 was presented by \citet{mavers40}, and covers an area of about 1.6 degrees diameter down to photographic magnitude 13. Based on a UBV photometric study, \citet{johnson53} determined a distance modulus of $7.2\pm0.2$, equivalent to a parallax of 3.6~mas. Around the same time, \citet{weaver53} presented spectra and photoelectric magnitudes for some 30 potential cluster members. \citet{meadows61} established that for the brightest members of M39 the rotational velocities are much like those of field stars of the same spectral types, a study continued and extended to radial velocities and a search for binary stars in the cluster by \citet{abt73}. A few more radial velocities of M39 members were presented by \citet{geyer85}. Using H$\beta$ photometry calibrated to the \citet{crawford73} relations, \citet{kilkenny75} derived individual distance moduli estimates for the brighter stars in the field, and determined, using UBV photometry and the zero-age main sequence by \citet{blaauw63}, a distance modulus of $6.87\pm0.25$, as well as a reddening of $\mathrm{E(B-V)}=0.04\pm0.03$. 

A new proper motion study by \citet{mcnamara77} seemed to indicate a rather low abundance of fainter stars in the cluster. From a total of 1710 stars studied, only 30 brighter stars were found to be probable members, and most of those are at $\mathrm{(B-V)}<0.3$. A distance modulus of $7.12$ was determined by fitting to the \citet{johnson57} calibration sequence, for which a reddening of 0.02 was assumed. The apparent sparsity of fainter members may have been caused by a magnitude equation in the proper motions, as it was contradicted later by \citet{platais84} in his proper motion study of the cluster \citep[see also][]{platais94}. This study was based on more homogeneous plate material. In it, 43 new (faint) candidate members were identified. The first distance-modulus determination that was at least indirectly based on parallax measurements \citep{nicol81} provided a value of $7.46\pm0.12$, equivalent to a parallax of 3.2~mas. A UBV photometric study by \citet{mohan85} based the distance modulus estimate of $7.4\pm0.2$ on a comparison with the zero-age main sequence as given by Schmidt-Kaler in \citep{voigt65}, and estimated the age of the cluster to be between 200 and 400 Myr. The interstellar reddening has been determined for the cluster by \citet{manteiga91}, based on IR photometry, at $\mathrm{E(B-V)=0.01}$. These authors put the age of M39 to be between 240 and 480 Myr. Finally, \citet{robic99} derived the parallax and distance modulus of M39 from the first release of the Hipparcos data as $3.22\pm0.29$~mas and $7.46\pm0.20$~magn.\ respectively. 

The new solution, based on 7 member stars, gives a parallax of $3.30\pm0.19$ and therefore a distance modulus of $7.41\pm0.12$, based on the abscissa residuals for 903 field-of view transits, of which only one was rejected. The standard deviation for the astrometric solution is, at 1.03, close to the expected value. The proper motion of this cluster is relatively large, which eases the member selection. Based on its proper motion, the brightest star in the field, HIP~106346 (HD~205210) has been excluded from the solution. It appears not to be a cluster member, with also its parallax outside the $3\sigma$ area. Including this star in the solution brings the parallax down to $3.04\pm0.17$~mas (a distance modulus of $7.59\pm0.12$), which demonstrates how sensitive these solutions can be to the selection of cluster members. Details of the solution are shown in Table~\ref{tab:distant}. The HR-diagram positions of the stars used in the astrometric solution are shown in Fig.~\ref{fig:HRD_N7092}, in comparison with similar data for the Praesepe and Coma~Ber. 

\subsection{NGC~2516}
\label{sec:n2516}
\begin{figure}[t]
\centering
\includegraphics[width=8cm]{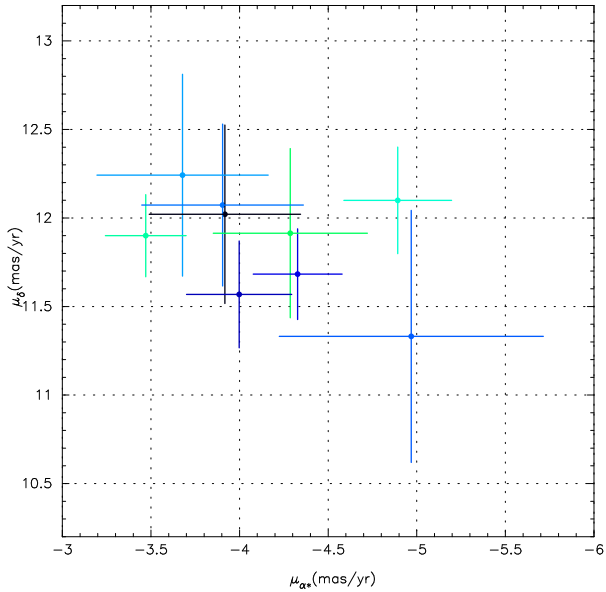}
\caption{Proper motion distribution of the stars identified as probable members of NGC~2516 in the Hipparcos catalogue}
\label{fig:pm_n2516}
\end{figure}
NGC~2516 has been described as a large and rich cluster. A study by \citet{cox55} provided the first reliable magnitudes and colours in the region of NGC~2516. From these data, Cox derived a distance estimate of $400\pm25$ pc, and a reddening of $\mathrm{E(B-V)}=0.11$. A spectroscopic study of cluster members by \citet{abt69} showed a rich variety of peculiar stars to be present in the cluster. \citet{dachs70} obtained UBV photometry for 70 stars in cluster, and derived a distance of $375\pm20$~pc (parallax of 2.7~mas) and a reddening of $\mathrm{E(B-V)}=0.09$. Dachs noted the presence of three evolved stars, and compared the cluster in age and content with the Pleiades, concluding that it is more than twice as rich. A similar study, but partially extended with measurements in RI and $\mathrm{H}\beta$ was presented by \citet{feinstein73}. In it, the distance of the cluster was estimated at $400\pm 40$~pc, and the reddening at 
$\mathrm{E(B-V)}=0.12$. The age of the cluster was estimated at $60$~Myr. \citet{snowden75}, using $ubvy\beta$ photometry, estimated the age of the cluster more than twice higher, at $135$~Myr, and re-emphasized its similarity with the Pleiades and IC~2602. The reddening was estimated at $\mathrm{E(b-y)}=0.088$. Based on Geneva photometry, \citet{nicol81} derived a distance modulus of $8.15\pm0.10$, equivalent to a distance of $427\pm22$~pc, or a parallax of 2.3~mas. A study of the cluster in Walraven VBLUW photometry by \citet{verschoor83} showed a reddening of the cluster by $\mathrm{E(B-V)}=0.127\pm0.005$, with a spread among individual cluster members at 0.043 magnitudes. The distance modulus of the cluster was derived as $7.85\pm0.10$ through a comparison with a ZAMS derived by \citet{the80}, which had been derived, through a transformation between photometric systems, from \citet{johnson66}. Based on Geneva photometric measurements for 67 stars, \citet{cameron85a} determined a distance modulus of 7.88 and a reddening of 0.12. A UBV study by \citet{dachs89} of 486 stars in the vicinity of the cluster indicated the presence of 362 cluster members in the sample. A distance modulus of $8.18\pm0.38$ and a reddening of 0.12 were derived. Spectroscopic observations for 36 bright stars in the cluster by \citet{gonzalez00} showed a radial velocity of $22.0\pm0.2$~km~s$^{-1}$. A CCD UBVI photometric study by \citet{sung02} provided a distance modulus of $7.77\pm0.11$ and reddening of $0.112\pm0.024$ through comparisons with theoretical isochrones. The age was determined as $160$~Myr.

\citet{dachs89} identified cluster members till at least 30' from the cluster centre. Within this range, 11 stars are found in the Hipparcos catalogue that very closely share the same proper motion (Fig.~\ref{fig:pm_n2516}); a further three stars showed similar but not sufficiently coinciding proper motions, and were not used as cluster members. The cluster proper motion and parallax have been estimated by means of the 1166 abscissa residuals of the selected stars, giving a value for the parallax of $2.92\pm0.10$~mas, equivalent to a distance modulus of $7.68\pm0.07$. As for the Pleiades, a shorter distance than was expected from the pre-Hipparcos estimates is found. It may be worth noting too that by extending the search radius to 1.3 degrees, 8 more stars are found that closely share the proper motion and distribution of parallaxes of the 11 selected cluster members. Scaled to the distance of the Pleiades, the radius of 1.3 degrees in NGC~2516 is equivalent to a radius of 3.7 degrees. Members of the Pleiades cluster are known at least up to 4 to 5 degrees from the cluster centre \citep{fvl86}. Including those stars in the cluster-parallax and proper motion determination gives a solution with very similar statistics as obtained with the initial selection, and a value for the parallax of $2.84\pm0.08$~mas instead (distance modulus at $7.73\pm0.06$). Details of the solution are shown in Table~\ref{tab:distant}. The HR-diagram positions of the stars used in the astrometric solution are shown in Fig.~\ref{fig:HRD_N2516}, in comparison with similar data for the Pleiades. 

\subsection{NGC~2232}
\label{sec:n2232}
The information on this cluster comes predominantly from \citet{claria72}, who measured 43 stars in its vicinity in UBV and 22 in H$\beta$. From these data Clari{\'a} derived a distance for the cluster of 360~pc, equivalent to a distance modulus of 7.78 and a parallax of 2.8~mas. The eight brightest stars in the list of cluster members by Clari{\'a} are included in the Hipparcos catalogue. They do not cluster as well in proper motion as could be expected for a cluster at 360~pc, and two stars in particular, HD~45153 (HIP~30580) and HD~45975 (HIP~31011) stand out in both proper motion and parallax. These stars have been excluded from determining the astrometric parameters of the cluster. The combined abscissa residuals for the six remaining stars give a cluster parallax of $2.83\pm0.18$~mas. Of the 522 residuals used, 2 were rejected. Including the two stars identified above as possible non-members puts the parallax at $3.27\pm0.18$~mas, and increases the standard deviation of the solution from 0.96 to 1.04. Details of the solution are shown in Table~\ref{tab:distant}. The HR-diagram positions of the stars used in the astrometric solution are shown in Fig.~\ref{fig:HRD_Cr140}, in comparison with similar data for NGC~2451 and $\alpha$~Per.

\subsection{IC~4665}
\label{sec:i4665}
Next to areas of the Pleiades and Praesepe clusters, the area of IC~4665 was one of the standard regions in the Johnson and Morgan UBV photometric system \citep{johnson54}. A study by \citet{hogg55} in UBV identified the cluster as somewhat younger than the Pleiades, at a parallax of about 2.3~mas (distance modulus of 8.2) and a reddening of $\mathrm{E(B-V)}=0.10$ to 0.18, based on a selection of 18 probable members. The membership selection was at least partly supported by the proper motion study of the cluster by \citet{vasil55}, though this study was of limited accuracy. The cluster main sequence was shown very clearly in a photographic photometry study by \citet{mccarthy69}, which incorporated further astrometric studies by Vasilevskis. Three studies in the early 70s significantly refined the information available on IC~4665, through astrometric data \citep{sanders72}, photometry in $ubvy\beta$ \citep{crawford72} and spectral classifications \citep{abtl75}. \citet{crawford72} derived a distance modulus of 7.5, and a reddening of $\mathrm{E(b-y)}=0.14\pm0.04$, where the error refers to the spread over individual cluster members. The proper motion of the cluster is quite small, and as a result there is increasing confusion between members and non-members for the fainter stars, which shows clearly from the HR diagrams of possible members by \citet{sanders72}. Based on data obtained in the Geneva photometric system, \citet{nicol81} derived a distance modulus of $7.90\pm0.22$, equivalent to a parallax of 2.6~mas. \citet{geyer85} determined the radial velocity of the cluster as $-12.5\pm4.7$, based on data for 5 stars. \citet{prosser93} combined existing and new astrometric, photometric and spectroscopic data for the cluster and identified numerous new candidate members. The age of the cluster was estimated to be between $30$~Myr and the age of the Pleiades. The radial velocity of the cluster was determined by \citet{prosser94} at about $-13$~km~s$^{-1}$.

The Hipparcos catalogue contains 6 or 7 stars that can be identified as probable members of the cluster. Four of these are contained in the astrometric study by \citet{sanders72}, and identified as probable members in their discussion: S110 (HIP~86954), S115 (HIP~86960), S159 (HIP~87002) and S181 (HIP~87032). One star can be identified in \citep{vasil55} as K~32 (HIP~86805), indicated as a possible member. Two more stars are just outside the areas covered by these studies, HIP~86993 and HIP~87592.  The mean parallax as derived from 856 field transit abscissa residuals for these 7 stars is $2.81\pm0.27$~mas, equivalent to a distance modulus of $7.75\pm0.21$. There were no rejected observations. Details of the solution are shown in Table~\ref{tab:distant}. The HR-diagram positions of the stars used in the astrometric solution are shown in Fig.~\ref{fig:HRD_N2547}, in comparison with similar data for the Pleiades, NGC~2451 and $\alpha$~Per.

\subsection{NGC~6633}
\label{sec:n6633}
\begin{figure*}[t]
\centering
\includegraphics[width=16cm]{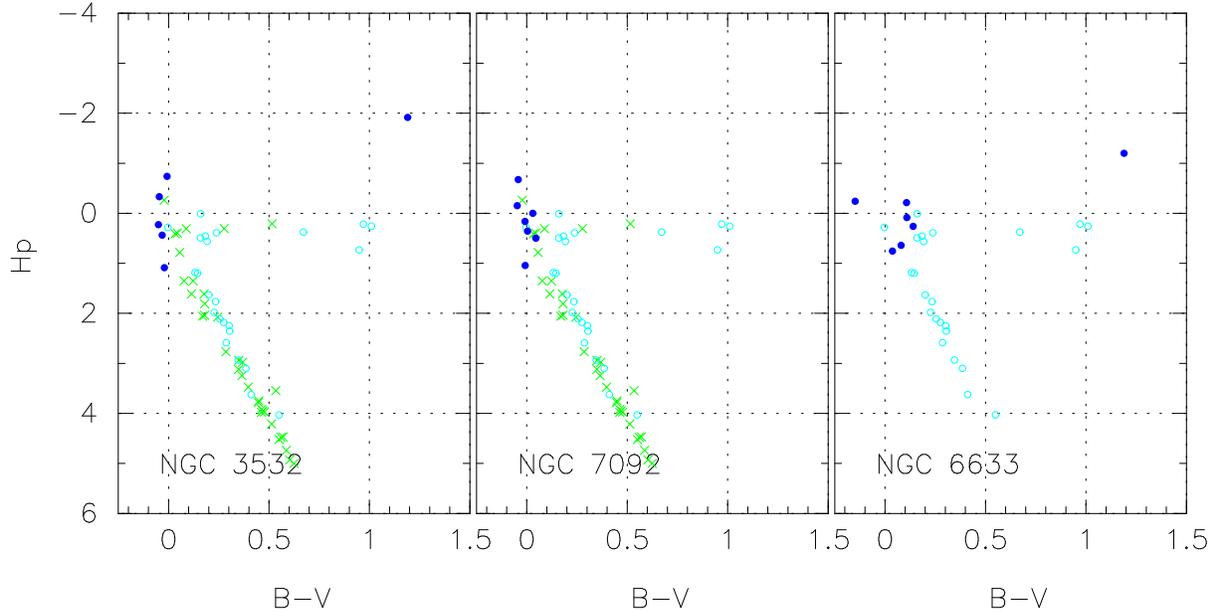}
\caption{The HR-diagram positions of the stars used in the astrometric solutions to NGC~3532 and NGC~7092 (each compared with Coma~Ber, crosses, and Praesepe, open circles), and NGC~6633 (compared with the Praesepe, open circles).}
\label{fig:HRD_N7092}
\end{figure*}
 A UBV photometric study of NGC~6633 by \citet{hiltner58} showed it to be similar in age to Praesepe, and subject to an interstellar reddening of $\mathrm{E(B-V)}=0.17$. They also derived a distance modulus of $7.50\pm0.25$, equivalent to a parallax of 3.2~mas. A proper motion study covering 207 stars down to photographic magnitude 13.5 in an area of $50'\times 40'$ centred on the cluster was presented by \citet{vasil58}, and identified some 90 potential cluster members. A study by \citet{sanders73} improved this still further, with 497 stars measured and 113 possible members. The problem faced by both these studies is the small size of the proper motions, which makes distinction from background stars difficult. The radial velocity for one of the giant members of the cluster was determined at $-22.0$~km~s$^{-1}$ by \citet{wallers63}, and later by \citet{geyer85} at $-24.8\pm1.0$. Photometric measurements in $ubvy\beta$ for 37 stars were obtained by \citet{schmidt76}, who noted a slightly variable extinction with an average value of $\mathrm{E(b-y)}=0.124$, and determined a distance modulus of 7.71, using the calibration by \citet{crawford73}. A study by \citet{levato77} determined spectral classifications for the brightest cluster members identified in the studies referred to above. Based on Geneva photometric measurements, \citet{nicol81} determined a distance modulus of $7.47\pm0.17$, while using UBV photometry by \citet{hiltner58}, \citet{cameron85a} derived a distance modulus of 7.63 and a reddening of $\mathrm{E(B-V)}=0.17\pm0.007$, based on 39 members. Distance fitting was done with respect to the ZAMS as defined by Schmidt-Kaler in \citet{voigt65}. Based on the first release of the Hipparcos data, \citet{robic99} determined a parallax of $2.70\pm0.70$~mas, using data for 4 stars. 

In the new solution, the abscissa residuals of 917 field-of-view transits for 6 stars was used, from which a parallax of $2.67\pm0.32$~mas was derived, equivalent to a distance modulus of $7.87\pm0.26$. Details of the solution are shown in Table~\ref{tab:distant}. The HR-diagram positions of the stars used in the astrometric solution are shown in Fig.~\ref{fig:HRD_N7092}, in comparison with similar data for Praesepe.

\begin{table*}[t]
\caption{Summary data for the distant clusters as derived from combined-abscissae fitting.}
\centering
\setlength{\extrarowheight}{2pt}
\begin{tabular}{|lrrr|rrr|rrr|rr|}
\hline
Cluster & $\alpha$, $\delta$ & $\log(\mathrm{Age})$ & r($^\circ$) & $\varpi$ & $\mu_{\alpha *}$ & $\mu_\delta$ & $\sigma_\mathrm{sol}$ & $\mathrm{N_{absc}}$ &  dm & $\rho_1$ & $\rho_2$ \\
 & degr. & $\mathrm{E(B-V)}$ &  $\mathrm{N_{stars}}$ & $\sigma(\varpi)$ & $\sigma(\mu_{\alpha *})$ & $\sigma(\mu_\delta$) & & $\mathrm{N_{rej}}$ & $\sigma$(dm) & & $\rho_3$ \\
\hline
NGC~6475& 268.4 & 8.22 & 1.9 &  3.70 &   2.06 &  -4.98 & 1.04 & 1328 &  7.16 &  -4 &   1 \\ 
C175$-$348 & -34.8 & 0.06 & 20 &  0.14 &   0.17 &   0.10 &  &    1 &  0.08 &     &  -8 \\ 
&&&&&&&&&&& \\
NGC~7092& 322.9 & 8.57 & 0.3 &  3.30 &  -8.02 & -20.36 & 1.03 &  903 &  7.41 &  -2 &  -6 \\ 
C2130$+$482 &  48.4 & 0.02 & 7 &  0.19 &   0.20 &   0.19 &    &    1 &  0.12 &     & -14 \\ 
&&&&&&&&&&& \\
NGC~2516& 119.4 & 8.08 & 0.8 &  2.92 &  -4.17 &  11.91 & 1.03 & 1166 &  7.68 &   0 &  -2 \\  
C0757$-$607 & -60.7 & 0.12 & 11 & 0.10 &   0.11 &   0.11 &    &    6 &  0.07 &     & -20 \\ 
&&&&&&&&&&& \\
NGC~2232&  96.9 & 7.49 & 0.8 &  2.84 &  -5.65 &  -2.56 & 0.96 &  522 &  7.73 &  -9 &   0 \\ 
C0624$-$047  &  -4.7 &  0.01 & 6 &  0.18 &   0.16 &   0.14 &    &    2 &  0.14 &     &   4 \\ 
&&&&&&&&&&& \\
IC~4665& 266.8 & 7.63 &1.8 &  2.81 &  -0.90 &  -8.27 & 1.04 &  856 &  7.75 &  -5 & -16 \\ 
C1743$+$057 &   5.6 &  0.19 & 7 &  0.27 &   0.24 &   0.16 &    &    0 &  0.21 &     &  -7 \\ 
&&&&&&&&&&& \\
NGC~6633& 276.5 & 8.76 & 0.9 &  2.67 &   1.01 &  -1.15 & 0.98 &  917 &  7.87 &  10 &  14 \\ 
C1825$+$065 &   6.7 & 0.17 & 6 &  0.32 &   0.31 &   0.28 &    &    0 &  0.26 &     &  25 \\ 
&&&&&&&&&&& \\
Coll~140& 111.0 & 7.57 & 0.9 &  2.66 &  -7.50 &   4.02 & 1.01 & 1413 &  7.88 &   6 &  -7 \\ 
C0722$-$321 & -32.1 & 0.03 & 9 &  0.13 &   0.09 &   0.13 &    &    4 &  0.11 &     &  -3 \\ 
&&&&&&&&&&& \\
Trumpler 10& 131.7 & 7.38 &1.4 &  2.58 & -12.48 &   6.81 & 1.07 & 1532 &  7.94 &   8 &   6 \\ 
C0646$-$423 & -42.5 &  0.06 & 12 &  0.16 &   0.14 &   0.14 &    &   9 &  0.14 &     &  -9 \\ 
&&&&&&&&&&& \\
NGC~2422& 114.1 & 8.12 & 0.3 &  2.52 &  -6.72 &   0.82 & 1.05 &  879 &  7.99 & -21 &  -6 \\  
C0734$-$143 & -14.6 & 0.12 & 7  & 0.21 &   0.16 &   0.14 &    &    0 &  0.18 &     &  11 \\
&&&&&&&&&&& \\
NGC~3532& 166.5 & 8.45 & 0.2 &  2.43 & -10.04 &   4.75 & 0.90 &  685 &  8.07 &   6 &   5 \\
C1104$-$584 & -58.7 & 0.02 & 6 &  0.24 &   0.24 &   0.21 &   &    0 &  0.22 &     &  33 \\ 
&&&&&&&&&&& \\
NGC~2547& 122.5 & 7.70 & 0.7 &  2.11 &  -9.34 &   4.18 & 0.99 & 1022 &  8.38 &  13 &  15 \\ 
C0809$-$491  & -49.2 & 0.05 & 8 &  0.17 &   0.18 &   0.16 &  &    2 &  0.17 &     &   0 \\  

\hline
\end{tabular}
\begin{list}{}{}
\item Ages have been obtained from \citet{khar05}, reddening values come from various papers.
\item The final two columns give the correlation coefficients between the proper motions and parallax determinations times 100 ($\rho_1\equiv\rho(\varpi,\mu_{\alpha,*})$; $\rho_2\equiv\rho(\varpi,\mu_\delta)$; $\rho_3\equiv\rho(\mu_{\alpha,*},\mu_\delta)$).
\end{list}
\label{tab:distant}
\end{table*}  

\subsection{Collinder~140}

A combined photometric and spectroscopic study of Collinder~140 was presented by \citet{claria78}, who derived a distance modulus of 7.78 (equivalent to a distance of 360~pc), reddening of $\mathrm{E(B-V)}=0.05$, and age of 21 to 23~Myr. From a UBV photometric study, \citet{williams78} placed the cluster at $420\pm20$~pc and at an age of 40~Myr, and very little reddening, at $\mathrm{E(B-V)}=0.01\pm0.01$. A spectroscopic study by \citet{fitzgerald80} derived a distance of $410\pm30$~pc, a reddening of $\mathrm{E(B-V)}=0.04$, and age of $20\pm6$ Myr. In his UBV photometric study of 38 clusters, \citet{cameron85a} determined a distance modulus of 7.96 (distance 391~pc) and reddening of $\mathrm{E(B-V)}=0.06$.

The distribution of proper motions and parallaxes in the region of Collinder~140 in the new Hipparcos catalogue shows the presence of 9 probable cluster members, for which the parallax, as derived from the 1413 individual field transits, is $2.66\pm0.13$~mas, equivalent to a distance modulus of $7.88\pm0.11$. Details of the solution are shown in Table~\ref{tab:distant}. The HR-diagram positions of the stars used in the astrometric solution are shown in Fig.~\ref{fig:HRD_Cr140}, in comparison with similar data for NGC~2451 and $\alpha$~Per.

\subsection{Trumpler 10}
\begin{figure*}[t]
\centering
\includegraphics[width=16cm]{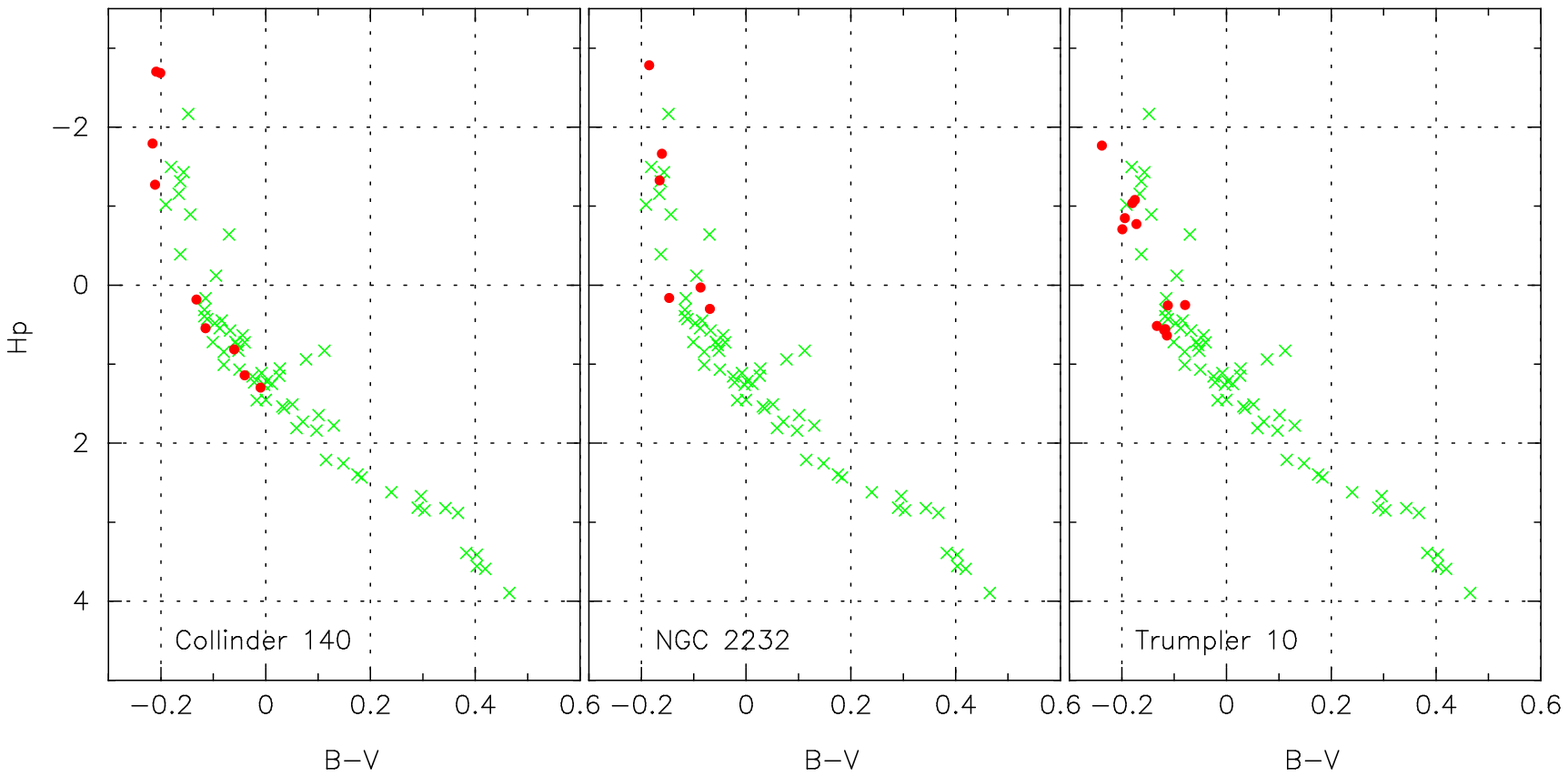}
\caption{The HR-diagram positions of the stars used in the astrometric solutions to Collinder~140, NGC~2232 and Trumpler~10 (each compared with $\alpha$~Per and NGC~2451, crosses).}
\label{fig:HRD_Cr140}
\end{figure*}
Trumpler~10 is a not very well studied, fairly young cluster. \citet{lynga60} did a UBV photometric study in the region of the cluster, measuring some 29 stars, for which a further analysis was provided by \citet{lynga62}, who derived an age estimate of $30$ to $60$~Myr. Spectroscopic types in the cluster were determined by \citet{levato75a}, who gave an estimate of the cluster distance of $440\pm50$~pc, equivalent to a distance modulus of 8.22, and reddening at $\mathrm{E(B-V)}=0.02\pm0.01$. A proper motion study by \citet{stock84} seemed to throw some doubt on the existence of the cluster, but the proper motions in this study were not of sufficient accuracy to recognize it.

A search for potential cluster members in the region of Trumpler~10 identified 12 likely cluster members, with a mean parallax of $2.58\pm0.16$~mas, equivalent to a distance modulus of $7.94\pm0.14$, close to the earlier ground-based estimate.  Details of the solution are shown in Table~\ref{tab:distant}. The HR-diagram positions of the stars used in the astrometric solution are shown in Fig.~\ref{fig:HRD_Cr140}, in comparison with similar data for NGC~2451 and $\alpha$~Per. 

\subsection{NGC~2422 (M47)}
\label{sec:n2422}

The only two proper motion studies on this cluster are by \citet{schewick66} and \citet{ishmuk66}, neither of which is easily accessible and played much of role in further investigations of this cluster. A UBV photometric study was presented by \citet{smyth62}, and a spectroscopic study by \citet{dworetsky75}. \citet{nicol81} determined a distance modulus of $8.66\pm0.16$ using Geneva photometry. An extensive study of the cluster in $ubvy\beta$ was carried out by \citet{rojo97}, who determined a distance of $470.8\pm4.8$ (distance modulus 8.36) and a reddening of $\mathrm{E(b-y)}=0.089\pm0.026$.  The age was estimated at $100$~Myr. Analysis of the first release of the Hipparcos data by \citet{robic99} showed a parallax of $2.01\pm0.43$. A CCD-photometry study in UBVRI by \citet{prisin03} was used to study the luminosity and mass functions of the cluster.

The Hipparcos catalogue contains at least 7 stars that are probable cluster members. The abscissa residuals for the 879 field transits of these stars show a cluster parallax of $2.52\pm0.21$~mas, equivalent to a distance modulus of $7.99\pm0.18$. There were no rejected observations. Details of the solution are shown in Table~\ref{tab:distant}. The HR-diagram positions of the stars used in the astrometric solution are shown in Fig.~\ref{fig:HRD_N2516}, in comparison with similar data for the Pleiades. 

\subsection{NGC~3532}
\label{sec:n3532}
This rich open cluster was studied in photometry and astrometry by \citet{koelbloed59}, who derived a distance of $432\pm40$~pc (a distance modulus of 8.2), a very small amount of reddening ($\mathrm{E(B-V)}=0.01$), and age of $100$~Myr. Further proper motions were obtained in an extensive study by \citet{king78}, who identified 382 likely members in a field of about 1 degree square, down to photographic magnitude 11.0. A UBV photometric study of 183 stars in the region of the cluster by \citet{fernandez80} indicated a slightly larger distance of $490\pm50$~pc (a distance modulus of $8.45\pm0.27$) and age of $200$~Myr. \citet{giesek81} determined a radial velocity of 4.2~km~s$^{-1}$ for the cluster, and identified and measured a number of spectroscopic binaries contained in it. The larger distance and higher age seemed to be confirmed by $ubvy\beta$ observations of 33 stars by \citet{eggen81}, who derived a distance modulus of $8.5\pm0.25$, a reddening of $\mathrm{E(B-V)}=0.023\pm0.014$, and age of $350$~Myr. A somewhat smaller distance modulus of $8.296\pm0.096$ was derived by \citet{nicol81} (distance 456~pc). Using photometric observations of 14 cluster stars in UBV and $ubvy\beta$, \citet{johansson81} derived a distance modulus of $8.38\pm0.51$, and a reddening of $\mathrm{E(B-V)}=0.10\pm0.04$. Further photometric observations for 68 stars, this time in UBVRI, were obtained by \citet{wizinow82}. This study showed some of the systematic differences that exist between different implementations of the same photometric system. Observations in UBV and DDO of 12 red giants in the cluster by \citet{claria88} showed a mean reddening of $\mathrm{E(B-V)}=0.07\pm0.02$. In their analysis of the first release of the Hipparcos data, \citet{robic99} derived a parallax of $2.47\pm0.39$~mas for the cluster, equivalent to a distance modulus of $8.04\pm0.35$.

In the new reduction of the Hipparcos data, 6 stars can be identified as very probable cluster members, having closely the same proper motion. Based on the 685 field-transit measurements for these stars, a cluster parallax of $2.43\pm0.24$~mas is derived, equivalent to a distance modulus of $8.07\pm0.22$.  Details of the solution are shown in Table~\ref{tab:distant}. The HR-diagram positions of the stars used in the astrometric solution are shown in Fig.~\ref{fig:HRD_N7092}, in comparison with similar data for Praesepe and Coma~Ber. 

\subsection{NGC~2547}
\label{sec:n2547}
\begin{figure}[t]
\centering
\includegraphics[width=9cm]{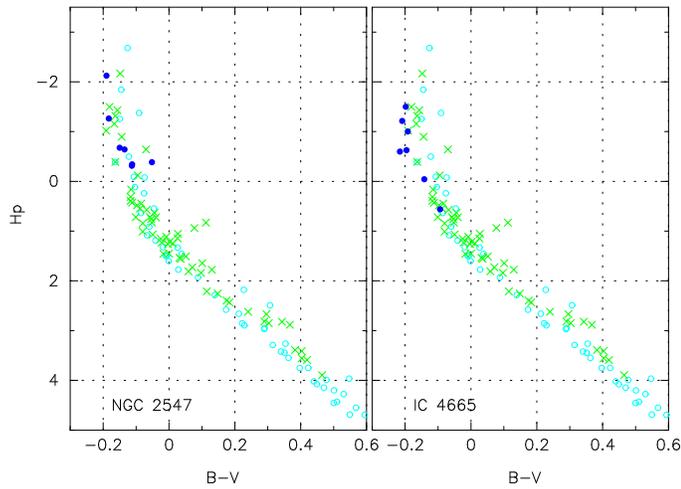}
\caption{The HR-diagram positions of the stars used in the astrometric solutions to NGC~2547 and IC~4665 (each compared with Pleiades, open symbols and $\alpha$~Per and NGC~2451, crosses).}
\label{fig:HRD_N2547}
\end{figure}
Photometric and astrometric studies of NGC~2547 were presented by \citet{fernie59, fernie60}, from which he derived an estimated distance of 360~pc (distance modulus 7.8) and reddening of $\mathrm{E(B-V)}=0.03$. \citet{nicol81} used Geneva photometry to derive a distance modulus of $8.22\pm0.13$ (distance 440~pc). Photoelectric data in UBV for 118 stars, and H$\beta$ for 23 stars in the region of NGC~2547 were obtained by \citet{claria82}, from which was derived a mean distance of 450~pc (distance modulus 8.3), a reddening of $\mathrm{E(B-V)}=0.06\pm0.02$ and age of $57$~Myr. Based on UBV data, \citet{cameron85a} derived a distance modulus of 8.1 and reddening of $\mathrm{E(B-V)}=0.05$. Analysis of the first release of the Hipparcos data by \citet{robic99} showed a parallax of $2.31\pm0.29$. A BVI CCD study by \citet{naylor02} gave a distance modulus between 8.00 and 8.15, and age between $20$ and $35$~Myr.

Eight stars were identified in the Hipparcos catalogue as probable cluster members based on their distribution of proper motions. The cluster parallax of $2.11\pm0.17$ (distance modulus $8.38\pm0.17$) was derived from the abscissa residuals of the 1022 field transits of these stars. There were 2 data points rejected. Details of the solution are shown in Table~\ref{tab:distant}. The HR-diagram positions of the stars used in the astrometric solution are shown in Fig.~\ref{fig:HRD_N2547}, in comparison with similar data for the Pleiades, NGC~2451 and $\alpha$~Per.

\subsection{NGC~2264}
\label{sec:n2264}
\begin{figure}[t]
\centering
\includegraphics[width=8cm]{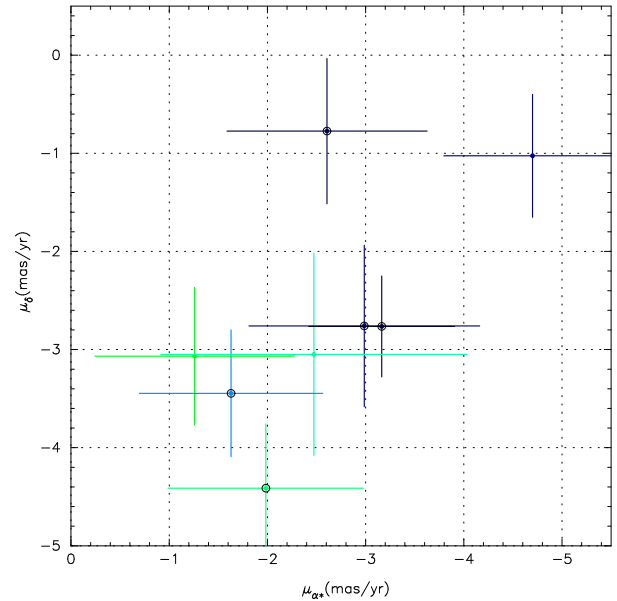}
\caption{Proper motions for nine stars in the field of NGC~2264. Data points indicates with a larger central dot represent the five stars selected as probable members based on a combination of proper motions, parallaxes and apparent magnitudes.}
\label{fig:pm_n2264}
\end{figure}
NGC~2264 is an extremely young cluster, which, probably for that reason, has been studied well. An extensive photometric survey in UBV was presented by \citet{walker56}, who identified the lower main sequence to lie above the expected locus when comparing his data with the "standard main sequence" of \citet{johnsonm53}, to which the data were fitted for the B stars. A distance modulus of 9.50 was derived this way, based on which it was concluded that stars of spectral types later than A0 were situated above the main sequence. A proper motion study by \citet{vasil65} identified 140 stars with membership probabilities greater than 0.5 among 245 stars brighter than photographic magnitude 13, in a $40'\times 40'$ field centred on the cluster. A strongly variable reddening for the cluster stars was reported by \citet{young78}, possibly related to the molecular cloud associated with the cluster \citep{crutcher78}, though much less variation was indicated by the spectroscopic study of \citet{barry79} and the UBVRI photometric study of \citet{mendoza80}. \citet{nicol81} determined a distance modulus of $9.64\pm0.67$, equivalent to a parallax of 1.2~mas. the radial velocity for the cluster was determined by \citet{liu89} as $24\pm8$~km~s$^{-1}$, based on observations of 12 early-type cluster members. Using a combination of photometric and spectroscopic data, \citet{perez87} derived a distance modulus of $9.88\pm0.17$ through a comparison with the main sequence as defined by Schmidt-Kaler in \citet{schaifers82}. 
\begin{table}[t]
\caption{Data on stars selected in the NGC~2264 field}
\centering
\begin{tabular}{|rlr|rrrr|}
\hline
HIP & id  &  Hp & $\varpi$ & $\mu_{\alpha,*}$ & $\mu_\delta$ & gof \\
HD  & p   &  B$-$V & $\sigma_\varpi$ & $\sigma_\mu$ & $\sigma_\mu$ & rej\\
\hline
 31917 &    & 8.717   & 2.12 & -2.57 & -0.86 &-1.44 \\
 47662 &   & -0.126   & 1.02 &  1.02 &  0.74 & 0 \\
&&&&&&\\
 31939 & V37 & 8.100  &  1.49 & -1.60 & -3.47 & 0.17 \\
 47731 & 0.94 & -0.143 &   0.93 &  0.93 &  0.65 & 0 \\
&&&&&&\\
 31951 &    & 8.786   & 1.98 & -2.96 & -2.84 &-1.17 \\
 47751 &   & -0.092   & 1.14 &  1.17  & 0.82 & 0 \\
&&&&&&\\
 31955 & V58 & 7.904  &  1.03 & -1.96 & -4.42 & 0.12 \\
 47777 & 0.98 & -0.162 &   0.86 &  1.00 &  0.66 & 0 \\
&&&&&&\\
 31978 & S~Mon & 4.555  &  3.55 & -2.61 & -1.61 & -0.82 \\
 47839 &   & -0.233   & 0.50 &  0.56  & 0.39 &1 \\
&&&&&&\\
 32030 & V168 & 7.470  &  2.22 & -3.16 & -2.79 & -0.18 \\
 47961 & 0.98  &  -0.143  &  0.69 &  0.74 &  0.51 &2 \\ 
&&&&&&\\
 32053 & V206 & 8.942   & 3.90 & -2.48 & -3.04 & 0.00 \\
 48055 & 0.94 & -0.133  &  1.16 &  1.56 &  1.03 &2 \\
&&&&&&\\
 32141 &    & 8.530  & 1.97  & -4.73 & -0.95 & -0.95 \\
 48232 &    & 0.035  &  0.95 &  0.90 &  0.63 & 0 \\
&&&&&&\\
 32245 &    & 8.373  & 3.94 & -1.32 & -3.01 & 1.61 \\
 48427 &    & 0.062  & 0.95 &  1.01 &  0.70 & 2 \\ 
\hline
\end{tabular}
\label{tab:n2264}
\end{table}

The Hipparcos catalogue contains at least 4 stars classified by \citet{vasil65} as highly probable members (see Table~\ref{tab:n2264}). In addition, the catalogue contains the brightest star in the field, S~Mon, which has a proper motion very similar to the four stars mentioned above. A little outside the field studied by \citet{vasil65}, four more stars can be identified to share the same proper motion, and are within the same range of parallax values. Details on all these stars are provided in Table~\ref{tab:n2264}, while their proper motions are shown in Fig.~\ref{fig:pm_n2264}. Taking the data for all 9 stars together, a parallax of $2.30\pm0.46$~mas is obtained, which is why this cluster became included in the initial selection of clusters within 500~pc. This solution shows, however, a relatively large standard deviation of 1.4, and also the combined positions of the selected stars in the HR diagram look suspicious. The stars that seem to contribute most to the large standard deviation are three in the south of the field and S~Mon, for which parallaxes near 3.5~mas are found. Eliminating these, a parallax of $1.81\pm0.40$~mas is found (from the abscissae solution), and a standard deviation of 1.01. Of the 375 abscissae used, one was rejected in iterations. Given its uncertainty in the parallax determination as well as the final value, the data for this cluster have not been included in Table~\ref{tab:distant}. They have been included in the discussion here to show the vulnerability of these solutions to more distant clusters to member selection criteria.
 
\section{Discussion}
\label{sec:disc}
One of the aims of this study is to provide for nearby clusters distance moduli that are determined completely independent from main-sequence fitting or other types of photometric calibration. Ultimately, when combined with homogeneous photometry for more extensive selections of cluster members, those data can provide a set of observational isochrones, to be compared both qualitatively and quantitatively with theoretical isochrones. A first, but very limited, combination with photometric data from the Hipparcos catalogue already shows some interesting relations as well as systematic differences. More extensive comparisons can be made when combining the cluster parallaxes with, for example, $ubvy\beta$ or Geneva photometry, which is available for many of the clusters discussed above \citep{mermilliod95}. 

\subsection{Distance modulus and parallax comparisons}

In this section the distance moduli as have been determined in various studies over the past 50 years are examined with respect to the new Hipparcos examinations. The comparisons are done as a function of cluster age, reflecting any systematic errors in the calibration sequences used. The ages used come from the catalogue by \citet{khar05}.

\begin{figure}[t]
\centering
\includegraphics[width=8cm]{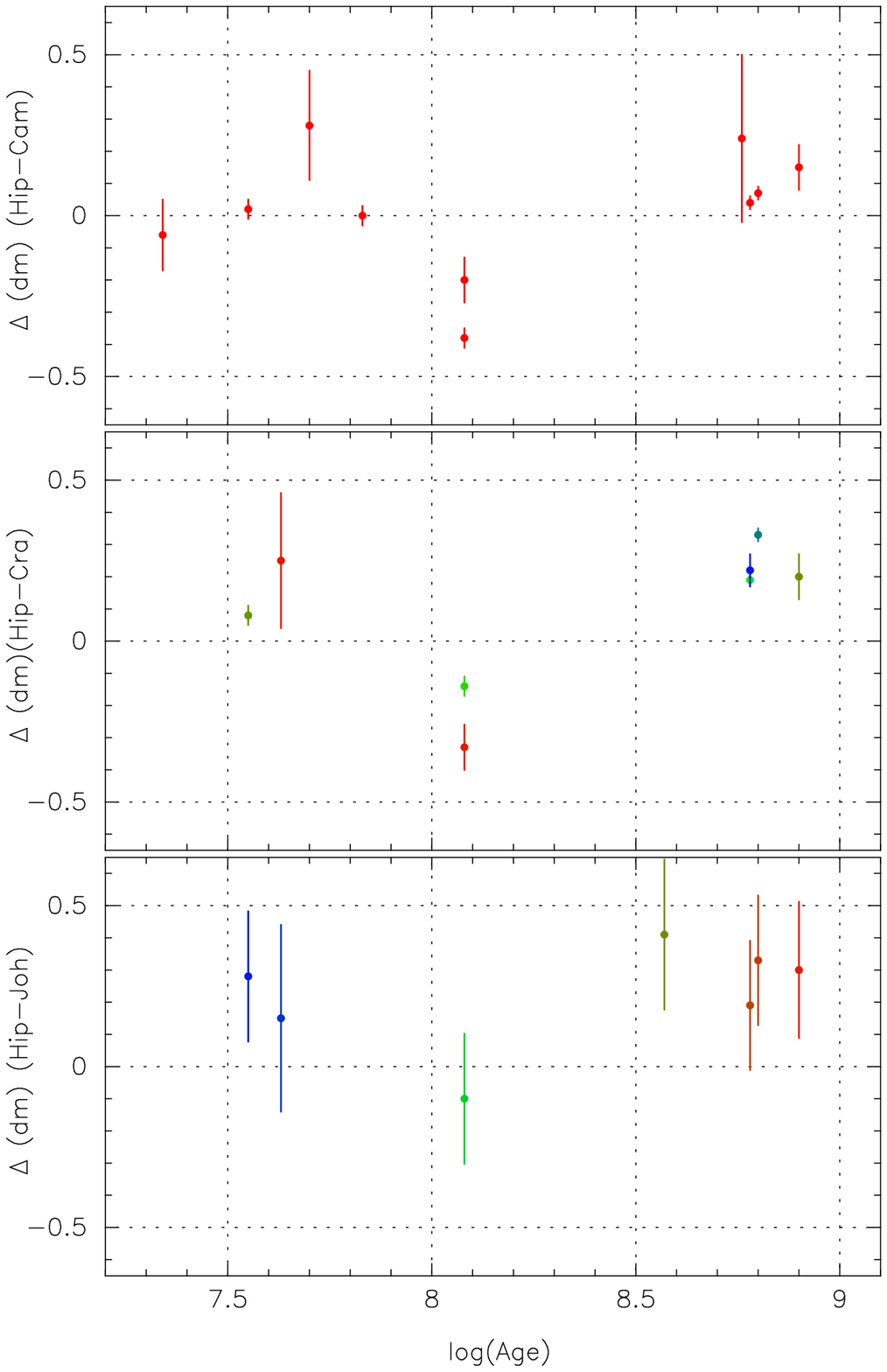}
\caption{Differences in distance moduli between values derived by \citet{johnson57} and the new Hipparcos reduction (bottom graph) and similarly (middle graph) for the data for the $ubvy\beta$ photometry \citep{crawford66AJ,crawford69,crawford72,crawford74,crawford76,crawfordb69}, and (top graph) data from \citet{cameron85a}. In the bottom graph the errors assigned to the distance moduli are mainly the uncertainty of 0.2 magnitudes indicated by Johnson. In the other two graphs the error bars represent only the Hipparcos measurements.}
\label{fig:johnson_dm}
\end{figure}
The calibration of the ZAMS by \citet{johnson57} has made a large impact, direct as well as indirect, on distance calibrations of open clusters for about half a century. It was built on three assumptions, the distance of the Hyades, assumed to be 40~pc (compared to the current estimate of $46.45\pm0.50$~pc), an area in the UBV colour-magnitude diagram ($\mathrm{0.55<B-V<0.85}$) where the Pleiades and Hyades would share the same locus, and a connection between NGC 2362 and the Pleiades around $\mathrm{(B-V)=0}$, which was based on a theoretical evolutionary correction for the magnitudes based on studies by \citet{tayler54, tayler56}. None of these assumptions could be verified at the time, as it is only by means of accurate parallaxes that such verification is made possible. Johnson estimated the uncertainty on the distance moduli determined this way to be of order 0.2 magnitudes. The distance moduli derived by Johnson played an important role in the early distance moduli determination through $ubvy\beta$ photometry \citep{crawford66AJ,crawford69,crawford72}. Figure~\ref{fig:johnson_dm} shows the observed differences between distance moduli for clusters based on these calibrations and the new Hipparcos determination. The distance of the Hyades as used by \citet{johnson57} was underestimated by 15~per~cent, resulting in errors of around 0.3 magnitudes in the distance estimates for similar clusters, such as Praesepe. The assumption Johnson made about the coincidence of the Pleiades and the Hyades is not supported by the Hipparcos data. In the $ubvy\beta$ photometry it would not be supported either if systematic differences in $c_1$ are taken into account. The details of the connection between the Pleiades and the youngest open clusters, as implemented by Johnson, are also in need of adjustment. 

The ZAMS definition by \citet{johnson57} played also an important role in the ZAMS as presented by Schmidt-Kaler in \citet{voigt65}. This was used by \citet{cameron85a} for the distance modulus determination of 38 open clusters. As can be observed in Fig.~\ref{fig:johnson_dm}, the same systematic differences  are observed as for the Johnson and Crawford calibrations.  

\begin{figure}[t]
\centering
\includegraphics[width=8cm]{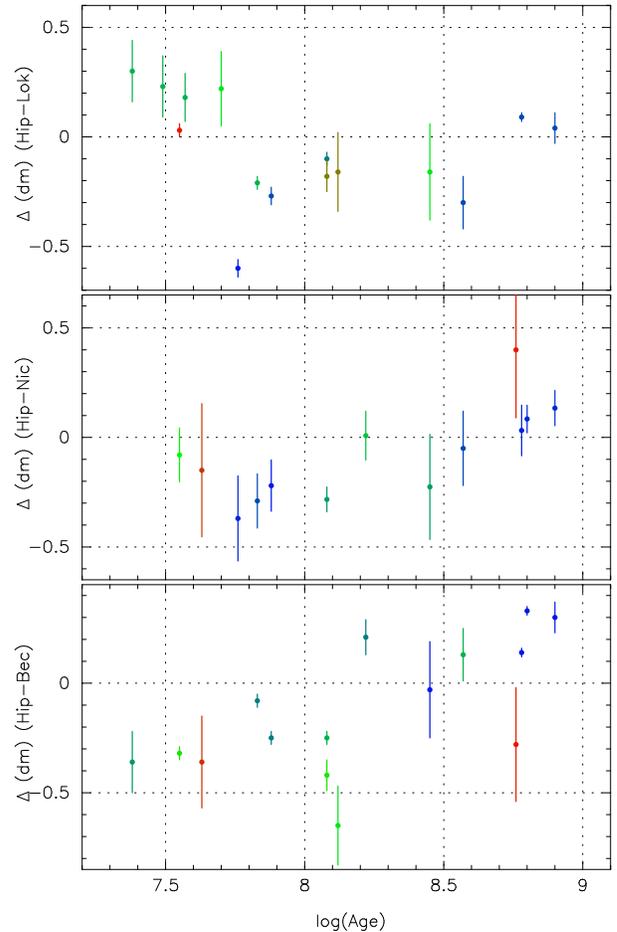}
\caption{Distance modulus differences. Bottom: between \citet{becker71} and Hipparcos for 15 clusters. Error bars represent Hipparcos determinations only. Middle: between \citet{nicol81} and the new reduction for 13 clusters. The error bars represent the combined errors from the two determinations. Top: between \citet{loktin94} and the new reduction for 10 clusters. The error bars represent the Hipparcos determinations only.}
\label{fig:becker_dm}
\end{figure}
The cluster distance determinations by \citet{becker71} are at least partly still based on the initial calibrations by \citet{johnson57}, but with some modifications that affect mainly the young clusters. These modifications, based on incorporating the $\mathrm{(B-U)}$ data, appear to have led to a systematic overestimate of the distances of clusters with ages less than about $100$~Myr.

A comparison with the distance moduli as determined by \cite{nicol81} (Section~\ref{sec:groundbased}) provides a test on whether, for example, stars  in young clusters are of the same luminosity as field stars when found in the same photometric boxes, or whether there are systematic differences, possibly related to age. The comparison, shown in Fig.~\ref{fig:becker_dm}, seems to indicate a systematic overestimate of distance moduli as obtained through the photometric boxes for young clusters, by 0.1 to 0.4 magnitudes, while the older clusters are not significantly offset. One may conclude that the random field stars with ground-based parallaxes as used by Nicolet more resemble the older than the younger cluster stars in luminosity at the same colours. However, the volume of stars with accurate parallax data now available could be used to check in much more detail what the actual luminosity variations of stars sharing photometric boxes are, before drawing any further conclusions. 

Clear systematic differences can be seen in Fig.~\ref{fig:becker_dm} for the comparison with the distance modulus determinations by \citet{loktin94}. These determinations are based \citep{loktin90} on theoretical isochrones by \citet{mermilliod86}, \citet{hejlesen80} and \citet{ vandenberg85}, describing evolutionary models for different ranges of age and mass. The systematic differences that can be seen in Fig.~\ref{fig:becker_dm} are likely to be a reflection of the offsets in the those models, and the way these affect clusters of different ages.

\begin{figure}[t]
\centering
\includegraphics[width=8cm]{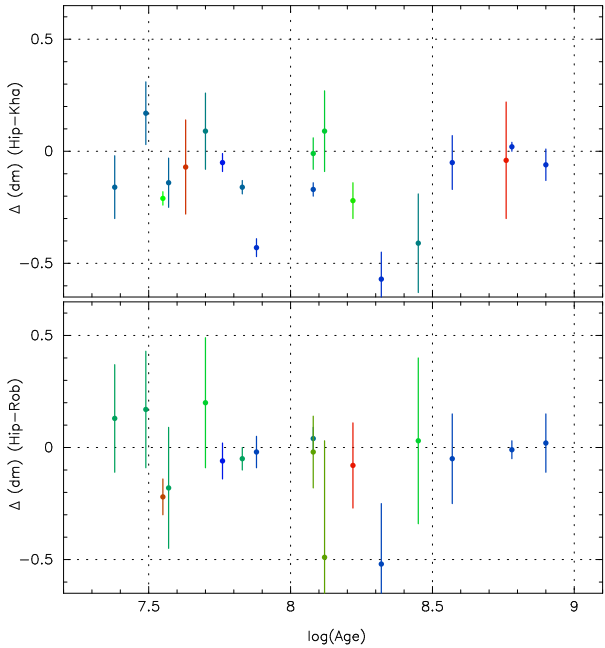}
\caption{Distance modulus differences. Bottom: between \citet{robic99} and Hipparcos for 17 clusters. Error bars represent the old Hipparcos determinations only. Top: between \citet{khar05} and the new reduction for 19 clusters. The error bars represent the new Hipparcos reduction only.}
\label{fig:robichon}
\end{figure}
An entirely different kind of comparison is that of the current determinations  with those based on the first Hipparcos data release. Here the differences for the nearby clusters are small, though the errors have been reduced by about a factor two. For the more distant clusters the membership selection often becomes the critical parameter. This is most clearly demonstrated by IC~4765, for which a parallax based on 7 member stars was presented by \citet{robic99}, but for which there is no solution in the current paper. The more accurate proper motions in the area of this cluster in the new catalogue show only three possible members, all of which appear to be red giants. The general agreement  for the nearby clusters between the current results and those derived using the 1997 catalogue is not entirely surprising. It relates to the size of the fields covered by most of these clusters compared to the radius within which correlations in the abscissa residuals were able to accumulate to correlations in astrometric parameters for neighbouring stars. Such accumulations have been shown to be only effective over distances of less than 1.5 degrees. For the Pleiades stars, spread over 8 degrees diameter on the sky, there were correlations for the brightest stars in the cluster centre, but all other stars remained largely unaffected. 

A comparison with the catalogue of \citet{khar05} was also made, but here conclusions are difficult to draw, as in particular for the range of clusters discussed here their methods of distance determination \citep[based amongst others on results by][]{robic99} varied. What is clear from the comparison is that the distance moduli by presented by \citet{khar05} for the current selection of clusters are on average too large by about 0.15 magnitudes.

In general, for many of the distance calibrations referred to above the same issue applied. Distance calibrations were presented as accurate, while not considering the inaccuracies and remaining noise levels of the actual calibration model itself. While a lot of attention was paid to exact reddening corrections, and sometimes metallicity variations, the most basic of errors as well as noise contribution, in the reference ZAMS, was generally not considered as a contribution to the final accuracies. Calibrations depended for a long time, directly or indirectly, on \citet{johnson57}, which itself was based on a distance estimate of the Hyades that was short by 15~per~cent, an unverified assumption on the relation between the Pleiades and Hyades main sequences, and a very early estimate of the evolutionary luminosity effects on B stars. It is only now, when distances can be measured with completely independent methods, that the limitations of these assumptions are exposed and can be corrected for.

\subsection{HR diagram comparisons}
\begin{figure*}[t]
\centering
\includegraphics[width=17cm]{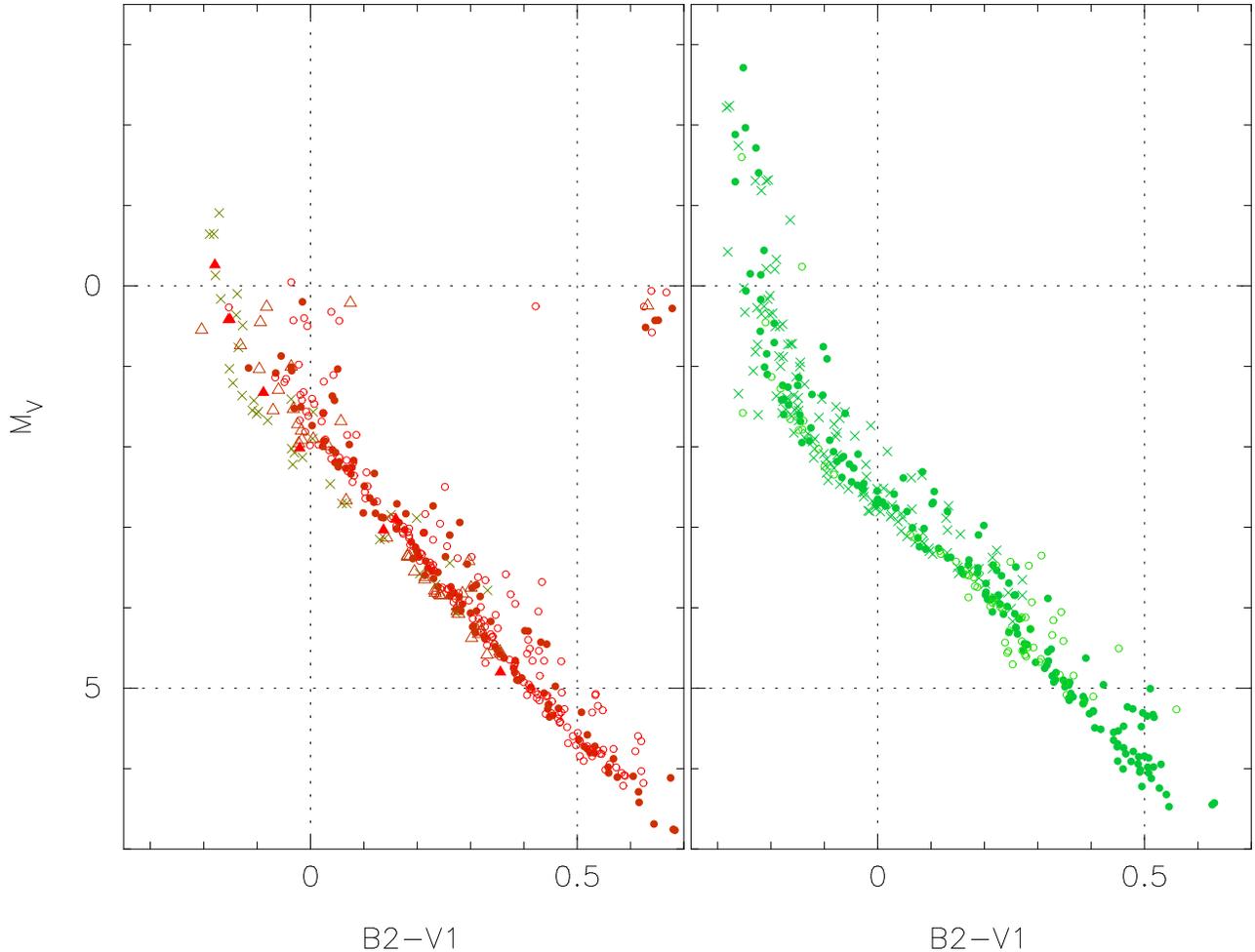}
\caption{The composite HR diagrams for groups 1 \& 2 (left)and group 3 (right). Symbols on the left: Hyades, closed circles; Praesepe, open circles; UMa, closed triangles; Coma~Ber, open triangles; NGC~7092, crosses. Symbols on the right: Pleiades, closed circles; Blanco~1, open circles; NGC~2516, crosses.}
\label{fig:gengr123}
\end{figure*}
When examining the composite HR diagram of the clusters a few distinct groups appear, for which the distributions of stars largely coincide in colour-magnitude and colour-colour diagrams. Photometric data in the $ubvy\beta$ and Geneva systems was extracted from WEBDA\footnote{http://www.univie.ac.at/webda/} to further examine these groups. Additional photometry for UMa stars was obtained from \citet{rufener88} and \citet{hauck98}. In most cases field stars had to be removed from the selected data, but this was only done when a cluster main sequence could be clearly recognized. Clusters for which this was not possible have been left out. For the Hyades only stars were selected for which a kinematic parallax had been determined based on the new Hipparcos reduction, as described in Section~\ref{sec:hyad}. Thus, individual parallaxes were available for the selected stars. Directly measured individual parallaxes were used for the UMa stars.

The (provisional) groups contain the following clusters for which more extensive photometry is available:
\begin{enumerate}
\item The Hyades and Praesepe
\item Coma~Ber, UMa and NGC~7092;
\item Pleiades, Blanco~1 and NGC~2516;
\item $\alpha$~Per, NGC~2451, IC~4665, IC~2391, IC~2602 and NGC~6475. 
\end{enumerate}
The groups contain mostly clusters of similar age, though NGC~6475 appears a little out of place in group 4. For display, groups 1 and 2 have been taken together, and are shown in Fig.~\ref{fig:gengr123}. A couple of observations can be made from this composite HR diagram:
\begin{itemize}
\item As was noted earlier, the Hyades and Praesepe data very closely overlap, and together form a very narrow sequence;
\item The Coma~Ber and UMa main sequences also overlap closely, and are sub luminous with respect to Hyades and Praesepe;
\item NGC~7092 appears to follow Coma~Ber and UMa for $0.1<\mathrm{(B2-V1)}<0.3$;
\item Increases in luminosity are observed for Hyades and Praesepe stars relative to Coma~Ber and UMa for $\mathrm{(B2-V1)}<0.2$;
\item Increases in luminosity are observed for Coma~Ber and UMa stars relative to NGC~7092 for $\mathrm{(B2-V1)}<0.0$;
\end{itemize}

\begin{figure}[t]
\centering
\includegraphics[width=8cm]{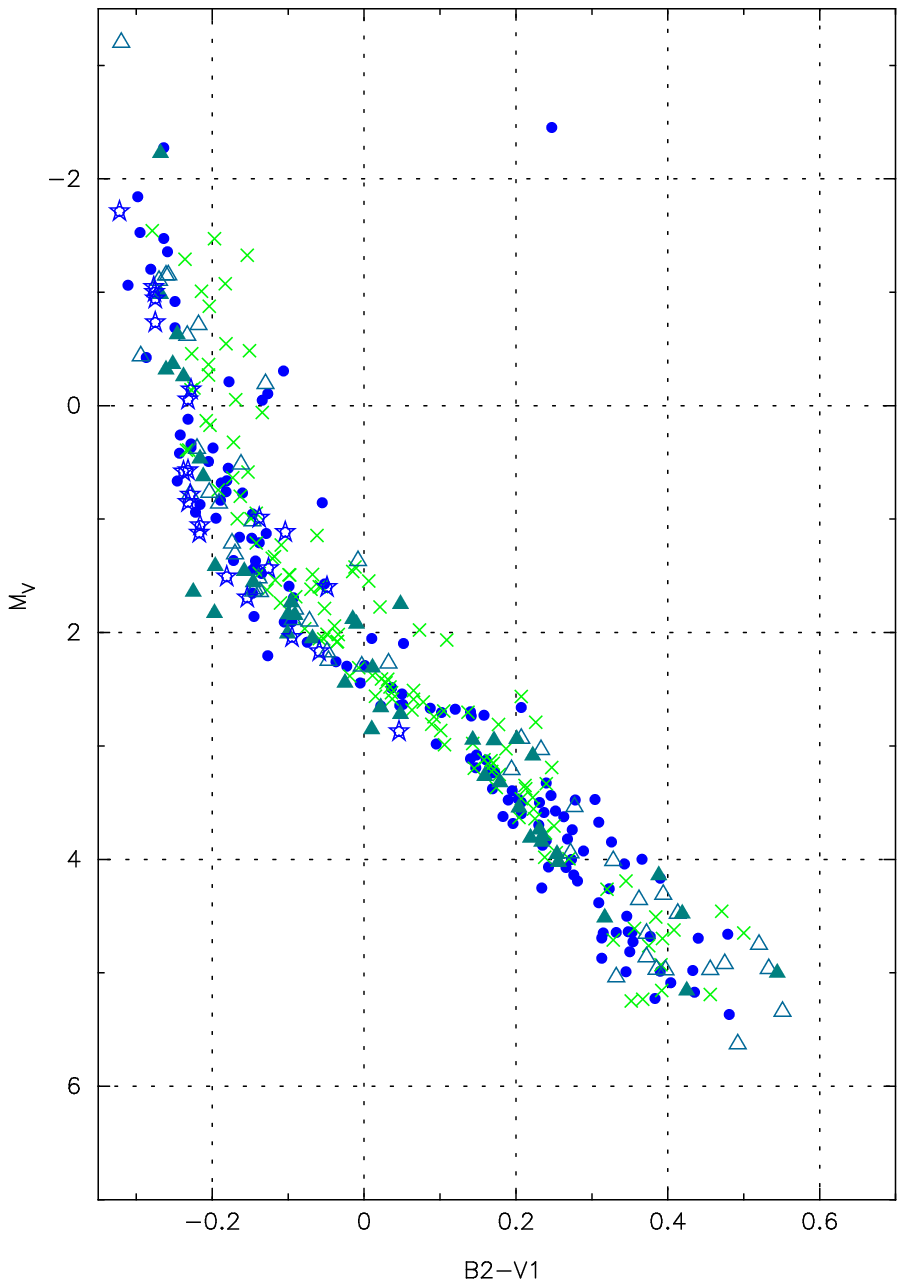}
\caption{The composite diagram for clusters in group 4 (see text).}
\label{fig:geneva_aper}
\end{figure} 
The third group contains three clusters which have been noted in the past as being very similar \citep{snowden75, mermilliod08}. They share the same sequences in the HR diagram when their distance moduli are obtained from the Hipparcos parallaxes, as can be observed in Fig.~\ref{fig:gengr123}. However, over the range $0.2<\mathrm{(B2-V1)}<0.65$ the combined sequence of the group-3 clusters is shifted by about 0.4 magnitudes with respect to the group-1 sequence, and by about 0.25 magnitudes with respect to the group-2 sequence. As current models providing theoretical isochrones appear not to accommodate for such shifts in luminosity, except through extreme assumptions on chemical composition, distance modulus estimates based on isochrones and those based on parallaxes remain systematically discrepant for clusters of the first and third groups. However, it should also be realized that isochrone calibration such as presented by \citep{pinsonn04} is based only on the position of the Hyades main sequence, and does not include an empirical calibration of the age-dependent luminosity effects. These are assumed to be all already fully covered by the theoretical models. The new result for the Pleiades, and its effective confirmation through NGC~2516 and Blanco~1, which both independently had been described to be very similar to the Pleiades, seem to show that this is probably not the case. The issue of the Pleiades distance is still far from "largely settled", as has been claimed by \citet{pinsonn04} and \citet{an07}. 

The clusters in group 4 are mostly young to very young, except for NGC~6475. In the region $0.15<\mathrm{(B2-V1)}<0.45$ the sequence fits well with the Coma~Ber and UMa sequence, and is offset by about 0.15 magnitudes from the group-1 sequence (Fig.~\ref{fig:geneva_aper}). For values of $\mathrm{(B2-V1)}<-0.1$ the sequences of group 3 and 4 coincide. One complication with the group-4 sequence is the lack of well-studied actual clusters. The sequence for $\alpha$~Per shows more variation than is observed for Pleiades, Hyades and Praesepe, while the clusters IC~2639 and IC~2602 are relatively sparse.  

\subsection{The Str{\"o}mgren $c_0$ index for F-type stars}
\begin{figure}[t]
\centering
\includegraphics[width=8.4cm]{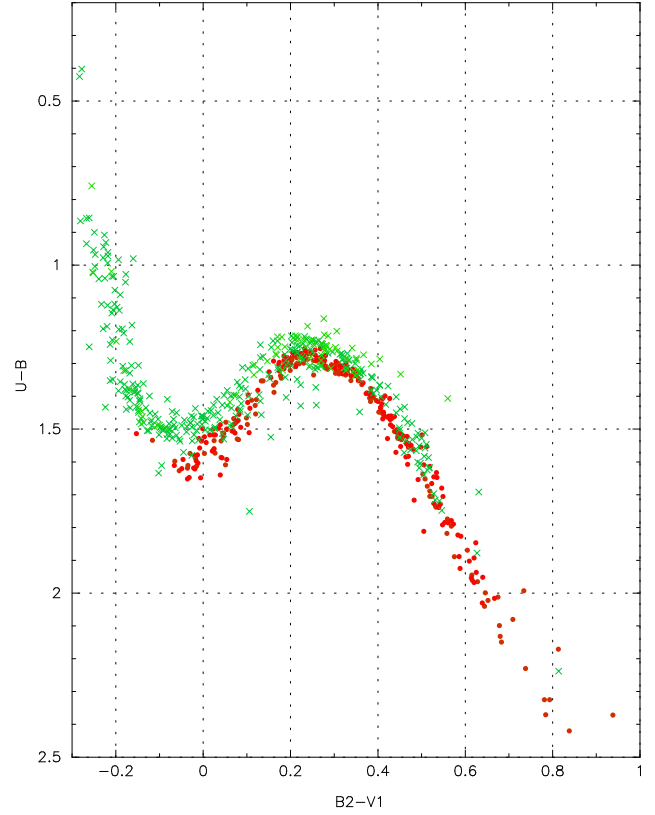}
\caption{The $\mathrm{U-B}$ versus $\mathrm{B2-V2}$ diagram in the Geneva photometry. The red dots represent data points in the Hyades and Praesepe clusters, the crosses for data from the Pleiades, Blanco~1 and NGC~2516. Some of the Pleiades points are offset because of localised differential reddening in the cluster. The difference in loci for the two groups of clusters is known as the Hyades anomaly. } 
\label{fig:UB_Gen}
\end{figure}

\begin{figure}[t]
\centering
\includegraphics[width=8.5cm]{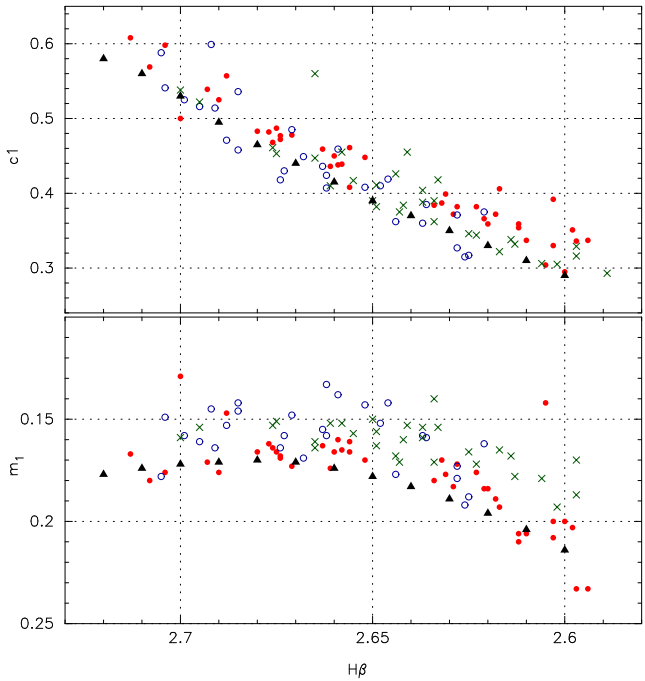}
\caption{Comparison of the $c_1$ (top) and $m_1$ (bottom) versus $\mathrm{H}\beta$ relations for F stars in the Hyades (filled circles), Coma~Ber (crosses) and the  Pleiades (open circles). The calibration relation from \citet{crawford75} is shown by filled triangles.}
\label{fig:betac1}
\end{figure}

The $c_0$ index ($c_1$ corrected for reddening) in the Str{\"o}mgren photometry was shown by \citet{crawford75} to be related to luminosity and surface gravity variations for F-type stars (see also Section~\ref{sec:crawford}).  
Variations in $c_0$ at a given value of $\beta$ or $b-y$ are given by $\Delta c_0$, i.e.\ $c_0$ relative to the calibration relation given in \citet{crawford75} (see also Fig.~\ref{fig:betac1}). This is the relation described in Section~\ref{sec:crawford} by Eq.~\ref{equ:mv_calibr_clust}.  It is further evaluated in the Appendix, Section~\ref{sec:provcalib}. The difference in the $\Delta c_0$ values for the Hyades and Praesepe with respect to, for example, the Pleiades, has been attributed in the past to increased helium abundances \citep{nissen88,dobson90} \citep[see also ][]{crawford69}, although isochrone fitting by \citet{an07} seems to suggest that this is not the case. It has been known for nearly half a century and has been referred to in the past as the Hyades anomaly. It is most clearly observed in the  $\mathrm{(U-B)}$ index, which is a measure of the Balmer jump for early-type stars. The significant difference in the Geneva $\mathrm{(U-B)}$ index for the clusters in group 1 and group 3 can be seen in Fig.~\ref{fig:UB_Gen}. The Str{\"o}mgren $c_0$ index is closely related to the $\mathrm{(U-B)}$ index.

 A comparison between the $c_0$ versus $b-y$ diagrams for clusters in the Pleiades and Hyades groups by \citet{nissen88} showed that the systematic difference in $c_0$ values between those clusters is about $0.033$ to $0.04$. Following Eq.~\ref{equ:dv}, this would correspond to an expected difference of 0.3 to 0.4 in absolute magnitude, which is close to what is observed in the HR diagram for F stars in these clusters when using the Hipparcos parallax-based distance moduli. It is in this context also interesting to note that \citet{nissen88} determined the difference between the true distance moduli, i.e.\ corrected for the $c_0$ effects, for the Pleiades and Hyades as 1.96 (which would imply a distance of 114~pc for the Pleiades), compared to $2.066\pm0.038$ determined by means of the Hipparcos data and 2.33 as derived from isochrone fitting \citep{an07}. In a way, by calibrating their isochrones on the Hyades main sequence, \citet{an07} have turned the "Hyades anomaly" into a "Pleiades anomaly". An extensive discussion on how the Hyades anomaly may be related to the differences in parallax and isochrone-based cluster distance determinations was presented by \citet{mermil97}. 

A systematic difference of 0.033 to 0.04 in $c_0$ values would indicate, following Eq.~\ref{equ:logg}, a lower surface gravity for the Hyades F stars by approximately 0.1, but that is also the noise level on that calibration, i.e.\ the same $\log g$ values can be observed for $\Delta c_1$ values that differ by 0.03 to 0.04. An explanation of the magnitude difference as due to metallicity variations would require an unrealistic, and also not observed, difference in the values for $\Delta m_1$ of around 0.15. 

It is not clear why \citet{stello01} made no reference to the $c_0$ index differences, as discussed earlier by \citet{nissen88}, in their discussion on the Str{\"o}mgren photometry and the Hipparcos parallax measurement for the Pleiades. From a purely observational point of view, a much smaller difference in absolute magnitudes between the Pleiades and Hyades stars (the result of a larger difference in distance moduli), as suggested by fits on theoretical isochrones, would be more difficult to explain, as this would contradict the general trend which can be clearly observed between the parameters $\Delta M_V$ and $\Delta c_0$ for F-type field stars (Fig.~\ref{fig:dvbeta}). A similar conclusion of internal observational consistency for the Hipparcos cluster parallaxes was reached by \citet{grenon02}, based on comparisons of K stars in the Hyades, Pleiades and Praesepe clusters. 

The relations used here show significant noise levels, and their applications do not provide exact results, they are merely an indication of trends and provide estimates of the effects differences in certain parameters may have. A comparison of F stars in the Hyades, Coma~Ber and Pleiades clusters, for example, shows that the overall differences in $c_1$ are in the same direction and at about the same amplitude as would be expected from the observed differences in the absolute magnitudes (Fig.~\ref{fig:betac1}), but it is difficult to assess whether there is not only a qualitative, but also a full quantitative agreement. 

\section{Space velocities}
\label{sec:spacevel}
\begin{figure}[t]
\centering
\includegraphics[width=8.5cm]{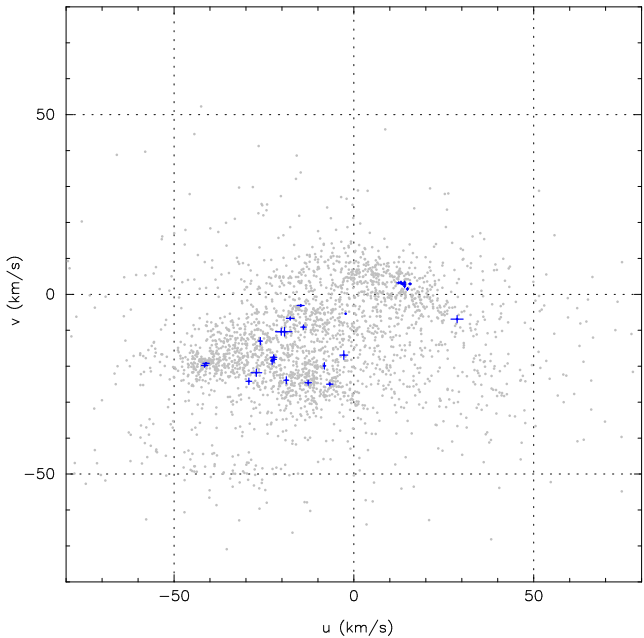}
\caption{Distribution of the velocities of 20 clusters and 8 UMa stars within the galactic plane, compared with the velocity distribution of A and F stars within 125~pc. The UMa stars form a small clump, and there are a few other coincidences in space velocity. The solar motion has not been subtracted.}
\label{fig:uv_clusters}
\end{figure}
For general reference, the positions and space velocities for all 20 clusters and the individual UMa stars have been derived in galactic coordinates:
\begin{equation}
\left[\begin{array}{c} X \\ Y \\ Z \end{array}\right] = R\cdot\left[\begin{array}{r} \cos l \cos b \\ \sin l \cos b \\ \sin b \end{array}\right],
\label{equ:galpos}
\end{equation}
and its derivative
\begin{equation}
\left[\begin{array}{c} u \\ v \\ w \end{array}\right] =
 V_R\cdot\left[\begin{array}{r} \cos l \cos b \\ \sin l \cos b \\ \sin b \end{array}\right] + 
\frac{\kappa\mu_{l*}}{\varpi}\cdot\left[\begin{array}{r} -\sin l \\ \cos l \\ 0 \end{array}\right] - 
\frac{\kappa\mu_{b}}{\varpi}\cdot\left[\begin{array}{r}\cos l \sin b \\ \sin l\sin b \\ -\cos b \end{array}\right],
\label{equ:galvel}
\end{equation}
where $\kappa=4.74047$ is the transformation factor for a proper motion of "1~mas~yr$^{-1}$ at 1~kpc" to "km~s$^{-1}$". The accumulated results for the positions and space velocities are shown in Table~\ref{tab:spacevel}. The space velocities are compared with those of the A and F stars within 125 ~pc \citep{fvl07} in Fig.~\ref{fig:uv_clusters}. 

The most striking of velocity coincidences is between the Hyades and Praesepe, which, within the accuracy of the measurements, have identical $u$ and $v$ velocities. Considering that these two clusters are also very similar in photometric aspects, it seems reasonable to assume a common origin, which would then also explain their very similar age and chemical composition. 

There are no such striking coincidences between the UMa stars and Coma~Ber, though both groups are somewhat anomalous in the distribution of velocities or positions. There is some similarity between the Pleiades and NGC~2516, but Blanco~1 does not seem to fit.

\begin{table*}[t]
\caption{Positions and velocities of 20 clusters in the galactic reference frame.}
\centering
\begin{tabular}{|lrr|rrr|rrr|}
\hline 
 Name     & \multicolumn{1}{c}{$l$} & \multicolumn{1}{c|}{$b$} & \multicolumn{1}{c}{$X$} & \multicolumn{1}{c}{$Y$} & \multicolumn{1}{c|}{$Z$} & \multicolumn{1}{c}{$u$}  & \multicolumn{1}{c}{$v$}  & \multicolumn{1}{c|}{$w$}  \\ 
  & \multicolumn{2}{c|}{degr.} & \multicolumn{3}{c|}{pc} & \multicolumn{3}{c|}{km~s$^{-1}$}   \\ 
\hline 
Hyades   & $179.1$ & $-21.9$ & $ -43.1$ & $   0.7$ & $ -17.3$ & $ -41.1\pm  0.9$ & $ -19.2\pm  0.2$ & $  -1.4\pm  0.4$ \\ 
Coma~Ber & $221.8$ & $ 83.8$ & $  -7.0$ & $  -6.3$ & $  86.2$ & $  -2.3\pm  0.1$ & $  -5.4\pm  0.1$ & $  -1.8\pm  1.0$ \\ 
Pleiades & $166.4$ & $-23.7$ & $-107.0$ & $  25.9$ & $ -48.3$ & $  -6.7\pm  0.9$ & $ -25.0\pm  0.5$ & $ -12.8\pm  0.5$ \\ 
Praesepe & $205.8$ & $ 32.4$ & $-138.4$ & $ -67.0$ & $  97.6$ & $ -41.5\pm  0.9$ & $ -19.8\pm  0.5$ & $  -9.7\pm  1.1$ \\ 
$\alpha$~Per & $147.8$ & $ -6.1$ & $-145.1$ & $  91.3$ & $ -18.2$ & $ -12.7\pm  0.9$ & $ -24.6\pm  0.7$ & $  -7.0\pm  0.2$ \\ 
IC~2391  & $270.4$ & $ -6.9$ & $   0.9$ & $-143.9$ & $ -17.5$ & $ -22.7\pm  0.5$ & $ -18.5\pm  1.0$ & $  -6.3\pm  0.3$ \\ 
IC~2602  & $289.4$ & $ -5.0$ & $  49.2$ & $-139.6$ & $ -13.0$ & $  -8.2\pm  0.4$ & $ -19.9\pm  0.9$ & $  -0.3\pm  0.2$ \\ 
Blanco~1 & $ 14.1$ & $-79.3$ & $  37.4$ & $   9.4$ & $-203.4$ & $ -17.7\pm  1.1$ & $  -6.7\pm  0.5$ & $  -7.1\pm  1.0$ \\ 
NGC~2451 & $252.5$ & $ -7.7$ & $ -54.8$ & $-173.4$ & $ -24.6$ & $ -26.0\pm  0.6$ & $ -13.0\pm  1.0$ & $ -12.6\pm  0.3$ \\ 
NGC~6475 & $355.8$ & $ -4.5$ & $ 268.7$ & $ -19.6$ & $ -21.0$ & $ -14.8\pm  1.0$ & $  -3.1\pm  0.3$ & $  -4.4\pm  0.3$ \\ 
NGC~7092 & $ 92.4$ & $ -2.2$ & $ -12.4$ & $ 302.5$ & $ -11.9$ & $  28.7\pm  1.7$ & $  -6.9\pm  1.0$ & $ -13.2\pm  0.8$ \\ 
NGC~2516 & $273.7$ & $-15.9$ & $  21.5$ & $-328.7$ & $ -93.7$ & $ -18.8\pm  0.7$ & $ -23.9\pm  1.0$ & $  -3.4\pm  0.3$ \\ 
NGC~2232 & $214.4$ & $ -7.4$ & $-288.0$ & $-197.4$ & $ -45.6$ & $ -14.0\pm  0.8$ & $  -9.1\pm  0.6$ & $ -12.6\pm  0.7$ \\ 
IC~4665  & $ 30.6$ & $ 16.8$ & $ 293.1$ & $ 173.5$ & $ 103.0$ & $  -2.8\pm  1.1$ & $ -16.9\pm  1.2$ & $  -8.5\pm  0.7$ \\ 
NGC~6633 & $ 36.0$ & $  8.7$ & $ 299.4$ & $ 217.8$ & $  56.6$ & $ -22.3\pm  0.9$ & $ -17.5\pm  0.7$ & $  -6.9\pm  0.6$ \\ 
Coll~140 & $245.1$ & $ -7.8$ & $-156.8$ & $-337.9$ & $ -51.0$ & $ -22.2\pm  0.7$ & $ -18.1\pm  0.9$ & $ -12.2\pm  0.5$ \\ 
Trump~10 & $262.7$ & $  0.5$ & $ -49.2$ & $-384.4$ & $   3.3$ & $ -27.1\pm  1.5$ & $ -21.8\pm  1.0$ & $  -9.8\pm  0.7$ \\ 
NGC~2422 & $231.0$ & $  3.0$ & $-249.2$ & $-308.1$ & $  21.0$ & $ -29.2\pm  0.8$ & $ -24.2\pm  0.9$ & $  -8.3\pm  0.9$ \\ 
NGC~3532 & $289.6$ & $  1.4$ & $ 137.9$ & $-387.6$ & $  10.2$ & $ -19.3\pm  2.1$ & $ -10.4\pm  1.2$ & $   0.9\pm  0.5$ \\ 
NGC~2547 & $264.4$ & $ -8.6$ & $ -45.4$ & $-466.4$ & $ -71.0$ & $ -20.2\pm  1.6$ & $ -10.4\pm  1.0$ & $ -14.8\pm  1.1$ \\ 
HIP~51814 & $152.2$ & $ 51.6$ & $ -14.6$ & $   7.7$ & $  20.8$ & $  12.5\pm  0.6$ & $   3.2\pm  0.3$ & $  -5.7\pm  0.8$ \\ 
HIP~53910 & $149.1$ & $ 54.8$ & $ -12.1$ & $   7.2$ & $  20.0$ & $  13.5\pm  0.5$ & $   2.9\pm  0.3$ & $  -7.5\pm  0.8$ \\ 
HIP~58001 & $140.8$ & $ 61.4$ & $  -9.5$ & $   7.7$ & $  22.4$ & $  15.6\pm  0.4$ & $   2.9\pm  0.3$ & $  -8.8\pm  0.9$ \\ 
HIP~59774 & $132.5$ & $ 59.5$ & $  -8.5$ & $   9.2$ & $  21.3$ & $  14.9\pm  0.3$ & $   1.5\pm  0.4$ & $ -10.3\pm  0.9$ \\ 
HIP~62956 & $122.2$ & $ 61.1$ & $  -6.5$ & $  10.3$ & $  22.2$ & $  14.1\pm  0.3$ & $   2.8\pm  0.4$ & $  -7.8\pm  0.9$ \\ 
HIP~63503 & $120.3$ & $ 60.7$ & $  -6.3$ & $  10.8$ & $  22.2$ & $  13.1\pm  0.3$ & $   3.3\pm  0.4$ & $  -9.1\pm  0.9$ \\ 
HIP~64532 & $116.8$ & $ 60.2$ & $  -5.7$ & $  11.2$ & $  21.9$ & $  14.1\pm  0.3$ & $   2.3\pm  0.5$ & $  -7.8\pm  0.9$ \\ 
HIP~65477 & $112.8$ & $ 61.5$ & $  -4.6$ & $  11.0$ & $  22.0$ & $  14.1\pm  0.2$ & $   3.3\pm  0.4$ & $  -8.8\pm  0.9$ \\ 
\hline 
\end{tabular}
\begin{list}{}{}
\item The 8 HIP numbers are all UMa members, treated on an individual basis.
\item The errors on the space velocities are in some cases uncertain as a result of uncertainty in the radial velocity accuracy.
\end{list}
\label{tab:spacevel}
\end{table*}

\section{Conclusions}
\label{sec:concl}

The new reduction has eliminated problems in the Hipparcos astrometric data that were caused by peculiarities in the satellite dynamics and an occasional  potential weakness in connectivity between the two fields of view. These were also issues raised by \citep{makar02, makar03, makarov06}, but these effects can not be eliminated at the intermediate-data level, as had been shown already when the idea of such corrections was first presented by \citet{fvl99b}. The new reduction dealt with these attitude problems there where they occurred (in the along-scan attitude reconstruction process) rather than with the indirect and diffuse effects they had on great-circle abscissa residuals. The attitude reconstruction is the only place in which they can be (and now have been) identified and corrected for. With also the connectivity issue having been addressed and tested in the new reduction, the current set of cluster parallax and proper motion values should not only be more precise, but also more accurate.

Differences between isochrone-fitting and parallax-based distance moduli for open clusters are not limited to the Pleiades, they are found to be systematic. The differences are observed between groups of clusters that show similar characteristics in for example the $c_0$ and $\mathrm{U-B}$ indices for F-type stars, and could therefore be related to systematic surface gravity differences. A closer examination of field star data confirms that such variations are correlated to differences in absolute magnitudes, though not in a simple one-to-one relation. Differences are also observed between the Hyades and Praesepe group and Coma~Ber and UMa, where none of the old "Hipparcos problems" ever applied, as concentrations of stars in these clusters are not dense enough. The notion that the parallax values for all these clusters have to be wrong because they do not fit the results obtained by traditional main sequence fitting, or the distance measurement for a single binary star in the cluster, such as expressed by \citet{trimble06}, gives the impression that reduced data has to confirm theoretical models to be acceptable as results, rather than that those data are judged upon the merits of the observations they have been derived from, and the quality of the data reductions. The Pleiades story over the past ten years has shown a number of examples where poor results, sometimes even based on flawed methods, were accepted (and quoted) without reservation, simply because they confirmed expectations. In this context it is worth while recalling a statement made by \citet{crawford73} in his Survey Lecture at the IAU General Assembly in 1970: \textit{In many cases, theory can help us observers quite a lot. We would be lost, or at least inefficient, without theory to guide us. However, we must be extremely careful not to force-fit or to let pre-conceived ideas mess us up. We are measuring observed parameters, and these we relate, or calibrate, to physical parameters.} 

\begin{acknowledgements}
Discussions with various colleagues over the past couple of years have contributed to this paper, in particular with Robin Catchpole, Derek Jones, and Simon Hodgkin. Carme Jordi, Mike Irwin, Bob Carswell, Jan Lub, Anthony Brown and George Wallerstein read earlier drafts and provided much appreciated comments and suggestions. 
 
\end{acknowledgements}

\appendix
\section{Provisional (re-)calibration of Str{\"o}mgren indices}
\label{sec:provcalib}

The currently available volume of information available on F stars (parallaxes, photometric and spectroscopic data), allows for a much more detailed calibration of the Str{\"o}mgren photometric indices with absolute magnitudes, metallicities, and spectroscopically determined effective temperatures and surface gravity values. Presented here are the provisional results of a more extensive study currently in preparation. It has been included here in order to see how the observed systematic differences in colour indices and absolute magnitudes can be related to physical parameters as calibrated from nearby field stars. 
\begin{figure}[t]
\centering
\includegraphics[width=8.5cm]{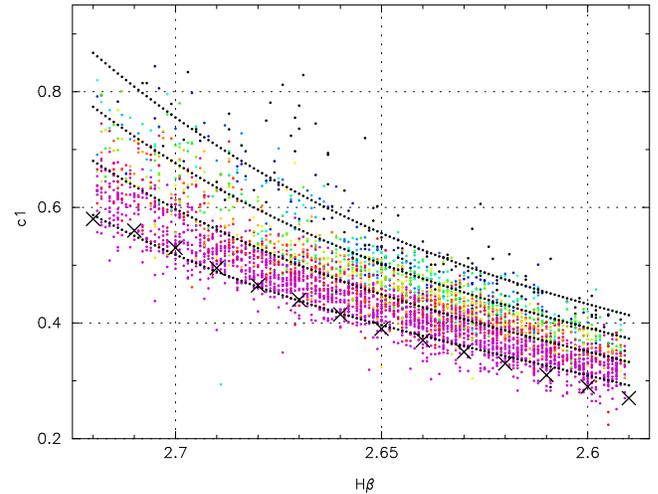}
\caption{The relation between $\mathrm{H}\beta$ and $c_1$ in the Str{\"o}mgren system for nearby F type stars with parallax errors below 10~per~cent. The curved lines of black dots represent from bottom to top the calibration relation at $\Delta v$ values of 1.2, 0.2, -0.8 and -1.8 magnitudes. The crosses show the calibration relation determined and used by \citet{crawford75}.}
\label{fig:beta_c1}
\end{figure}
The photometric data have been extracted from the $ubvy\beta$ catalogue by \citet{hauck97, hauck98}. The spectroscopic data has been extracted from \citet{cayrel01}. For this exploratory exercize the data for multiple entries in the latter catalogue were simply averaged. The facilities of the Vizier\footnote{http://webviz.u-strasbg.fr/viz-bin/VizieR} web page were used to extract samples of F stars, which were later further "cleaned" from luminosity class I and II stars and binaries. Only data with Hipparcos parallax determinations better than 10~per~cent were kept, and all stars fainter than apparent magnitude about 10 were left out. This criterion is only approximate, as close to the limit a specific star may be fainter in one catalogue, and still just bright enough in another.

The first calibration done was a more detailed repeat of the luminosity calibrations of \citep{crawford75}. This involved two linked calibrations, namely between absolute $v$ magnitude (in what follows, $v$ is always assumed to be corrected for the parallax-based distance modulus) and $b-y$, which determined $\mathrm{d}v$ values, and between the observed values of $c_1$ and $\mathrm{H}\beta$ as a function of $\mathrm{d}v$. The iterations use what is observed in the data, and was also implemented by Crawford, that stars occupying the lower boundary in the $v$ versus $b-y$ diagram also occupy the lower envelope in the  $c_1$ versus $\mathrm{H}\beta$ diagram. Thus, the following relations were obtained, using only stars with $\Delta c_1$ values less than 0.02:
\begin{eqnarray}
v_0(b-y) &=& (3.891\pm0.010) \nonumber \\
&+& (10.66\pm0.22)\,(b-y - 0.32),
\end{eqnarray}
where the standard deviation was found to be 0.27 magnitudes. The range in $b-y$ covered is 0.21 to 0.43. The values of $\Delta c_1$ were determined relative to the following calibrated relation, where $\mathrm{d}\beta\equiv(\mathrm{H}\beta-2.64)\cdot 10$ ($\mathrm{d}\beta$ thus ranges from $-0.5$ to $+0.8$), and $\Delta v\equiv v - v_0(b-y) +0.75$:
\begin{eqnarray}
c_1 &=& (0.4370\pm0.0007) + (-0.049\pm0.010)\,\Delta v \nonumber \\
&+& (0.2205\pm0.0029)\,\mathrm{d}\beta + (-0.032\pm0.003)\,\Delta v\,\mathrm{d}\beta \nonumber \\
&+& (0.086\pm0.007)\,\mathrm{d}\beta^2 + (-0.029\pm0.007)\,\Delta v\,\mathrm{d}\beta^2 \nonumber \\
&+& (0.060\pm0.013)\,\mathrm{d}\beta^3,
\end{eqnarray}
with a standard deviation of 0.029 magnitudes. This relation is shown in Fig.~\ref{fig:beta_c1} for four different values of $\Delta v$ superimposed on the actual data points. The lower curve, for $\Delta v=1.2$, serves as the reference values $c_1(r)$ for calculating the $\Delta c_1$ values
\begin{equation}
c_1(r) \equiv 0.3782 + 0.1833\,\mathrm{d}\beta +0.0512\,\mathrm{d}\beta^2 + 0.060\,\mathrm{d}\beta^3,
\end{equation}
and $\Delta c_1\equiv c_1-c_1(r)$.

\begin{figure}[t]
\centering
\includegraphics[width=8.4cm]{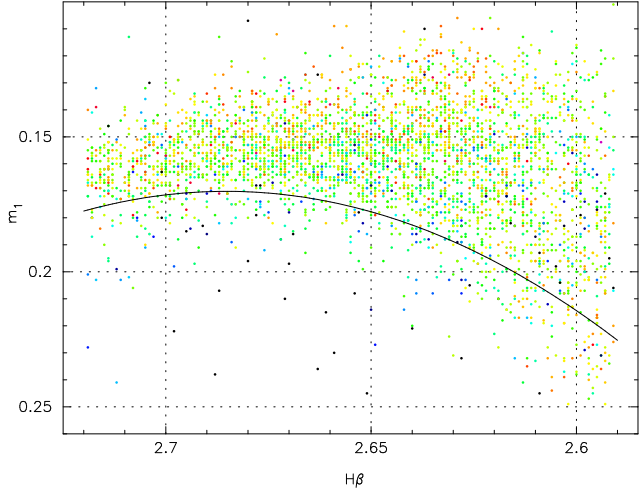}
\caption{The relation between $\mathrm{H}\beta$ and $m_1$ for nearby F dwarfs with parallax accuracies better than 10~per~cent. The line represents the definition of the lower envelope position by \citet{crawford75}.}
\label{fig:beta_m1}
\end{figure}
For determining the $\Delta m_1$ values the reference function as presented by \citet{crawford75} is used, which accurately fits a second-order polynomial
\begin{equation}
m_1(r) \equiv 0.1826 - 0.0550\,\mathrm{d}\beta +0.0607\,\mathrm{d}\beta^2,
\end{equation}  
where the standard deviation with respect to the discrete points defined by Crawford is 0.0005 magnitudes (Fig.~\ref{fig:beta_m1}). As was done above for $\Delta c_1$, we define $\Delta m_1 \equiv m_1 - m_1(r)$. Contrary the situation for $\Delta c_1$, there is no easily recognized relation between $\Delta m_1$ and the absolute magnitude differences $\mathrm{d}v$.

For completeness also the relation between $b-y$, $\Delta c_1$, $\Delta m_1$ and $\mathrm{H}\beta$ has been derived:
\begin{eqnarray}
b-y &=& (0.3376\pm0.0005) + (-0.1695\pm0.0013)\,\mathrm{d}\beta \nonumber \\
&+& (0.0446\pm0.0021)\,\mathrm{d}\beta^2 +(-0.0506\pm0.0048)\,\Delta c_1 \nonumber \\
&+& (-0.589\pm0.012)\,\Delta m_1 +(0.453\pm0.036)\,\Delta m_1\,\mathrm{d}\beta,
\end{eqnarray}
with a standard deviation of 0.014 magnitudes. To compare with Crawford's original calibration, which was expressed in $\delta\beta = 2.72-\beta$, the following equivalent relation is obtained: 
\begin{eqnarray}
b-y &=& (0.231\pm0.001) + (0.981\pm0.028)\,\delta\beta \nonumber \\
&+& (4.46\pm0.21)\,\delta\beta^2 +(-0.0506\pm0.0048)\,\Delta c_1 \nonumber \\
&+& (-0.227\pm0.035)\,\Delta m_1 +(-4.53\pm0.36)\,\Delta m_1\,\delta\beta.
\end{eqnarray}
The main differences with respect to Crawford's original calibration are in the higher order terms, which are less well determined for the much smaller volume of data that was available to Crawford at the time of the calibration.
\begin{figure}[t]
\centering
\includegraphics[width=8.4cm]{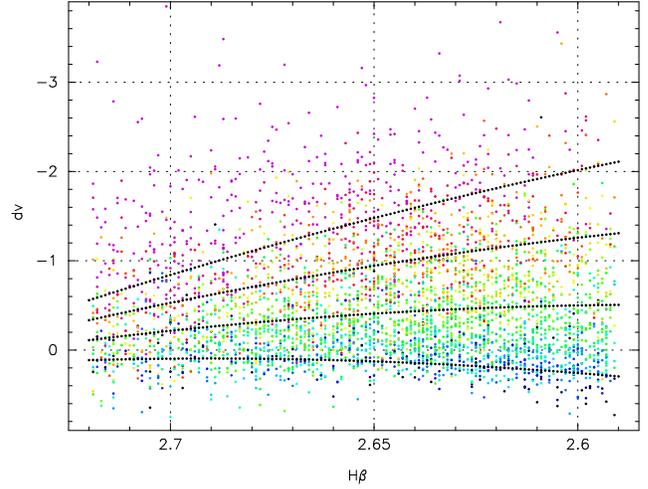}
\caption{Calibration of $\Delta v$. The four lines of black dots show (from bottom to top) the calibration relation for values of $\Delta c_1=-0.02,~0.04, ~0.10,~0.16$, and $\Delta m_1=0.015$.}
\label{fig:dvbeta}
\end{figure} 

The calibration in terms of absolute magnitudes has been done for $\Delta v$ as a function of $\mathrm{d}\beta$, $\Delta c_1$ and $\Delta m_1$:
\begin{eqnarray}
\Delta v &=& (-0.085\pm0.014)+(-0.053\pm0.029)\,\mathrm{d}\beta \nonumber \\
&+& (-9.66\pm0.17)\,\Delta c_1 + (7.42\pm0.41)\,\mathrm{d}\beta\,\Delta c_1 \nonumber \\
&+& (0.201\pm0.55)\,\mathrm{d}\beta^2 + (2.43\pm0.31)\,\Delta m_1,
\label{equ:dv}
\end{eqnarray}
with a standard deviation of 0.34 magnitudes. Important here is that we are now able to recognize the effect of variation in the metallicity index $\Delta m_1$ on the absolute magnitudes. The calibration relation is shown in Fig.~\ref{fig:dvbeta}.
\begin{figure}[t]
\centering
\includegraphics[width=8.4cm]{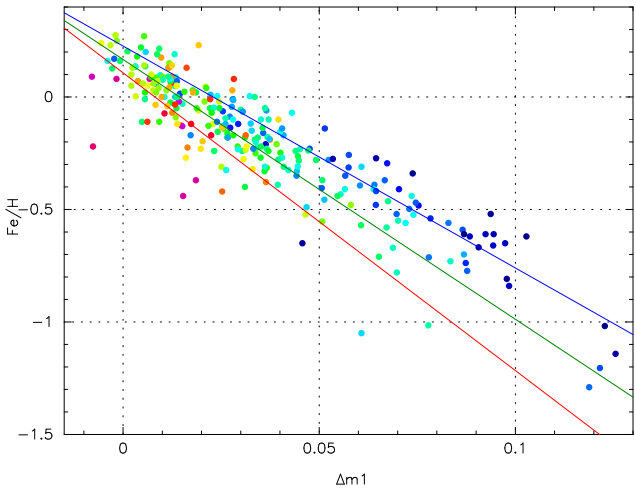}
\caption{The relation between $\Delta m_1$ and $\mathrm{Fe/H}$ for 271 F stars. The calibration for $\mathrm{H}\beta=2.60,~2.65, ~2.70$ is shown by the diagonal lines, with the lowest value of $\mathrm{H}\beta$ as the top line.}
\label{fig:dm1_beta}
\end{figure}

Spectroscopic data for 271 F dwarfs with $ubvy\beta$ photometry were selected from \citet{cayrel01}. For each star also spectroscopically determined effective temperatures and surface gravity values are provided, allowing for a direct comparison between these physical parameters and the Str{\"o}mgren indices. The first of those relations is between $\mathrm{d}\beta$ and the $\mathrm{Teff}$:
\begin{equation}
\mathrm{Teff} = (6294\pm4) +(861\pm16)\,\mathrm{d}\beta~~\mathrm{K},
\end{equation}
with a dispersion of 69~K. Thus, the stars examined here would range in temperature from $5862\pm9$ to $6983\pm13$~K. 

The situation for the calibration of $\Delta m_1$ as a function of $\mathrm{Fe/H}$ is less simple due to the dependence on $\mathrm{H}\beta$ and the uneven distribution and relatively small number of calibration points. The relation used in the final fit is as follows
\begin{eqnarray}
\mathrm{Fe/H}&=& (0.179\pm0.012)+(-0.119\pm0.038)\,\mathrm{d}\beta \nonumber \\
&+& (-11.21\pm0.38)\,\Delta m_1 +(-3.37\pm1.05)\,\Delta m_1\,\mathrm{d}\beta,
\end{eqnarray}
with a standard deviation of 0.10. The data and the calibration are shown in Fig.~\ref{fig:dm1_beta}. These results are in general agreement with similar calibrations presented by \citet{crawford75} and \citet{gustafsson72}.
\begin{figure}[t]
\centering
\includegraphics[width=8.4cm]{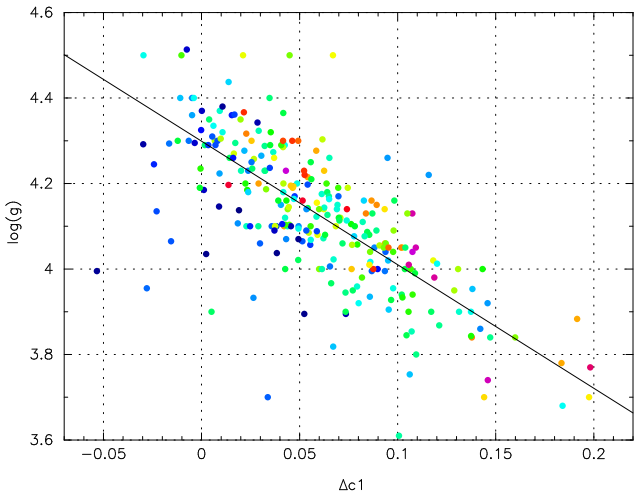}
\caption{The relation between $\Delta c_1$ and $\log g$ for 271 F stars of luminosity types III, IV and V. The diagonal line is the calibration curve.}
\label{fig:c1_logg}
\end{figure}
The final calibration concerns the surface gravity $\log g$ as a function of $\Delta c_1$. There is a generally well-defined relation:
\begin{equation}
\log g = (4.294\pm0.010)+(-2.91\pm0.14)\,\Delta c_1,
\label{equ:logg}
\end{equation}
with a standard deviation of 0.10. It is noted, however, that for the later-type stars more data points are found with significant disagreement between the $\log g$ and $\Delta c_1$ values. The data and the calibration are shown in Fig.~\ref{fig:c1_logg}.

In general, when applying any of the above relations the standard deviation of the fit should be used as a measure of the uncertainty in the result obtained.

\bibliographystyle{aa}
\bibliography{11382}

\begin{thebibliography}{213}
\expandafter\ifx\csname natexlab\endcsname\relax\def\natexlab#1{#1}\fi

\bibitem[{{Abt}(1975)}]{abt75}
{Abt}, H.~A. 1975, \pasp, 87, 417

\bibitem[{{Abt}(1978)}]{abt78}
{Abt}, H.~A. 1978, \pasp, 90, 692

\bibitem[{{Abt} \& {Levato}(1975)}]{abtl75}
{Abt}, H.~A. \& {Levato}, H. 1975, \pasp, 87, 849

\bibitem[{{Abt} \& {Morgan}(1969)}]{abt69}
{Abt}, H.~A. \& {Morgan}, W.~W. 1969, \aj, 74, 813

\bibitem[{{Abt} \& {Sanders}(1973)}]{abt73}
{Abt}, H.~A. \& {Sanders}, W.~L. 1973, \apj, 186, 177

\bibitem[{{An} {et~al.}(2007){An}, {Terndrup}, {Pinsonneault}, {Paulson},
  {Hanson}, \& {Stauffer}}]{an07}
{An}, D., {Terndrup}, D.~M., {Pinsonneault}, M.~H., {et~al.} 2007, \apj, 655,
  233

\bibitem[{{Barry} {et~al.}(1979){Barry}, {Cromwell}, \& {Schoolman}}]{barry79}
{Barry}, D.~C., {Cromwell}, R.~H., \& {Schoolman}, S.~A. 1979, \apjs, 41, 119

\bibitem[{{Baumgardt}(1998)}]{baumg98}
{Baumgardt}, H. 1998, \aap, 340, 402

\bibitem[{{Becker} \& {Fenkart}(1971)}]{becker71}
{Becker}, W. \& {Fenkart}, R. 1971, \aaps, 4, 241

\bibitem[{{Bertiau}(1958)}]{bertiau58}
{Bertiau}, F.~C. 1958, \apj, 128, 533

\bibitem[{{Blaauw}(1963)}]{blaauw63}
{Blaauw}, A. 1963, {The Calibration of Luminosity Criteria} (University of
  Chicago Press, Chicago), 383--+

\bibitem[{{Blanco}(1949)}]{blanco49}
{Blanco}, V.~M. 1949, \pasp, 61, 183

\bibitem[{{Braes}(1961)}]{braes61}
{Braes}, L.~L.~E. 1961, Monthly Notes of the Astronomical Society of South
  Africa, 20, 7

\bibitem[{{Braes}(1962)}]{braes62}
{Braes}, L.~L.~E. 1962, \bain, 16, 297

\bibitem[{{Buscombe} \& {Kennedy}(1968)}]{buscombe68}
{Buscombe}, W. \& {Kennedy}, P.~M. 1968, \mnras, 139, 215

\bibitem[{{Cameron}(1985)}]{cameron85a}
{Cameron}, L.~M. 1985, \aap, 147, 39

\bibitem[{{Cayrel de Strobel} {et~al.}(2001){Cayrel de Strobel}, {Soubiran}, \&
  {Ralite}}]{cayrel01}
{Cayrel de Strobel}, G., {Soubiran}, C., \& {Ralite}, N. 2001, \aap, 373, 159

\bibitem[{{Chen}(1999)}]{chen99}
{Chen}, B. 1999, \aap, 344, 494

\bibitem[{{Clari{\'a}}(1972)}]{claria72}
{Clari{\'a}}, J.~J. 1972, \aap, 19, 303

\bibitem[{{Clari{\'a}}(1982)}]{claria82}
{Clari{\'a}}, J.~J. 1982, \aaps, 47, 323

\bibitem[{{Clari{\'a}} \& {Lapasset}(1988)}]{claria88}
{Clari{\'a}}, J.~J. \& {Lapasset}, E. 1988, \mnras, 235, 1129

\bibitem[{{Clari{\'a}} \& {Rosenzweig}(1978)}]{claria78}
{Clari{\'a}}, J.~J. \& {Rosenzweig}, P. 1978, \aj, 83, 278

\bibitem[{{Constantine} {et~al.}(1969){Constantine}, {Harris}, \&
  {Nikoloff}}]{constantine69}
{Constantine}, S.~M., {Harris}, B.~J., \& {Nikoloff}, I. 1969, Proceedings of
  the Astronomical Society of Australia, 1, 207

\bibitem[{{Cox}(1955)}]{cox55}
{Cox}, A.~N. 1955, \apj, 121, 628

\bibitem[{{Crawford}(1973)}]{crawford73}
{Crawford}, D.~L. 1973, in IAU Symposium, Vol.~54, Problems of Calibration of
  Absolute Magnitudes and Temperature of Stars, ed. B.~{Hauck} \& B.~E.
  {Westerlund}, 93--113

\bibitem[{{Crawford}(1975)}]{crawford75}
{Crawford}, D.~L. 1975, \aj, 80, 955

\bibitem[{{Crawford}(1978)}]{crawford78}
{Crawford}, D.~L. 1978, \aj, 83, 48

\bibitem[{{Crawford}(1979)}]{crawford79}
{Crawford}, D.~L. 1979, \aj, 84, 1858

\bibitem[{{Crawford} \& {Barnes}(1969{\natexlab{a}})}]{crawford69}
{Crawford}, D.~L. \& {Barnes}, J.~V. 1969{\natexlab{a}}, \aj, 74, 818

\bibitem[{{Crawford} \& {Barnes}(1969{\natexlab{b}})}]{crawfordb69}
{Crawford}, D.~L. \& {Barnes}, J.~V. 1969{\natexlab{b}}, \aj, 74, 407

\bibitem[{{Crawford} \& {Barnes}(1972)}]{crawford72}
{Crawford}, D.~L. \& {Barnes}, J.~V. 1972, \aj, 77, 862

\bibitem[{{Crawford} \& {Barnes}(1974)}]{crawford74}
{Crawford}, D.~L. \& {Barnes}, J.~V. 1974, \aj, 79, 687

\bibitem[{{Crawford} \& {Perry}(1966)}]{crawford66AJ}
{Crawford}, D.~L. \& {Perry}, C.~L. 1966, \aj, 71, 206

\bibitem[{{Crawford} \& {Perry}(1976)}]{crawford76}
{Crawford}, D.~L. \& {Perry}, C.~L. 1976, \aj, 81, 419

\bibitem[{{Creze} {et~al.}(1980){Creze}, {Turon Lacarrieu}, {Golay}, \&
  {Mandwewala}}]{creze80}
{Creze}, M., {Turon Lacarrieu}, C., {Golay}, M., \& {Mandwewala}, N. 1980,
  \aap, 85, 311

\bibitem[{{Crutcher} {et~al.}(1978){Crutcher}, {Hartkopf}, \&
  {Giguere}}]{crutcher78}
{Crutcher}, R.~M., {Hartkopf}, W.~I., \& {Giguere}, P.~T. 1978, \apj, 226, 839

\bibitem[{{Dachs}(1970)}]{dachs70}
{Dachs}, J. 1970, \aap, 5, 312

\bibitem[{{Dachs} \& {Kabus}(1989)}]{dachs89}
{Dachs}, J. \& {Kabus}, H. 1989, \aaps, 78, 25

\bibitem[{{Dambis}(1999)}]{dambis99}
{Dambis}, A.~K. 1999, Astronomy Letters, 25, 7

\bibitem[{{de Bruijne} {et~al.}(2001){de Bruijne}, {Hoogerwerf}, \& {de
  Zeeuw}}]{bruijne01}
{de Bruijne}, J.~H.~J., {Hoogerwerf}, R., \& {de Zeeuw}, P.~T. 2001, \aap, 367,
  111

\bibitem[{{de Epstein} \& {Epstein}(1985)}]{epstein85}
{de Epstein}, A.~E.~A. \& {Epstein}, I. 1985, \aj, 90, 1211

\bibitem[{{de Zeeuw} {et~al.}(1999){de Zeeuw}, {Hoogerwerf}, {de Bruijne},
  {Brown}, \& {Blaauw}}]{zeeuw99}
{de Zeeuw}, P.~T., {Hoogerwerf}, R., {de Bruijne}, J.~H.~J., {Brown}, A.~G.~A.,
  \& {Blaauw}, A. 1999, \aj, 117, 354

\bibitem[{{Dobson}(1990)}]{dobson90}
{Dobson}, A.~K. 1990, \pasp, 102, 88

\bibitem[{{Domingo} \& {Figueras}(1999)}]{domingo99}
{Domingo}, A. \& {Figueras}, F. 1999, \aap, 343, 446

\bibitem[{{Dworetsky}(1975)}]{dworetsky75}
{Dworetsky}, M.~M. 1975, \aj, 80, 131

\bibitem[{{Ebbighausen}(1939)}]{ebbig39}
{Ebbighausen}, E.~G. 1939, \apj, 89, 431

\bibitem[{{Ebbighausen}(1940)}]{ebbig40}
{Ebbighausen}, E.~G. 1940, \apj, 92, 434

\bibitem[{{Eggen}(1951)}]{eggen51}
{Eggen}, O.~J. 1951, \apj, 113, 657

\bibitem[{{Eggen}(1970)}]{eggen70}
{Eggen}, O.~J. 1970, \apj, 161, 159

\bibitem[{{Eggen}(1981)}]{eggen81}
{Eggen}, O.~J. 1981, \apj, 246, 817

\bibitem[{ESA(1992)}]{esa92}
ESA, ed. 1992, The Hipparcos Input Catalogue, SP No. 1136 (ESA)

\bibitem[{ESA(1997)}]{esa97}
ESA, ed. 1997, The Hipparcos and Tycho Catalogues, SP No. 1200 (ESA)

\bibitem[{{Feinstein} {et~al.}(1973){Feinstein}, {Marraco}, \&
  {Mirabel}}]{feinstein73}
{Feinstein}, A., {Marraco}, H.~G., \& {Mirabel}, I. 1973, \aaps, 9, 233

\bibitem[{{Fernandez} \& {Salgado}(1980)}]{fernandez80}
{Fernandez}, J.~A. \& {Salgado}, C.~W. 1980, \aaps, 39, 11

\bibitem[{{Fernie}(1959)}]{fernie59}
{Fernie}, J.~D. 1959, Monthly Notes of the Astronomical Society of South
  Africa, 18, 57

\bibitem[{{Fernie}(1960)}]{fernie60}
{Fernie}, J.~D. 1960, Monthly Notes of the Astronomical Society of South
  Africa, 19, 120

\bibitem[{{Fernie}(1965)}]{fernie65}
{Fernie}, J.~D. 1965, \aj, 70, 575

\bibitem[{{Fitzgerald} {et~al.}(1980){Fitzgerald}, {Miller}, \&
  {Harris}}]{fitzgerald80}
{Fitzgerald}, M.~P., {Miller}, M., \& {Harris}, G.~L.~H. 1980, \mnras, 191, 95

\bibitem[{{Fox Machado} {et~al.}(2006){Fox Machado}, {P{\'e}rez Hern{\'a}ndez},
  {Su{\'a}rez}, {Michel}, \& {Lebreton}}]{fox06}
{Fox Machado}, L., {P{\'e}rez Hern{\'a}ndez}, F., {Su{\'a}rez}, J.~C.,
  {Michel}, E., \& {Lebreton}, Y. 2006, Memorie della Societa Astronomica
  Italiana, 77, 455

\bibitem[{{Gatewood}(1995)}]{gatew95}
{Gatewood}, G. 1995, \apj, 445, 712

\bibitem[{{Gatewood} {et~al.}(1990){Gatewood}, {Castelaz}, {Han}, {Persinger},
  {Stein}, {Stephenson}, \& {Tangren}}]{gatew90}
{Gatewood}, G., {Castelaz}, M., {Han}, I., {et~al.} 1990, \apj, 364, 114

\bibitem[{{Gatewood} \& {de Jonge}(1994)}]{gatew94}
{Gatewood}, G. \& {de Jonge}, J.~K. 1994, \apj, 428, 166

\bibitem[{Gatewood {et~al.}(2000)Gatewood, de~Jonge, \& Han}]{gatew00}
Gatewood, G., de~Jonge, J.~K., \& Han, I. 2000, \apj, 533, 938

\bibitem[{{Gatewood} {et~al.}(1998){Gatewood}, {Kiewiet de Jonge}, \&
  {Persinger}}]{gatew98}
{Gatewood}, G., {Kiewiet de Jonge}, J., \& {Persinger}, T. 1998, \aj, 116, 1501

\bibitem[{{Geyer} \& {Nelles}(1985)}]{geyer85}
{Geyer}, E.~H. \& {Nelles}, B. 1985, \aaps, 62, 301

\bibitem[{{Giannuzzi}(1995)}]{giann95}
{Giannuzzi}, M.~A. 1995, \aap, 293, 360

\bibitem[{{Gieseking}(1981)}]{giesek81}
{Gieseking}, F. 1981, \aap, 99, 155

\bibitem[{{Gieseking}(1985)}]{giesek85}
{Gieseking}, F. 1985, \aaps, 61, 75

\bibitem[{{Gonz{\'a}lez} \& {Lapasset}(2000)}]{gonzalez00}
{Gonz{\'a}lez}, J.~F. \& {Lapasset}, E. 2000, \aj, 119, 2296

\bibitem[{{Grenon}(2002)}]{grenon02}
{Grenon}, M. 2002, Highlights of Astronomy, 12, 680

\bibitem[{{Gustafsson} \& {Nissen}(1972)}]{gustafsson72}
{Gustafsson}, B. \& {Nissen}, P.~E. 1972, \aap, 19, 261

\bibitem[{{Hauck} \& {Mermilliod}(1997)}]{hauck97}
{Hauck}, B. \& {Mermilliod}, M. 1997, VizieR Online Data Catalog, 2215, 0

\bibitem[{{Hauck} \& {Mermilliod}(1998)}]{hauck98}
{Hauck}, B. \& {Mermilliod}, M. 1998, \aaps, 129, 431

\bibitem[{{Hejlesen}(1980)}]{hejlesen80}
{Hejlesen}, P.~M. 1980, \aaps, 39, 347

\bibitem[{{Hill} \& {Perry}(1969)}]{hill69}
{Hill}, G. \& {Perry}, C.~L. 1969, \aj, 74, 1011

\bibitem[{{Hiltner} {et~al.}(1958){Hiltner}, {Iriarte}, \&
  {Johnson}}]{hiltner58}
{Hiltner}, W.~A., {Iriarte}, B., \& {Johnson}, H.~L. 1958, \apj, 127, 539

\bibitem[{{Hogg}(1960)}]{hogg60}
{Hogg}, A.~R. 1960, \pasp, 72, 85

\bibitem[{{Hogg} \& {Kron}(1955)}]{hogg55}
{Hogg}, A.~R. \& {Kron}, G.~E. 1955, \aj, 60, 365

\bibitem[{{Ishmukhamedov}(1966)}]{ishmuk66}
{Ishmukhamedov}, K.~Z. 1966, Tsirkulyar Tashkentskoj Astronomicheskoj
  Observatorii, 346, 16

\bibitem[{{Johansson}(1981)}]{johansson81}
{Johansson}, K.~L.~V. 1981, \aaps, 43, 421

\bibitem[{{Johnson}(1953)}]{johnson53}
{Johnson}, H.~L. 1953, \apj, 117, 353

\bibitem[{{Johnson}(1954)}]{johnson54}
{Johnson}, H.~L. 1954, \apj, 119, 181

\bibitem[{{Johnson}(1957)}]{johnson57}
{Johnson}, H.~L. 1957, \apj, 126, 121

\bibitem[{{Johnson}(1966)}]{johnson66}
{Johnson}, H.~L. 1966, \araa, 4, 193

\bibitem[{{Johnson} {et~al.}(1961){Johnson}, {Hoag}, {Iriarte}, {Mitchell}, \&
  {Hallam}}]{johnson61}
{Johnson}, H.~L., {Hoag}, A.~A., {Iriarte}, B., {Mitchell}, R.~I., \& {Hallam},
  K.~L. 1961, Lowell Observatory Bulletin, 5, 133

\bibitem[{{Johnson} \& {Morgan}(1953)}]{johnsonm53}
{Johnson}, H.~L. \& {Morgan}, W.~W. 1953, \apj, 117, 313

\bibitem[{{Jordi} {et~al.}(2002){Jordi}, {Luri}, {Masana}, {Torra}, {Figueras},
  {Domingo}, {G{\'o}mez}, \& {Mennessier}}]{jordi02}
{Jordi}, C., {Luri}, X., {Masana}, E., {et~al.} 2002, Highlights of Astronomy,
  12, 684

\bibitem[{{Jordi} {et~al.}(1997){Jordi}, {Masana}, {Figueras}, \&
  {Torra}}]{jordi97}
{Jordi}, C., {Masana}, E., {Figueras}, F., \& {Torra}, J. 1997, \aaps, 123, 83

\bibitem[{{Kharchenko} {et~al.}(2005){Kharchenko}, {Piskunov}, {R{\"o}ser},
  {Schilbach}, \& {Scholz}}]{khar05}
{Kharchenko}, N.~V., {Piskunov}, A.~E., {R{\"o}ser}, S., {Schilbach}, E., \&
  {Scholz}, R.-D. 2005, \aap, 438, 1163

\bibitem[{{Kilkenny} {et~al.}(1975){Kilkenny}, {Hilditch}, {Hilditch}, {Hill},
  \& {Lynas-Gray}}]{kilkenny75}
{Kilkenny}, D., {Hilditch}, C.~E.~J., {Hilditch}, R.~W., {Hill}, P.~W., \&
  {Lynas-Gray}, A.~E. 1975, \mnras, 172, 5P

\bibitem[{{King}(1978)}]{king78}
{King}, D.~S. 1978, J.~Proc.~R.~Soc.~N.S.W., Vol.~111, p.~1 - 12 = Sydney
  Obs.~Pap.~No.~79, 111, 1

\bibitem[{{Koelbloed}(1959)}]{koelbloed59}
{Koelbloed}, D. 1959, \bain, 14, 265

\bibitem[{{Kovalevsky} {et~al.}(1992){Kovalevsky}, {Falin}, {Pieplu},
  {Bernacca}, {Donati}, {Froeschle}, {Galligani}, {Mignard}, {Morando},
  {Perryman}, {Schrijver}, {van Daalen}, {van der Marel}, {Villenave},
  {Walter}, {Badiali}, {Borriello}, {Brouw}, {Canuto}, {Guerry}, {Hering},
  {Huc}, {Iorio-Fili}, {Lacroute}, {Lattanzi}, {Le Poole}, {Murgolo},
  {Preston}, {R{\" o}ser}, {Sanso}, {Wielen}, {Belforte}, {Bernstein},
  {Bucciarelli}, {Cardini}, {Emanuele}, {Fassino}, {Lenhardt}, {Lestrade},
  {Prezioso}, \& {Tommasini Montanari}}]{koval92}
{Kovalevsky}, J., {Falin}, J.~L., {Pieplu}, J.~L., {et~al.} 1992, \aap, 258, 7

\bibitem[{{Levato} \& {Abt}(1977)}]{levato77}
{Levato}, H. \& {Abt}, H.~A. 1977, \pasp, 89, 274

\bibitem[{{Levato} \& {Malaroda}(1975{\natexlab{a}})}]{levato75}
{Levato}, H. \& {Malaroda}, S. 1975{\natexlab{a}}, \pasp, 87, 823

\bibitem[{{Levato} \& {Malaroda}(1975{\natexlab{b}})}]{levato75a}
{Levato}, H. \& {Malaroda}, S. 1975{\natexlab{b}}, \pasp, 87, 173

\bibitem[{Lindegren {et~al.}(1992)Lindegren, Hoeg, van Leeuwen, Murray, Evans,
  Penston, Perryman, Petersen, Ramamani, \& Snijders}]{lindeg92}
Lindegren, L., Hoeg, E., van Leeuwen, F., {et~al.} 1992, \aap, 258, 18

\bibitem[{Lindegren {et~al.}(2000)Lindegren, Madsen, \& Dravins}]{lmd00}
Lindegren, L., Madsen, S., \& Dravins, D. 2000, \aap, 356, 1119

\bibitem[{{Liu} {et~al.}(1989){Liu}, {Janes}, \& {Bania}}]{liu89}
{Liu}, T., {Janes}, K.~A., \& {Bania}, T.~M. 1989, \aj, 98, 626

\bibitem[{{Loden} {et~al.}(1980){Loden}, {Lindblad}, {Schober}, \&
  {Urban}}]{loden80}
{Loden}, K., {Lindblad}, P.~O., {Schober}, J., \& {Urban}, A. 1980, \aaps, 41,
  85

\bibitem[{{Loden}(1984)}]{loden84}
{Loden}, L.~O. 1984, \aaps, 58, 595

\bibitem[{{Loktin} \& {Matkin}(1990)}]{loktin90}
{Loktin}, A.~V. \& {Matkin}, N.~V. 1990, Soviet Astronomy, 34, 571

\bibitem[{{Loktin} \& {Matkin}(1994)}]{loktin94}
{Loktin}, A.~V. \& {Matkin}, N.~V. 1994, Astronomical and Astrophysical
  Transactions, 4, 153

\bibitem[{Lutz \& Kelker(1973)}]{lutzk73}
Lutz, T.~E. \& Kelker, D.~E. 1973, \pasp, 85, 573

\bibitem[{{Lyng{\aa}}(1960)}]{lynga60}
{Lyng{\aa}}, G. 1960, Arkiv for Astronomi, 2, 379

\bibitem[{{Lyng{\aa}}(1962)}]{lynga62}
{Lyng{\aa}}, G. 1962, Arkiv for Astronomi, 3, 65

\bibitem[{{Madsen}(1999)}]{madsen99}
{Madsen}, S. 1999, in Astronomical Society of the Pacific Conference Series,
  Vol. 167, Harmonizing Cosmic Distance Scales in a Post-HIPPARCOS Era, ed.
  D.~{Egret} \& A.~{Heck}, 78--83

\bibitem[{Madsen {et~al.}(2002)Madsen, Dravins, \& Lindegren}]{mdl02}
Madsen, S., Dravins, D., \& Lindegren, L. 2002, \aap, 381, 446

\bibitem[{{Madsen} {et~al.}(2000){Madsen}, {Lindegren}, \&
  {Dravins}}]{madsen00}
{Madsen}, S., {Lindegren}, L., \& {Dravins}, D. 2000, in ASP Conf. Ser. 198:
  Stellar Clusters and Associations: Convection, Rotation, and Dynamos, ed.
  R.~{Pallavicini}, G.~{Micela}, \& S.~{Sciortino}, 137--140

\bibitem[{{Madsen} {et~al.}(2001){Madsen}, {Lindegren}, \&
  {Dravins}}]{madsen01}
{Madsen}, S., {Lindegren}, L., \& {Dravins}, D. 2001, in ASP Conf. Ser. 228:
  Dynamics of Star Clusters and the Milky Way, ed. S.~{Deiters}, B.~{Fuchs},
  A.~{Just}, R.~{Spurzem}, \& R.~{Wielen}, 506--508

\bibitem[{{Maitzen} \& {Catalano}(1986)}]{maitzen86}
{Maitzen}, H.~M. \& {Catalano}, F.~A. 1986, \aaps, 66, 37

\bibitem[{Makarov(2002)}]{makar02}
Makarov, V. 2002, \aj, 124, 3299

\bibitem[{{Makarov}(2003)}]{makar03}
{Makarov}, V.~V. 2003, \aj, 126, 2408

\bibitem[{{Makarov}(2006)}]{makarov06}
{Makarov}, V.~V. 2006, \aj, 131, 2967

\bibitem[{{Mamajek}(2006)}]{mamajek06}
{Mamajek}, E.~E. 2006, \aj, 132, 2198

\bibitem[{{Manteiga} {et~al.}(1991){Manteiga}, {Martinez-Roger}, {Morales}, \&
  {Sabau}}]{manteiga91}
{Manteiga}, M., {Martinez-Roger}, C., {Morales}, C., \& {Sabau}, L. 1991,
  \aaps, 87, 419

\bibitem[{{M{\"a}vers}(1940)}]{mavers40}
{M{\"a}vers}, F.-W. 1940, Astronomische Nachrichten, 270, 201

\bibitem[{{McCarthy} \& {O'Sullivan}(1969)}]{mccarthy69}
{McCarthy}, M.~F. \& {O'Sullivan}, S. 1969, Ricerche Astronomiche, 7, 483

\bibitem[{{McNamara} \& {Sanders}(1977)}]{mcnamara77}
{McNamara}, B.~J. \& {Sanders}, W.~L. 1977, \aaps, 30, 45

\bibitem[{{Meadows}(1961)}]{meadows61}
{Meadows}, A.~J. 1961, \apj, 133, 907

\bibitem[{{Mendoza V.} \& {Gomez}(1980)}]{mendoza80}
{Mendoza V.}, E.~E. \& {Gomez}, T. 1980, \mnras, 190, 623

\bibitem[{{Mermilliod}(1995)}]{mermilliod95}
{Mermilliod}, J.-C. 1995, in Information, On-Line Data in Astronomy, ed.
  D.~{Egret} \& M.~A. {Albrecht}, 127

\bibitem[{{Mermilliod} \& {Maeder}(1986)}]{mermilliod86}
{Mermilliod}, J.-C. \& {Maeder}, A. 1986, \aap, 158, 45

\bibitem[{{Mermilliod} {et~al.}(2008){Mermilliod}, {Platais}, {James},
  {Grenon}, \& {Cargile}}]{mermilliod08}
{Mermilliod}, J.-C., {Platais}, I., {James}, D.~J., {Grenon}, M., \& {Cargile},
  P.~A. 2008, \aap, 485, 95

\bibitem[{{Mermilliod} {et~al.}(1997){Mermilliod}, {Turon}, {Robichon},
  {Arenou}, \& {Lebreton}}]{mermil97}
{Mermilliod}, J.-C., {Turon}, C., {Robichon}, N., {Arenou}, F., \& {Lebreton},
  Y. 1997, in ESA Special Publication, Vol. 402, Hipparcos - Venice '97, ed.
  R.~M. {Bonnet}, E.~{H{\o}g}, P.~L. {Bernacca}, L.~{Emiliani}, A.~{Blaauw},
  C.~{Turon}, J.~{Kovalevsky}, L.~{Lindegren}, H.~{Hassan}, M.~{Bouffard},
  B.~{Strim}, D.~{Heger}, M.~A.~C. {Perryman}, \& L.~{Woltjer}, 643--650

\bibitem[{{Mitchell} \& {Johnson}(1957)}]{mitchel57}
{Mitchell}, R.~I. \& {Johnson}, H.~L. 1957, \apj, 125, 414

\bibitem[{{Mohan} \& {Sagar}(1985)}]{mohan85}
{Mohan}, V. \& {Sagar}, R. 1985, \mnras, 213, 337

\bibitem[{{Munari} {et~al.}(2004){Munari}, {Dallaporta}, {Siviero}, {Soubiran},
  {Fiorucci}, \& {Girard}}]{munar04}
{Munari}, U., {Dallaporta}, S., {Siviero}, A., {et~al.} 2004, \aap, 418, L31

\bibitem[{Narayanan \& Gould(1999)}]{naray99}
Narayanan, V.~K. \& Gould, A. 1999, \apj, 523, 328

\bibitem[{{Naylor} {et~al.}(2002){Naylor}, {Totten}, {Jeffries}, {Pozzo},
  {Devey}, \& {Thompson}}]{naylor02}
{Naylor}, T., {Totten}, E.~J., {Jeffries}, R.~D., {et~al.} 2002, \mnras, 335,
  291

\bibitem[{{Nicolet}(1981)}]{nicol81}
{Nicolet}, B. 1981, \aap, 104, 185

\bibitem[{{Nissen}(1970{\natexlab{a}})}]{nissen70b}
{Nissen}, P.~E. 1970{\natexlab{a}}, \aap, 8, 476

\bibitem[{{Nissen}(1970{\natexlab{b}})}]{nissen70a}
{Nissen}, P.~E. 1970{\natexlab{b}}, \aap, 6, 138

\bibitem[{{Nissen}(1988)}]{nissen88}
{Nissen}, P.~E. 1988, \aap, 199, 146

\bibitem[{{O'Mullane} {et~al.}(2007){O'Mullane}, {Lammers}, {Bailer-Jones},
  {Bastian}, {Brown}, {Drimmel}, {Eyer}, {Huc}, {Katz}, {Lindegren},
  {Pourbaix}, {Luri}, {Torra}, {Mignard}, \& {van Leeuwen}}]{omullane07}
{O'Mullane}, W., {Lammers}, U., {Bailer-Jones}, C., {et~al.} 2007, in
  Astronomical Society of the Pacific Conference Series, Vol. 376, Astronomical
  Data Analysis Software and Systems XVI, ed. R.~A. {Shaw}, F.~{Hill}, \& D.~J.
  {Bell}, 99--108

\bibitem[{Pan {et~al.}(2004)Pan, Shao, \& Kulkarni}]{pan04}
Pan, X., Shao, M., \& Kulkarni, S.~R. 2004, Nature, 427, 326

\bibitem[{{Patenaude}(1978)}]{patenaude78}
{Patenaude}, M. 1978, \aap, 66, 225

\bibitem[{{Percival} {et~al.}(2005){Percival}, {Salaris}, \&
  {Groenewegen}}]{perciv05}
{Percival}, S.~M., {Salaris}, M., \& {Groenewegen}, M.~A.~T. 2005, \aap, 429,
  887

\bibitem[{Percival {et~al.}(2003)Percival, Salaris, \& Kilkenny}]{perciv03}
Percival, S.~M., Salaris, M., \& Kilkenny, D. 2003, \aap, 400, 541

\bibitem[{{Perez} {et~al.}(1987){Perez}, {The}, \& {Westerlund}}]{perez87}
{Perez}, M.~R., {The}, P.~S., \& {Westerlund}, B.~E. 1987, \pasp, 99, 1050

\bibitem[{{Perry} \& {Hill}(1969)}]{perry69}
{Perry}, C.~L. \& {Hill}, G. 1969, \aj, 74, 899

\bibitem[{{Perry} {et~al.}(1978){Perry}, {Walter}, \& {Crawford}}]{perry78}
{Perry}, C.~L., {Walter}, D.~K., \& {Crawford}, D.~L. 1978, \pasp, 90, 81

\bibitem[{{Perryman} {et~al.}(1998){Perryman}, {Brown}, {Lebreton}, {Gomez},
  {Turon}, {de Strobel}, {Mermilliod}, {Robichon}, {Kovalevsky}, \&
  {Crifo}}]{perry98}
{Perryman}, M.~A.~C., {Brown}, A.~G.~A., {Lebreton}, Y., {et~al.} 1998, \aap,
  331, 81

\bibitem[{{Perryman} {et~al.}(1997){Perryman}, {Lindegren}, {Kovalevsky}, Hoeg,
  {Bastian}, {Bernacca}, {Cr{\' e}z{\' e}}, {Donati}, {Grenon}, {van Leeuwen},
  {van der Marel}, {Mignard}, {Murray}, {Le Poole}, {Schrijver}, {Turon},
  {Arenou}, {Froeschl{\' e}}, \& {Petersen}}]{perry97L}
{Perryman}, M.~A.~C., {Lindegren}, L., {Kovalevsky}, J., {et~al.} 1997, \aap,
  323, L49

\bibitem[{{Petrie} \& {Heard}(1969)}]{petrie69}
{Petrie}, R.~M. \& {Heard}, J.~F. 1969, Publications of the Dominion
  Astrophysical Observatory Victoria, 13, 329

\bibitem[{{Pinsonneault} {et~al.}(1998){Pinsonneault}, {Stauffer}, {Soderblom},
  {King}, \& {Hanson}}]{pinso98}
{Pinsonneault}, M.~H., {Stauffer}, J., {Soderblom}, D.~R., {King}, J.~R., \&
  {Hanson}, R.~B. 1998, \apj, 504, 170

\bibitem[{{Pinsonneault} {et~al.}(2004){Pinsonneault}, {Terndrup}, {Hanson}, \&
  {Stauffer}}]{pinsonn04}
{Pinsonneault}, M.~H., {Terndrup}, D.~M., {Hanson}, R.~B., \& {Stauffer}, J.~R.
  2004, \apj, 600, 946

\bibitem[{{Platais}(1994)}]{platais94}
{Platais}, I. 1994, Bulletin d'Information du Centre de Donnees Stellaires, 44,
  9

\bibitem[{{Platais} {et~al.}(2001){Platais}, {Kozhurina-Platais}, {Barnes},
  {Girard}, {Demarque}, {van Altena}, {Deliyannis}, \& {Horch}}]{platais01}
{Platais}, I., {Kozhurina-Platais}, V., {Barnes}, S., {et~al.} 2001, \aj, 122,
  1486

\bibitem[{{Platais}(1984)}]{platais84}
{Platais}, I.~K. 1984, Soviet Astronomy Letters, 10, 84

\bibitem[{{Prisinzano} {et~al.}(2003){Prisinzano}, {Micela}, {Sciortino}, \&
  {Favata}}]{prisin03}
{Prisinzano}, L., {Micela}, G., {Sciortino}, S., \& {Favata}, F. 2003, \aap,
  404, 927

\bibitem[{{Prosser}(1993)}]{prosser93}
{Prosser}, C.~F. 1993, \aj, 105, 1441

\bibitem[{{Prosser} \& {Giampapa}(1994)}]{prosser94}
{Prosser}, C.~F. \& {Giampapa}, M.~S. 1994, \aj, 108, 964

\bibitem[{Robichon {et~al.}(1999)Robichon, Arenou, Mermilliod, \&
  Turon}]{robic99}
Robichon, N., Arenou, F., Mermilliod, J.~C., \& Turon, C. 1999, \aap, 345, 471

\bibitem[{{Rojo Arellano} {et~al.}(1997){Rojo Arellano}, {Pena}, \&
  {Gonzalez}}]{rojo97}
{Rojo Arellano}, E., {Pena}, J.~H., \& {Gonzalez}, D. 1997, \aaps, 123, 25

\bibitem[{{R{\"o}ser} \& {Bastian}(1994)}]{roeser94}
{R{\"o}ser}, S. \& {Bastian}, U. 1994, \aap, 285, 875

\bibitem[{{Rufener}(1988)}]{rufener88}
{Rufener}, F. 1988, {Catalogue of stars measured in the Geneva Observatory
  photometric system : 4 : 1988} (Sauverny: Observatoire de Geneve, 1988)

\bibitem[{{Sanders}(1973)}]{sanders73}
{Sanders}, W.~L. 1973, \aaps, 9, 213

\bibitem[{{Sanders} \& {van Altena}(1972)}]{sanders72}
{Sanders}, W.~L. \& {van Altena}, W.~F. 1972, \aap, 17, 193

\bibitem[{{Schaifers} {et~al.}(1982){Schaifers}, {Voigt}, {Landolt},
  {Boernstein}, \& {Hellwege}}]{schaifers82}
{Schaifers}, K., {Voigt}, H.~H., {Landolt}, H., {Boernstein}, R., \&
  {Hellwege}, K.~H. 1982, {Astronomy and Astrophysics. C: Interstellar Matter,
  Galaxy, Universe} (Landolt-Boernstein: Numerical Data and functional
  Relationships in Science and Technology.~New Series, Berlin: Springer)

\bibitem[{{Schmidt}(1976)}]{schmidt76}
{Schmidt}, E.~G. 1976, \pasp, 88, 63

\bibitem[{{Smyth} \& {Nandy}(1962)}]{smyth62}
{Smyth}, M.~J. \& {Nandy}, K. 1962, Publications of the Royal Observatory of
  Edinburgh, 3, 24

\bibitem[{{Snowden}(1975)}]{snowden75}
{Snowden}, M.~S. 1975, \pasp, 87, 721

\bibitem[{{Snowden}(1976)}]{snowden76}
{Snowden}, M.~S. 1976, \pasp, 88, 174

\bibitem[{Soderblom {et~al.}(1998)Soderblom, King, Hanson, Jones, Fischer, \&
  Stauffer}]{soder98}
Soderblom, D.~R., King, J.~R., Hanson, R.~B., {et~al.} 1998, \apj, 504, 192

\bibitem[{{Soderblom} {et~al.}(2005){Soderblom}, {Nelan}, {Benedict},
  {McArthur}, {Ramirez}, {Spiesman}, \& {Jones}}]{soder05}
{Soderblom}, D.~R., {Nelan}, E., {Benedict}, G.~F., {et~al.} 2005, \aj, 129,
  1616

\bibitem[{{Southworth} {et~al.}(2005){Southworth}, {Maxted}, \&
  {Smalley}}]{southw05}
{Southworth}, J., {Maxted}, P.~F.~L., \& {Smalley}, B. 2005, in IAU Colloq.
  196: Transits of Venus: New Views of the Solar System and Galaxy, ed. D.~W.
  {Kurtz}, 361--376

\bibitem[{Stello \& Nissen(2001)}]{stello01}
Stello, D. \& Nissen, P.~E. 2001, \aap, 374, 105

\bibitem[{{Stock}(1984)}]{stock84}
{Stock}, J. 1984, Revista Mexicana de Astronomia y Astrofisica, 9, 127

\bibitem[{{Str{\"o}mgren}(1966)}]{stroemg66}
{Str{\"o}mgren}, B. 1966, \araa, 4, 433

\bibitem[{{Sung} {et~al.}(2002){Sung}, {Bessell}, {Lee}, \& {Lee}}]{sung02}
{Sung}, H., {Bessell}, M.~S., {Lee}, B.-W., \& {Lee}, S.-G. 2002, \aj, 123, 290

\bibitem[{{Tayler}(1954)}]{tayler54}
{Tayler}, R.~J. 1954, \apj, 120, 332

\bibitem[{{Tayler}(1956)}]{tayler56}
{Tayler}, R.~J. 1956, \mnras, 116, 25

\bibitem[{{The} {et~al.}(1980){The}, {Bakker}, \& {Antalova}}]{the80}
{The}, P.~S., {Bakker}, R., \& {Antalova}, A. 1980, \aaps, 41, 93

\bibitem[{{Trimble} {et~al.}(2006){Trimble}, {Aschwanden}, \&
  {Hansen}}]{trimble06}
{Trimble}, V., {Aschwanden}, M.~J., \& {Hansen}, C.~J. 2006, \pasp, 118, 947

\bibitem[{{Trumpler}(1928)}]{trumpler28}
{Trumpler}, R. 1928, \pasp, 40, 265

\bibitem[{{Turon} {et~al.}(1992){Turon}, {Gomez}, {Crifo}, {Creze}, {Perryman},
  {Morin}, {Arenou}, {Nicolet}, {Chareton}, \& {Egret}}]{turon92}
{Turon}, C., {Gomez}, A., {Crifo}, F., {et~al.} 1992, \aap, 258, 74

\bibitem[{{Upgren} {et~al.}(1979){Upgren}, {Weis}, \& {Deluca}}]{upgren79}
{Upgren}, A.~R., {Weis}, E.~W., \& {Deluca}, E.~E. 1979, \aj, 84, 1586

\bibitem[{{Valls-Gabaud}(2007)}]{vallsg07}
{Valls-Gabaud}, D. 2007, in IAU Symposium, Vol. 240, IAU Symposium, ed. W.~I.
  {Hartkopf}, E.~F. {Guinan}, \& P.~{Harmanec}, 281--289

\bibitem[{{van Altena}(1966)}]{altena66}
{van Altena}, W.~F. 1966, \aj, 71, 482

\bibitem[{van~der Marel(1988)}]{vdmar88}
van~der Marel, H. 1988, PhD thesis, Technische Universiteit Delft

\bibitem[{van~der Marel \& Petersen(1992)}]{vdmar92}
van~der Marel, H. \& Petersen, C.~S. 1992, \aap, 258, 60

\bibitem[{{van Leeuwen}(1980)}]{fvl80}
{van Leeuwen}, F. 1980, in IAU Symp. 85: Star Formation, 157--162

\bibitem[{{van Leeuwen}(1983)}]{fvl83}
{van Leeuwen}, F. 1983, Ph.D.~Thesis

\bibitem[{van Leeuwen(1999{\natexlab{a}})}]{fvl99a}
van Leeuwen, F. 1999{\natexlab{a}}, \aap, 341, L71

\bibitem[{van Leeuwen(1999{\natexlab{b}})}]{fvl99b}
van Leeuwen, F. 1999{\natexlab{b}}, in Harmonizing cosmic distance scales in a
  post-Hipparcos era, ed. D.~Egret \& A.~Heck, Vol. 167 (PASPC), 52--71

\bibitem[{{van Leeuwen}(2005{\natexlab{a}})}]{fvl05a}
{van Leeuwen}, F. 2005{\natexlab{a}}, \aap, 439, 805

\bibitem[{{van Leeuwen}(2005{\natexlab{b}})}]{fvl05c}
{van Leeuwen}, F. 2005{\natexlab{b}}, in IAU Colloq. 196: Transits of Venus:
  New Views of the Solar System and Galaxy, 347--360

\bibitem[{{van Leeuwen}(2007{\natexlab{a}})}]{fvl07}
{van Leeuwen}, F. 2007{\natexlab{a}}, {Hipparcos, the new reduction of the raw
  data} (Dordrecht: Springer)

\bibitem[{{van Leeuwen}(2007{\natexlab{b}})}]{fvl07Val}
{van Leeuwen}, F. 2007{\natexlab{b}}, \aap, 474, 653

\bibitem[{{van Leeuwen}(2008)}]{fvl08}
{van Leeuwen}, F. 2008, in IAU Symposium, Vol. 248, IAU Symposium, 82--88

\bibitem[{{van Leeuwen} \& {Alphenaar}(1982)}]{fvl82}
{van Leeuwen}, F. \& {Alphenaar}, P. 1982, The Messenger, 28, 15

\bibitem[{{van Leeuwen} {et~al.}(1986){van Leeuwen}, {Alphenaar}, \&
  {Brand}}]{fvl86}
{van Leeuwen}, F., {Alphenaar}, P., \& {Brand}, J. 1986, \aaps, 65, 309

\bibitem[{{van Leeuwen} {et~al.}(1987){van Leeuwen}, {Alphenaar}, \&
  {Meys}}]{fvl87}
{van Leeuwen}, F., {Alphenaar}, P., \& {Meys}, J.~J.~M. 1987, \aaps, 67, 483

\bibitem[{van Leeuwen \& Evans(1998)}]{vLDWE}
van Leeuwen, F. \& Evans, D.~W. 1998, \aap, 323, 157

\bibitem[{van Leeuwen \& Fantino(2003)}]{paper4}
van Leeuwen, F. \& Fantino, E. 2003, \ssr, 108, 537

\bibitem[{{van Schewick}(1966)}]{schewick66}
{van Schewick}, H. 1966, Veroeffentlichungen des Astronomisches Institute der
  Universitaet Bonn, 74, 1

\bibitem[{{Vandenberg}(1985)}]{vandenberg85}
{Vandenberg}, D.~A. 1985, \apjs, 58, 711

\bibitem[{{Vasilevskis}(1955)}]{vasil55}
{Vasilevskis}, S. 1955, \aj, 60, 384

\bibitem[{{Vasilevskis} {et~al.}(1958){Vasilevskis}, {Klemola}, \&
  {Preston}}]{vasil58}
{Vasilevskis}, S., {Klemola}, A., \& {Preston}, G. 1958, \aj, 63, 387

\bibitem[{{Vasilevskis} {et~al.}(1965){Vasilevskis}, {Sanders}, \&
  {Balz}}]{vasil65}
{Vasilevskis}, S., {Sanders}, W.~L., \& {Balz}, A.~G.~A. 1965, \aj, 70, 797

\bibitem[{{Verschoor} \& {van Genderen}(1983)}]{verschoor83}
{Verschoor}, J.~N. \& {van Genderen}, A.~M. 1983, \aaps, 53, 419

\bibitem[{{Voigt}(1965)}]{voigt65}
{Voigt}, H.~H. 1965, {Landolt-B{\"o}rnstein: Numerical Data and Functional
  Relationships in Science and Technology - New Series '' Gruppe/Group 6}
  (Landolt-Bornstein)

\bibitem[{{Walker}(1956)}]{walker56}
{Walker}, M.~F. 1956, \apjs, 2, 365

\bibitem[{{Wallerstein} {et~al.}(1963){Wallerstein}, {Westbrooke}, \&
  {Hannibal}}]{wallers63}
{Wallerstein}, G., {Westbrooke}, W., \& {Hannibal}, D. 1963, \pasp, 75, 522

\bibitem[{{Weaver}(1953)}]{weaver53}
{Weaver}, H.~F. 1953, \apj, 117, 366

\bibitem[{{Westerlund} {et~al.}(1988){Westerlund}, {Lundgren}, {Pettersson},
  {Garnier}, \& {Breysacher}}]{westerl88}
{Westerlund}, B.~E., {Lundgren}, K., {Pettersson}, B., {Garnier}, R., \&
  {Breysacher}, J. 1988, \aaps, 76, 101

\bibitem[{{Whiteoak}(1961)}]{whiteoak61}
{Whiteoak}, J.~B. 1961, \mnras, 123, 245

\bibitem[{{Williams}(1978)}]{williams78}
{Williams}, P.~M. 1978, \mnras, 183, 49

\bibitem[{{Wizinowich} \& {Garrison}(1982)}]{wizinow82}
{Wizinowich}, P. \& {Garrison}, R.~F. 1982, \aj, 87, 1390

\bibitem[{{Young}(1978)}]{young78}
{Young}, A. 1978, \pasp, 90, 144

\bibitem[{{Zentelis}(1983)}]{zentelis83}
{Zentelis}, N. 1983, \aaps, 53, 445

\bibitem[{{Zwahlen} {et~al.}(2004){Zwahlen}, {North}, {Debernardi}, {Eyer},
  {Galland}, {Groenewegen}, \& {Hummel}}]{zwahl04}
{Zwahlen}, N., {North}, P., {Debernardi}, Y., {et~al.} 2004, \aap, 425, L45

\end{thebibliography}
\end{document}